\documentclass[floatfix,twocolumn,showpacs,preprintnumbers,amsmath,amssymb,pra,superscriptaddress,longbibliography]{revtex4-1}
\usepackage{color}
\usepackage[usenames,dvipsnames,svgnames,table]{xcolor}
\usepackage[colorlinks=true,linkcolor=blue,urlcolor=blue,citecolor=blue]{hyperref}
\usepackage{mathtools}
\usepackage{graphicx}
\usepackage{dcolumn}
\usepackage{array}
\usepackage{lipsum}
\usepackage{bm}
\usepackage{subfigure}
\usepackage{amssymb}
\usepackage{multirow}
\usepackage{tabularx}
\usepackage{amsmath}
\usepackage{braket}
\graphicspath{{plots/}}
 \usepackage{lipsum}
\usepackage{mathrsfs}

\newcommand{\beq}{\begin{equation}}
\newcommand{\eeq}{\end{equation}}
\newcommand{\bea}{\begin{eqnarray}}
\newcommand{\eea}{\end{eqnarray}}

\begin{document}
\title{
Momentum distribution of the Uniform Electron Gas at finite temperature: Effects of spin-polarization
}

\author{Tobias Dornheim}
\email{t.dornheim@hzdr.de}

\affiliation{Center for Advanced Systems Understanding (CASUS), D-02826 G\"orlitz, Germany}
\affiliation{Helmholtz-Zentrum Dresden-Rossendorf (HZDR), D-01328 Dresden, Germany}

\author{Jan Vorberger}
\affiliation{Helmholtz-Zentrum Dresden-Rossendorf (HZDR), D-01328 Dresden, Germany}

\author{Burkhard Militzer}
\affiliation{Department of Earth and Planetary Science, University of California, Berkeley, California 94720, USA}
\affiliation{Department of Astronomy, University of California, Berkeley, California 94720, USA}

\author{Zhandos A.~Moldabekov}

\affiliation{Center for Advanced Systems Understanding (CASUS), D-02826 G\"orlitz, Germany}
\affiliation{Helmholtz-Zentrum Dresden-Rossendorf (HZDR), D-01328 Dresden, Germany}

\begin{abstract}
We carry out extensive direct path integral Monte Carlo (PIMC) simulations of the uniform electron gas (UEG) at finite temperature for different values of the spin-polarization $\xi$. This allows us to unambiguously quantify the impact of spin-effects on the momentum distribution function $n(\mathbf{k})$ and related properties. We find that interesting physical effects like the interaction-induced increase in the occupation of the zero-momentum state $n(\mathbf{0})$ substantially depend on $\xi$.
Our results further advance the current understanding of the UEG as a fundamental model system, and are of practical relevance for the description of transport properties of warm dense matter in an external magnetic field. 

All PIMC results are freely available online and can be used as a benchmark for the development of new methods and applications.
\end{abstract}

\maketitle

\section{Introduction\label{sec:introduction}}

The uniform electron gas (UEG)~\cite{loos,quantum_theory} is one of the most fundamental model systems in theoretical physics and related disciplines. In particular, the UEG has been pivotal for the development for groundbreaking concepts such as Fermi liquid theory~\cite{quantum_theory}, the Bardeen-Cooper-Schrieffer (BCS) theory of superconductivity~\cite{Bardeen_PhysRev_1957}, and the quasi-particle picture of collective excitations~\cite{pines}.
Despite its apparent simplicity, it exhibits a wealth of interesting physical effects such as Wigner crystallization~\cite{Drummond_PRB_2004,Ceperley_Wigner_PRL}, a potentially incipient excitonic mode at low density~\cite{Takada_PRL_2002,Takada_PRB_2016,dornheim_dynamic}, and the possibility of a charge-density or spin-density wave~\cite{schweng,quantum_theory,dornheim_electron_liquid}.

From a theoretical perspective, the accurate description of the UEG is quite challenging and requires the application of advanced numerical techniques. Following the seminal studies by Ceperley and Alder~\cite{Ceperley_UEG_1978,Ceperley_Alder_PRL_1980}, the \emph{ab initio} quantum Monte Carlo (QMC) approach~\cite{Foulkes_RevModPhys_2001} has emerged as the method of choice for the description of the UEG and has allowed for unprecedented insights into important properties like the (exchange--correlation) energy~\cite{Ceperley_UEG_1978,Ceperley_Alder_PRL_1980,Ortiz_Ballone_PRB_1994,Ortiz_Harris_Ballone_PRL_1999,Spink_PRB_2013,Holzmann_PRL_2011}, static structure factor~\cite{Ortiz_Ballone_PRB_1994,Spink_PRB_2013}, or linear response properties like the electronic local field correction~\cite{moroni,moroni2,bowen2}. These data have been subsequently used as input for different analytical parametrizations~\cite{vwn,Perdew_Zunger_PRB_1981,Perdew_Wang_PRB_1992,Perdew_Wang_PDF_1992,farid,Gori_Giorgi_PRB_2000,Gori_Giorgi_PRB_2001,Gori_Giorgi_PRB_2002,Ziesche_Gori_Giorgi_PRB_2002,cdop}, which form the basis for many theoretical studies. Most notably, the accurate parametrization of the exchange--correlation energy of the UEG has facilitated the possibly unrivaled success of density functional theory regarding the description of real materials~\cite{Jones_RevModPhys_2015}.

Over the last decades, there has emerged a growing interest in so-called \emph{warm dense matter} (WDM), an exotic state characterized by extreme temperatures and densities~\cite{fortov_review,Benuzzi_Mounaix_2014}. In nature, these conditions occur in astrophysical objects such as giant-planet interiors~\cite{Nettelmann2008,Militzer_2008,militzer1,Benuzzi_Mounaix_2014}, brown dwarfs~\cite{saumon1,becker}, and neutron star crusts~\cite{Chamel2008}. Furthermore, WDM plays an important role in cutting-edge technological applications such as inertial confinement fusion~\cite{hu_ICF}, hot-electron chemistry~\cite{Brongersma2015}, and the discovery of novel materials~\cite{Kraus2016,Kraus2017,Lazicki2021}.
From a theory point of view, WDM is defined by two characteristic parameters that are simultaneously of the order of unity: 1) the density parameter (also known as the Wigner-Seitz radius or quantum coupling parameter~\cite{Ott2018}) $r_s=\overline{r}/a_\textnormal{B}$, with $\overline{r}$ and $a_\textnormal{B}$ being the average inter-particle distance and first Bohr radius, and 2) the reduced temperature $\theta=k_\textnormal{B}T/E_\textnormal{F}$, with $E_\textnormal{F}$ being the usual Fermi energy, c.f.~Eq.~(\ref{eq:Efermi}).

Phenomenologically, these conditions manifest as the highly nontrivial interplay of quantum effects, Coulomb coupling, and thermal excitations, which renders WDM theory a most formidable challenge~\cite{new_POP,wdm_book}. Coming back to the UEG, it has become clear that previous ground-state descriptions of the UEG are not sufficient for applications in the WDM regime~\cite{karasiev_importance,kushal,Sjostrom_PRB_2014}.
This has sparked a surge of developments in the field of fermionic QMC simulations at finite temperatures~\cite{Brown_PRL_2013,Blunt_PRB_2014,Schoof_PRL_2015,Malone_JCP_2015,Dornheim_NJP_2015,dornheim_jcp,Groth_PRB_2016,Malone_PRL_2016,Rubenstein_auxiliary_finite_T,lee2020phaseless,dornheim_prl,Yilmaz_JCP_2020} (see Ref.~\cite{dornheim_POP} for a review), which has culminated in the first accurate parametrizations of the exchange--correlation free energy of the UEG at WDM conditions~\cite{ksdt,groth_prl,review,status}. In particular, this allows for thermal density functional theory simulations on the level of the local density approximation~\cite{karasiev_importance,kushal}, and constitutes the basis for the development of more sophisticated functionals that explicitly take into account the temperature~\cite{Karasiev_PRL_2018,Karasiev_PRB_2020}.
Further progress on the UEG at WDM conditions includes the characterization of linear-response properties such as the static local field correction~\cite{dynamic_folgepaper,dornheim_ML,Dornheim_PRL_2020_ESA,Dornheim_PRB_2021}, numerical and theoretical results for the nonlinear electronic density response~\cite{Dornheim_PRL_2020,Dornheim_PRR_2021}, and even the study of dynamic quantities based on the analytic continuation of imaginary-time correlation functions~\cite{dornheim_dynamic,dynamic_folgepaper,Dornheim_PRE_2020,Hamann_CPP_2020,Hamann_PRB_2020}.

Another fundamental property of the UEG is the momentum distribution function $n(\mathbf{k})$, which is highly important for the development of accurate models for the computation of different transport properties of WDM~\cite{rightley2021kinetic}. At finite temperatures, the first results have been presented by Militzer \emph{et al.}~\cite{Militzer_PRL_2002,Militzer_momentum_HEDP_2019} based on approximate restricted path integral Monte Carlo (PIMC) simulations. More recently, the momentum distribution of the unpolarized UEG has been revisited by Hunger~\emph{et al.}~\cite{Hunger_PRE_2021}, who have carried out exact configuration PIMC simulations at high densities ($r_s\lesssim1$), and by Dornheim \emph{et al.}~\cite{dornheim2021ab} on the basis of exact direct PIMC simulations going from metallic densities $r_s\sim2$ to the strongly coupled electron liquid regime ($r_s\sim10-100$). More specifically, these studies have addressed interesting phenomena such as the interaction-induced increase in the occupation of the zero-momentum state~\cite{Militzer_PRL_2002,Kraeft_PRE_2002}, the related negative exchange--correlation part to the kinetic energy $K_\textnormal{xc}$, and the algebraic tail in the limit of large momenta $k$, cf.~Eq.~(\ref{eq:k8-g0}) below.

At the same time, a thorough investigation of the dependence of the momentum distribution function on the spin polarization $\xi=(N^\uparrow - N^\downarrow)/N$ (with $N^\uparrow$, $N^\downarrow$, and $N$ being the number of spin-up electrons, spin-down electrons, and all electrons, respectively) has yet been missing. In the present work, we aim to fill this gap by presenting extensive new direct PIMC results for $n(\mathbf{k})$ for different values of $\xi$. Firstly, we mention that such an investigation is interesting in its own right, and helps to significantly advance our current understanding of the UEG as a fundamental model system~\cite{review,status}.
Secondly, spin-polarized systems are ubiquitous in quantum chemistry for atoms and molecules like oxygen, and the properties of the spin-polarized UEG are, consequently, important for the construction of exchange--correlation functionals for density functional theory.
Finally, the impact of the spin-polarization is of central importance for the properties of WDM in an external magnetic field.
In the case of strong electronic degeneracy ($\theta\ll1$) in a non-quantizing magnetic field, the effect of the spin polarization on transport properties is negligible as it scales as $\xi\sim \mathcal{O}(\theta)$ \cite{Lifshitz}. Similarly, quantum effects are not relevant due to strong thermal excitations in the regime of very high temperatures, $\theta\gg1$. In stark contrast, the effects of the spin polarization  play an important role precisely in the WDM regime ($\theta\sim1$) in an external magnetic field, as both aforementioned conditions do not apply here. More specifically, the condition for a non-quantizing magnetic field follows from the requirement that  the electron cyclotron energy $\hbar \omega_c$ is much smaller than the characteristic quantum kinetic energy $\sqrt{E_F^2+T^2}$ \cite{Haensel}. From this condition, one can find that the range  of non-quantizing magnetic fields is given by $B/B_0\ll \frac{18.4}{r_s^2}\sqrt{\theta^2+1}$ (where $B_0\simeq 2.25\times 10^5 {\rm T}$). For example, at  $\theta \sim 1$ and $r_s \sim 1$, non-quantizing strong magnetic fields $B\sim 10~{\rm T}-10^4~{\rm T}$ can be generated in experiments related to inertial confinement fusion~\cite{Perkins, Appelbe}. Yet, the physical properties of WDM in such strong magnetic fields remain largely unknown. We are, thus, convinced that our new results for the impact of the spin-polarization on the momentum distribution are of direct importance for the future exploration of WDM at these extreme conditions.


The paper is organized as follows: In Sec.~\ref{sec:theory}, we introduce the required theoretical background including the PIMC method (\ref{sec:PIMC}), the corresponding estimation of the momentum distribution function (\ref{sec:nk_theory}), and the spin-resolved reduced system parameters (\ref{sec:parameters}). Sec.~\ref{sec:results} is devoted to the presentation of our new simulation results, starting with PIMC data for the fully spin-polarized case (\ref{sec:spin-polarized}). Subsequently, we extend these consideration to arbitrary values of the spin-polarization in Sec.~\ref{sec:intermediate}. The paper is concluded by a brief summary and discussion in Sec.~\ref{sec:summary}.

\section{Theory\label{sec:theory}}
We assume Hartree atomic units throughout this work.

\subsection{The path integral Monte Carlo method\label{sec:PIMC}}

Throughout this work, we simulate $N=N^\uparrow + N^\downarrow$ spin-restricted electrons in a cubic simulation box of constant volume $V=L^3$, and at a fixed temperature $T=1/\beta$. Further, we restrict ourselves to the case of thermodynamic equilibrium, and the expectation value of a physical observable $\hat O$ is given by
\begin{eqnarray}\label{eq:expectation_value}
\braket{\hat O} = \frac{1}{Z} \textnormal{Tr}\left(
\hat\rho\hat O
\right)\ .
\end{eqnarray}
Here $\hat\rho=\textnormal{exp}(-\beta\hat H)$ denotes the (unnormalized) canonical density operator, and the normalization is given by the corresponding canonical partition function $Z$.
The basic idea of the path integral Monte Carlo method~\cite{cep,Berne_JCP_1982,Takahashi_Imada_PIMC_1984} is the stochastic evaluation of the matrix elements of $\hat\rho$. More specifically, this requires the evaluation of extremely high-dimensional integrals, which is accomplished efficiently using variations of the Metropolis Monte Carlo method~\cite{metropolis}. While being computationally involved, the PIMC method is, in principle, capable to provide a quasi-exact solution of the quantum $N$-body problem of interest. The term \emph{quasi-exact} implies that PIMC estimations of Eq.~(\ref{eq:expectation_value}) can be made arbitrarily accurate in a controlled way when the convergence parameters (typically the number of imaginary-time slices $P$ and the number of Monte Carlo samples $N_\textnormal{MC})$ are increased. A more detailed introduction to the PIMC method is beyond the scope of the current work, and the interested reader is referred to Refs.~\cite{cep,boninsegni1,review}.

An additional problem arises due to the fermionic nature of electrons, which manifests as an anti-symmetry under the exchange of particle coordinates in Eq.~(\ref{eq:expectation_value}). This is the origin of the notorious fermion sign problem~\cite{Loh_sign_problem,troyer,dornheim_sign_problem}, which leads to an exponential increase in computation time with increasing the system size $N$ or decreasing the temperature $T$; see Refs.~\cite{dornheim_sign_problem,dornheim2021fermion} for an accessible topical discussion. A popular approach to circumvent this issue is the application of the fixed-node approximation~\cite{Ceperley1991} (commonly known as \emph{restricted} PIMC or RPIMC), which formally removes the sign problem for simulations in the canonical ensemble. Indeed, the RPIMC method constitutes at present the only QMC approach that is capable to simulate real materials in the WDM regime, e.g., Refs.~\cite{Militzer_PRL_2012,Militzer_PRL_2015,Militzer_PRE_2018}.
Unfortunately, this great advantage comes at the cost of an uncontrolled approximation, as the exact nodes of correlated quantum many-body systems are a-priori unknown. More specifically, Schoof \textit{et al.}~\cite{Schoof_PRL_2015} have shown that RPIMC leads to errors of up to $10\%$ in the description of exchange--correlation properties of electrons in the WDM regime.

For this reason, we do not impose any nodal restriction in the present work. Consequently, our \emph{direct} PIMC simulations are computationally extremely demanding, and we spend up to $\mathcal{O}\left(10^5\right)$ CPUh for a single density--temperature combination in the most challenging regime. Still, the sign problem constitutes the main limitation of our approach, and prevents us from accessing the low-temperature regime ($\theta < 1$) except for very strong coupling.

\subsection{PIMC estimation of the momentum distribution\label{sec:nk_theory}}

The momentum distribution of $N_\sigma$ (with $\sigma\in[\uparrow,\downarrow]$ denoting the spin) electrons is defined as~\cite{Militzer_momentum_HEDP_2019}
\begin{eqnarray}\label{eq:momentum_distribution}
n_\sigma(\mathbf{k}) = \frac{(2\pi)^d}{V} \left<\sum_{l=1}^{N_\sigma}\delta\left({\mathbf{\hat{k}_l}}-\mathbf{k}\right) \right>\ ,
\end{eqnarray}
with the corresponding normalization
\begin{eqnarray}\label{eq:normalization}
\sum_\mathbf{k}n_\sigma(\mathbf{k}) = N_\sigma\ .
\end{eqnarray}
In addition, we mention that Eq.~(\ref{eq:momentum_distribution}) is related to the off-diagonal density matrix in coordinate space $n_\sigma(s)\coloneqq n_\sigma(\mathbf{r},\mathbf{r}')$, with $s=|\mathbf{r}-\mathbf{r}'|$, by
\begin{eqnarray}\label{eq:offdiagonal}
n_\sigma(\mathbf{k}) = \int \textnormal{d}\mathbf{s}\ n_\sigma(\mathbf{s}) e^{-i\mathbf{s}\cdot\mathbf{k}}\ .
\end{eqnarray}

For an ideal (i.e., noninteracting) Fermi system, Eq.~(\ref{eq:momentum_distribution}) is given by the well-known Fermi distribution
\begin{eqnarray}\label{eq:Fermi}
n_0(\mathbf{k}) = \frac{1}{1+\textnormal{exp}\left(\beta(E_\mathbf{k}-\mu)\right)}\ ,
\end{eqnarray}
with $\mu$ being the usual chemical potential~\cite{quantum_theory}, and the ideal dispersion relation
\begin{eqnarray}
E_\mathbf{k} = \frac{\mathbf{k}^2}{2}\ .
\end{eqnarray}

\begin{figure}\centering
\includegraphics[width=0.475\textwidth]{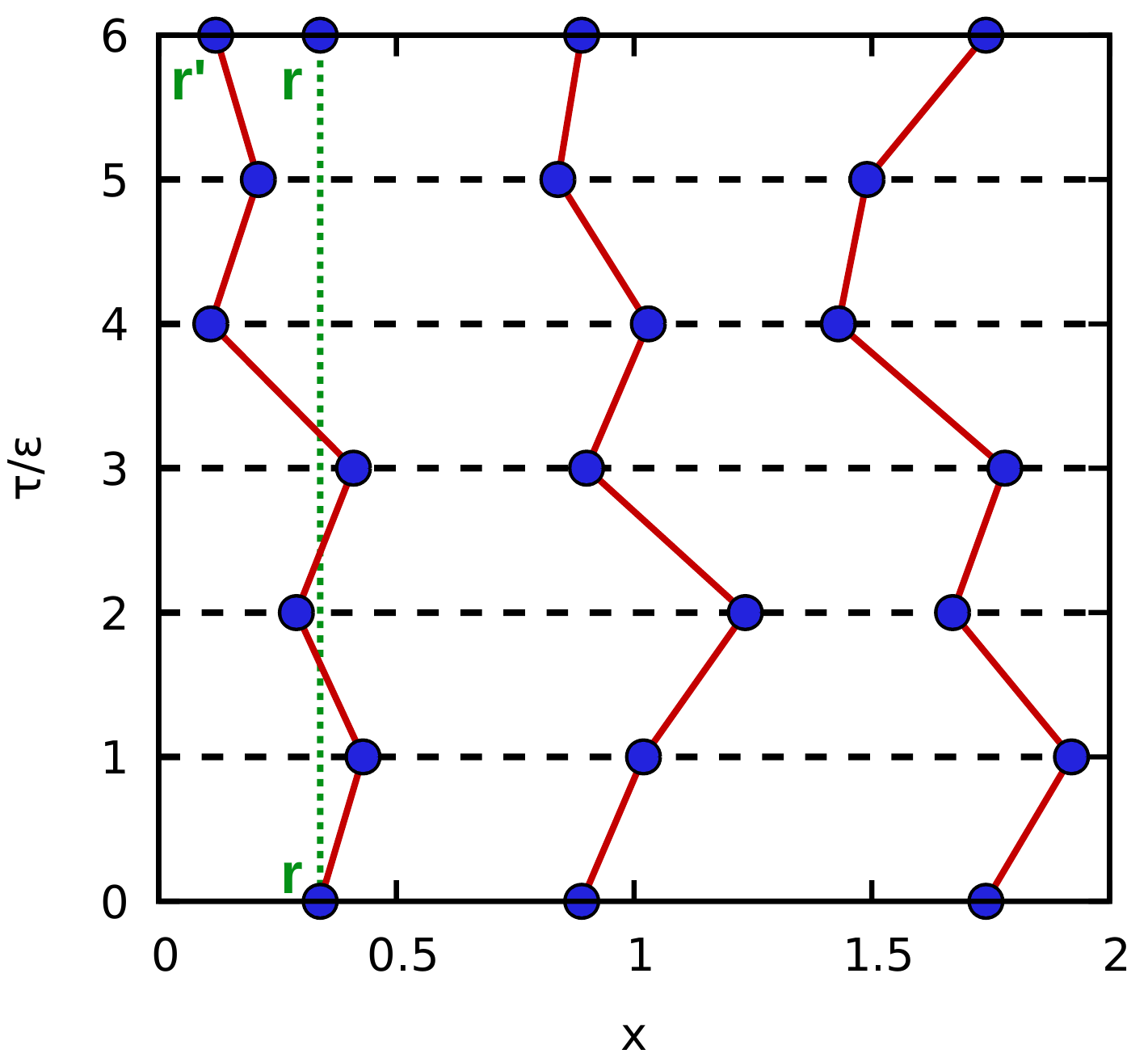}
\caption{\label{fig:illustration}
Schematic illustration of the off-diagonal configuration space $Z_{\mathbf{r},\mathbf{r'};\sigma}$ including a single open trajectory with different start- and end-points $\mathbf{r}$ and $\mathbf{r}'$. Adapted from Ref.~\cite{dornheim2021ab} with the permission of the authors.
}
\end{figure}

Interestingly, the evaluation of Eq.~(\ref{eq:momentum_distribution}) is not straightforward in the PIMC method, as it constitutes an off-diagonal observable in the underlying coordinate representation. More specifically, each particle is represented as a closed path over different coordinates in the imaginary time within the PIMC method. 
In contrast, the estimation of $n(\mathbf{k})$ requires the presence of a single \emph{open} path within the PIMC simulation, thereby resulting in a modified configuration space. This is illustrated in Fig.~\ref{fig:illustration} for a schematic configuration of $N=3$ electrons on $P=6$ imaginary-time slices (with $\epsilon=\beta/P$ being a discretized time step), depicted in the $\tau$-$x$-plane. While the two right-most paths exhibit the same coordinates for $\tau=0$ and $\tau=\beta$, the electronic path on the left is \emph{open} and has different coordinates $\mathbf{r}$ and $\mathbf{r}'$ at its start and end.

The expression for Eq.~(\ref{eq:momentum_distribution}) in the path-integral picture is then given by~\cite{cep,Militzer_momentum_HEDP_2019}
\begin{eqnarray}\label{eq:nk_formula}
n_\sigma(\mathbf{k}) = \frac{1}{V} \frac{Z_{\mathbf{r},\mathbf{r'};\sigma}}{Z} \left<
e^{i\mathbf{k}(\mathbf{r}-\mathbf{r}')}
\right>_{\mathbf{r},\mathbf{r'};\sigma}\ ,
\end{eqnarray}
where the sub-script of the angular brackets indicates this modified configuration space, and $Z_{\mathbf{r},\mathbf{r'};\sigma}$ its corresponding normalization. In practice, we use the extended-ensemble approach presented in Ref.~\cite{dornheim2021ab} which is based on the worm algorithm by Boninsegni \textit{et al.}~\cite{boninsegni1,boninsegni2}. One particular strength of this scheme compared to earlier works~\cite{cep,Militzer_momentum_HEDP_2019} is the possibility to directly compute the normalization of Eq.~(\ref{eq:nk_formula}) without the need for a subsequent fitting of the off-diagonal density matrix $n(\mathbf{r},\mathbf{r}')$ or an artificial imposition of the condition in Eq.~(\ref{eq:normalization}).
The practical implications of this advantage are discussed in Sec.~\ref{sec:spin-polarized} below.

\subsection{Reduced system parameters for arbitrary spin-polarizations\label{sec:parameters}}

To understand the effect of an arbitrary spin-polarization $\xi\in[0,1]$ on physical observables, it is helpful to consider modified, explicitly spin-resolved reduced parameters.
To this end, we introduce the spin-resolved density parameter $r_s^\sigma$ via the relation
\begin{eqnarray}
\frac{4}{3}\pi \left(r_s^\sigma\right)^3 = \frac{V}{N^\sigma}\ ,
\end{eqnarray}
which immediately gives
\begin{eqnarray}\label{eq:rs_spin_resolved}
r_s^\sigma = r_s \left( \frac{N}{N^\sigma} \right)^{1/3}\ .
\end{eqnarray}
Furthermore, Eq.~(\ref{eq:rs_spin_resolved}) can be expressed in terms of the spin-polarization $\xi$ as
\begin{eqnarray}\label{eq:rs_up_down}
r_s^\uparrow &=& r_s \left( \frac{1+\xi}{2} \right)^{-1/3}\ \textnormal{and} \\ \nonumber
r_s^\downarrow &=& r_s \left( \frac{1-\xi}{2} \right)^{-1/3}\ .
\end{eqnarray}

\begin{figure}\centering
\includegraphics[width=0.475\textwidth]{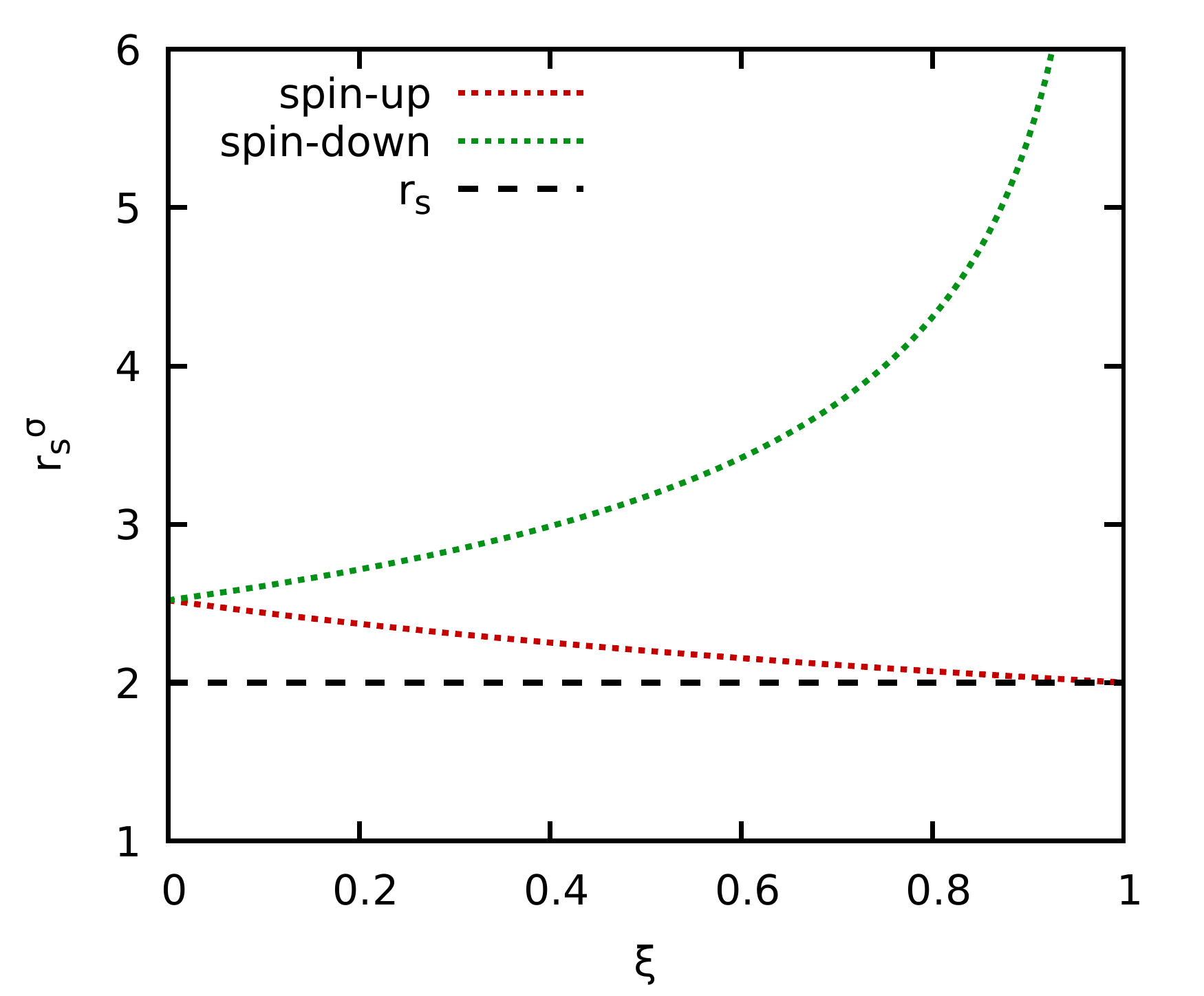}\\\vspace*{-1.02cm}\includegraphics[width=0.475\textwidth]{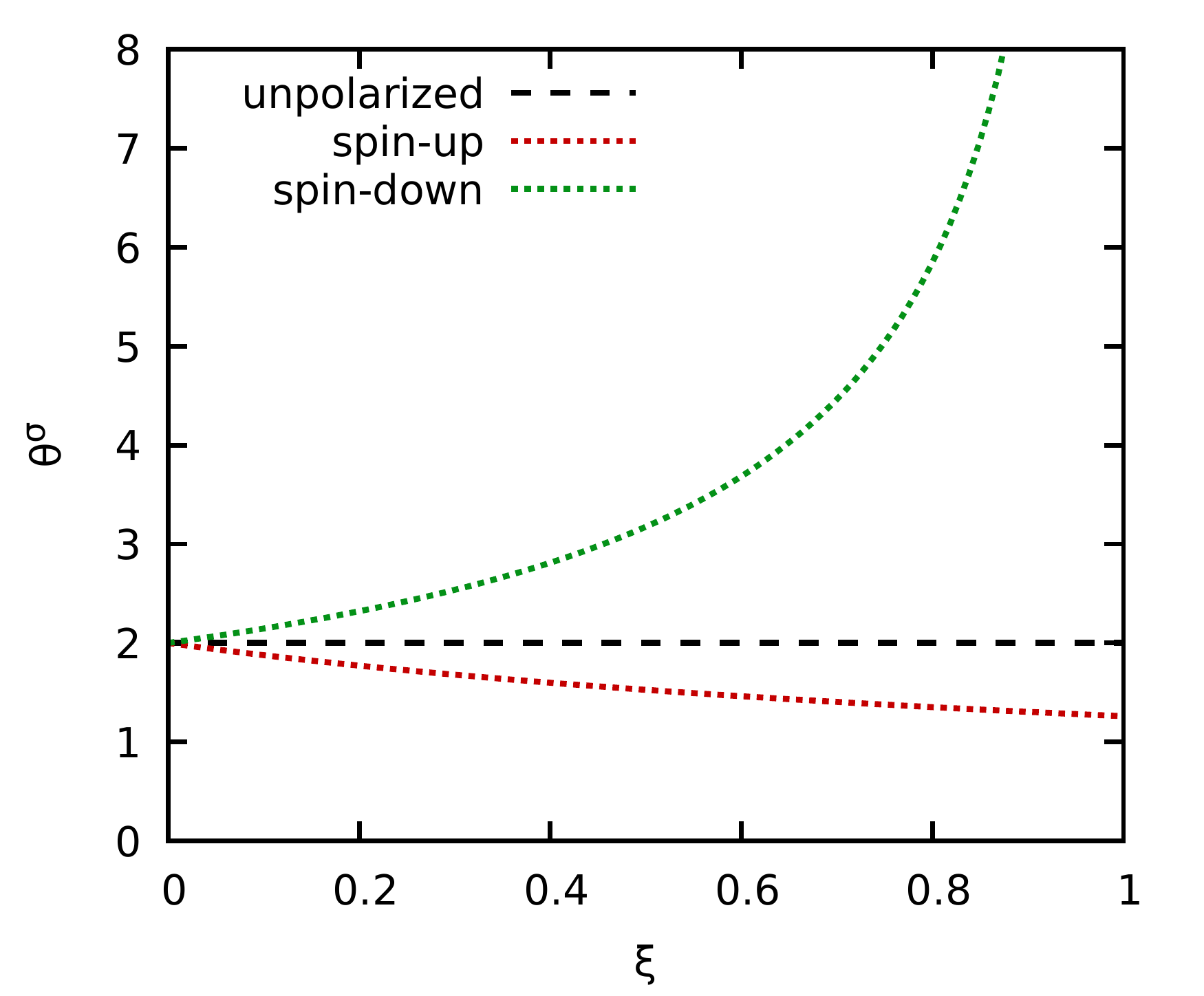}
\caption{\label{fig:xi_scale}
Polarization dependence of the spin-resolved density parameter $r_s^\sigma$ (top panel) and reduced temperature $\theta^\sigma$ (bottom panel) for $r_s=2$ and $\theta=2$ (using the Fermi energy of the unpolarized system as a reference).
}
\end{figure}

The dependence of Eq.~(\ref{eq:rs_up_down}) on $\xi$ is shown in the top panel of Fig.~\ref{fig:xi_scale}, using a total density parameter $r_s=2$ as a reference (dashed black line). For $\xi=0$, the system is fully unpolarized, i.e., $N^\uparrow=N^\downarrow=N/2$, which immediately gives $r_s^\uparrow=r_s^\downarrow=r_s\cdot2^{1/3}$. Upon increasing $\xi$, the fraction of spin-up electrons increases, and, consequently, $r_s^\uparrow$ converges towards the full density parameter $r_s$ in the limit of $\xi=1$. In contrast, the number density of spin-down electrons decreases with $\xi$ and eventually attains zero. Thus, $r_s^\downarrow$ actually diverges towards the fully spin-polarized case.

A second parameter that is relevant to the present study is given by the spin-resolved reduced temperature, which we simply express as a function of $r_s^\sigma$,
\begin{eqnarray}\label{eq:theta_up_down}
\theta^\sigma(\beta) = \frac{1}{\beta E^\sigma_\textnormal{F}(r_s^\sigma)}\ ,
\end{eqnarray}
with $E^\sigma_\textnormal{F}(r_s^\sigma)$ being the Fermi energy for a fully spin-polarized system with the density parameter $r_s^\sigma$,
\begin{eqnarray}\label{eq:Efermi}
E_\textnormal{F}^\sigma(r_s^\sigma) = \frac{(\mathbf{k}_\textnormal{F}^\sigma)^2}{2}\ ,
\end{eqnarray}
with the corresponding spin dependent Fermi wave number
\begin{eqnarray}
k_\textnormal{F}^\sigma = \left(
6\pi^2 n^\sigma
\right)^{1/3}\ .
\end{eqnarray}

The results for Eq.~(\ref{eq:theta_up_down}) are shown in the bottom panel of Fig.~\ref{fig:xi_scale}, using $\theta=2$ as a reference for the unpolarized case. With increasing $\xi$, the number density of the spin-up electrons is increased, which results in a larger Fermi energy. Therefore, the reduced temperature is decreased when $\beta$ (or, equivalently, $T$) are being kept constant.
Conversely, both the number density of the spin-down electrons and the corresponding Fermi energy vanish towards $\xi=1$, which means that $\theta^\downarrow$ diverges.

\section{Results\label{sec:results}}

\subsection{The spin-polarized electron gas\label{sec:spin-polarized}}

\subsubsection{Density dependence}

\begin{figure*}\centering
\includegraphics[width=0.475\textwidth]{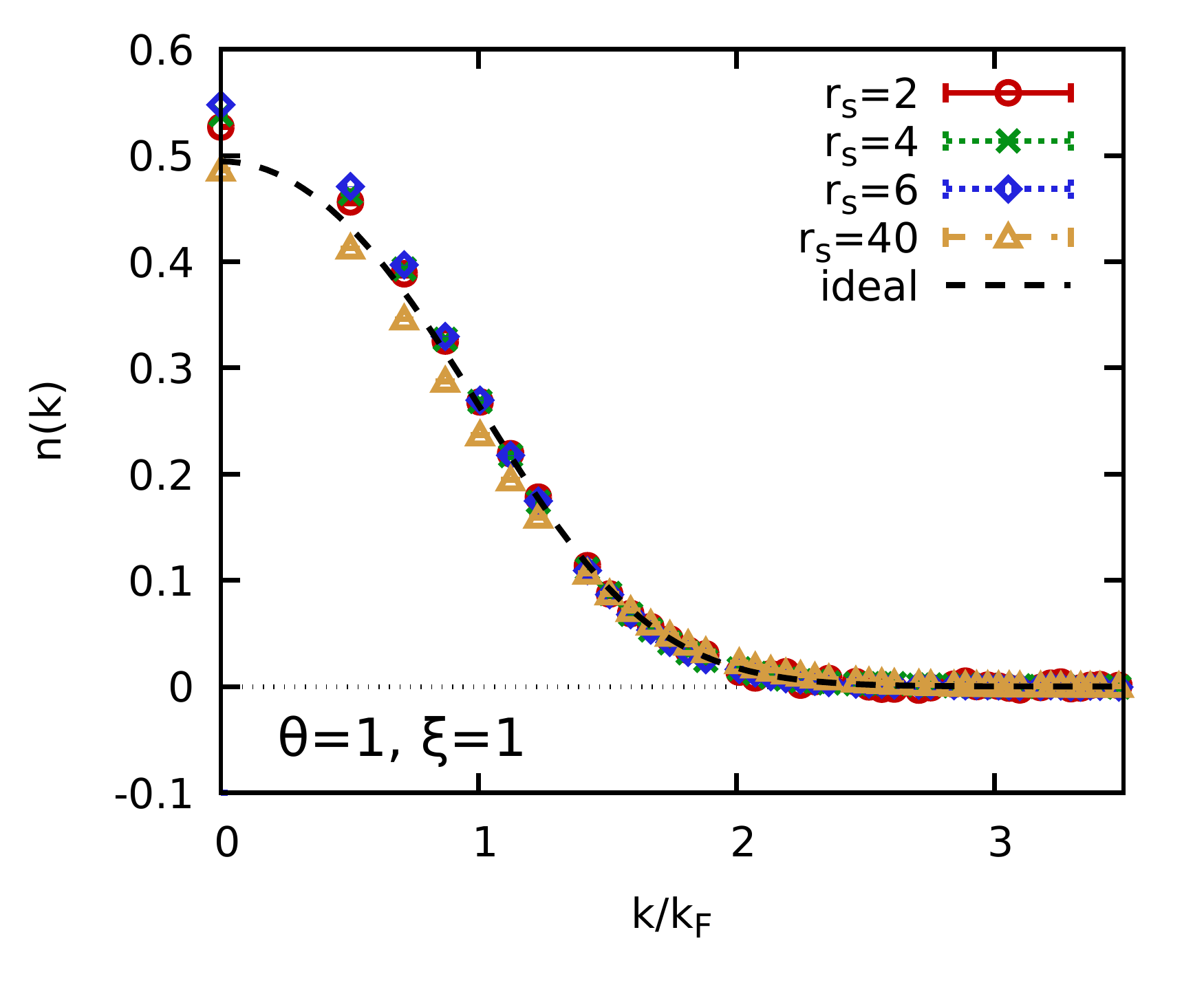}\includegraphics[width=0.475\textwidth]{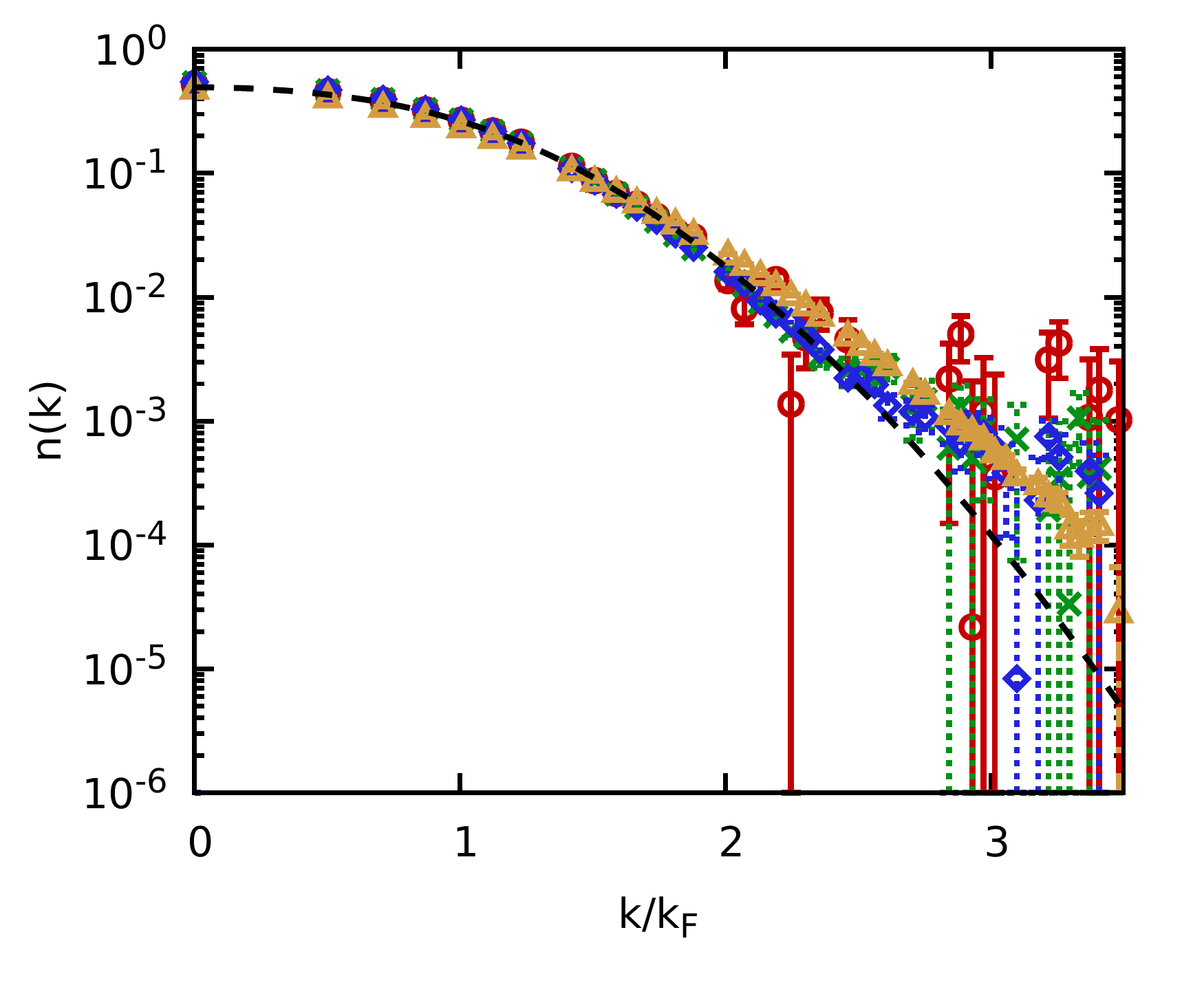}
\caption{\label{fig:polarized_nk_theta}
Momentum distribution of the spin-polarized ($\xi=1$) electron gas for $\theta^\uparrow=1$. The dashed black curve shows results for the ideal Fermi gas [Eq.~(\ref{eq:Fermi})], and the data points correspond to our PIMC data for different $r_s$. The left and right panel shows $n(\mathbf{k})$ on a linear and logarithmic scale, respectively.
}
\end{figure*}

Let us start our investigation of spin-effects on the momentum distribution with an analysis of the fully spin-polarized case, $\xi=1$. To this end, we show the density dependence of $n(\mathbf{k})$ for $\theta^\uparrow=1$ in Fig.~\ref{fig:polarized_nk_theta}. More specifically, the dashed black curve shows results for the ideal Fermi gas [cf.~Eq.~(\ref{eq:Fermi})], which are independent of the density when the reduced temperature is being kept constant. In addition, the different symbols show our new PIMC results that have been obtained for $N^\uparrow=33$ electrons for different values of $r_s$. For completeness, we mention that the left and right panels correspond to a linear and logarithmic scale, which allows to focus on different features of $n(\mathbf{k})$ at different $k$. Furthermore, an extensive analysis of finite-size effects for different densities is shown in Sec.~\ref{sec:FSC} below.

First and foremost, we find that all depicted data sets are qualitatively quite similar to the ideal Fermi distribution, which is substantially broadened at these conditions due to the comparably large thermal energy. Remarkably, the momentum distribution at zero-momentum, $n(0)$, is increased compared to $n_0(0)$ for $r_s=2$ (red circles), $r_s=4$ (green crosses), and $r_s=6$ (blue diamonds), and this trend even increases with $r_s$ for these three cases. This fairly counter-intuitive phenomenon was first reported by Militzer and Pollock~\cite{Militzer_PRL_2002}, and can be explained in terms of a negative mean-field contribution to the single-particle dispersion. A more systematic investigation of this trend and its relation to the kinetic energy is shown in Fig.~\ref{fig:polarized_rs_dependence} below. For $r_s=40$ (yellow triangles), the system becomes strongly correlated and we find $n(0)<n_0(0)$, although both are comparable in magnitude. 

The logarithmic depiction of $n(\mathbf{k})$ shown in the right panel of Fig.~\ref{fig:polarized_nk_theta} allows to study the behaviour of the momentum distribution at large momenta, where it rapidly decays. Consequently, this regime is hard to resolve with the PIMC method, and the relative noise level increases and eventually surpasses $100\%$.
For the unpolarized UEG, it can be shown analytically that $n(\mathbf{k})$ decays algebraically in the limit of large $k$, and the exact relation is given by~\cite{hofmann_short-distance_2013,yasuhara_note_1976}
\begin{align}
    \lim_{k\to \infty} n(k)  = \frac{4}{9}\left(\frac{4}{9\pi}\right)^{2/3}\left( \frac{r_s}{\pi}\right)^2 \frac{k^8_F}{k^8}g^{\uparrow\downarrow}(0)\,,
      \label{eq:k8-g0}
\end{align}
with $g^{\uparrow\downarrow}(0)$ being the pair correlation function between electrons of opposite spin, which has been parametrized in Ref.~\cite{Dornheim_PRL_2020_ESA} for different values of $r_s$ and $\theta$.
This relation has recently been verified on the basis of highly accurate numerical data by Hunger \textit{et al.}~\cite{Hunger_PRE_2021}. For the spin-polarized case, on the other hand, Eq.~(\ref{eq:k8-g0}) cannot hold as there is only a single spin-component, and the on-top pair correlation function is always zero.
Instead, Rajagopal \textit{et al.}~\cite{rajagopal_short_ranged_1978} have found the relation
\begin{equation}
    \label{eq:g0_k10}
     \lim_{k\to \infty}    n(k) =
    \frac{4}{3} \frac{8}{9\pi^2}(\alpha r_s)^2\, \frac{g^{\uparrow\uparrow ''}(0)}{2} \left(\frac{k_F}{k}\right)^{10}
\end{equation}
with $\alpha=(\frac{4}{9\pi})^{1/3}$, which depends on the second derivative of the pair correlation function between electrons of equal spin, again at zero distance.
While being asymptotically exact, Eqs.~(\ref{eq:k8-g0}) and (\ref{eq:g0_k10}) do not give any information about the particular values of $k$ for which these limits are attained in practice. The empirical findings for the unpolarized case in Ref.~\cite{Hunger_PRE_2021} indicate that this happens for $k\gtrsim5k_\textnormal{F}$ at these conditions. Therefore, resolving the asymptotic tail would require to accurately estimate the momentum distribution over at least eight orders of magnitude in $n(\mathbf{k})$ itself, which is beyond the capability of PIMC methods operating in coordinate space. The same issue has been reported by Dornheim \textit{et al.}~\cite{dornheim2021ab} for the unpolarized case, too.

\begin{figure}\centering
\includegraphics[width=0.475\textwidth]{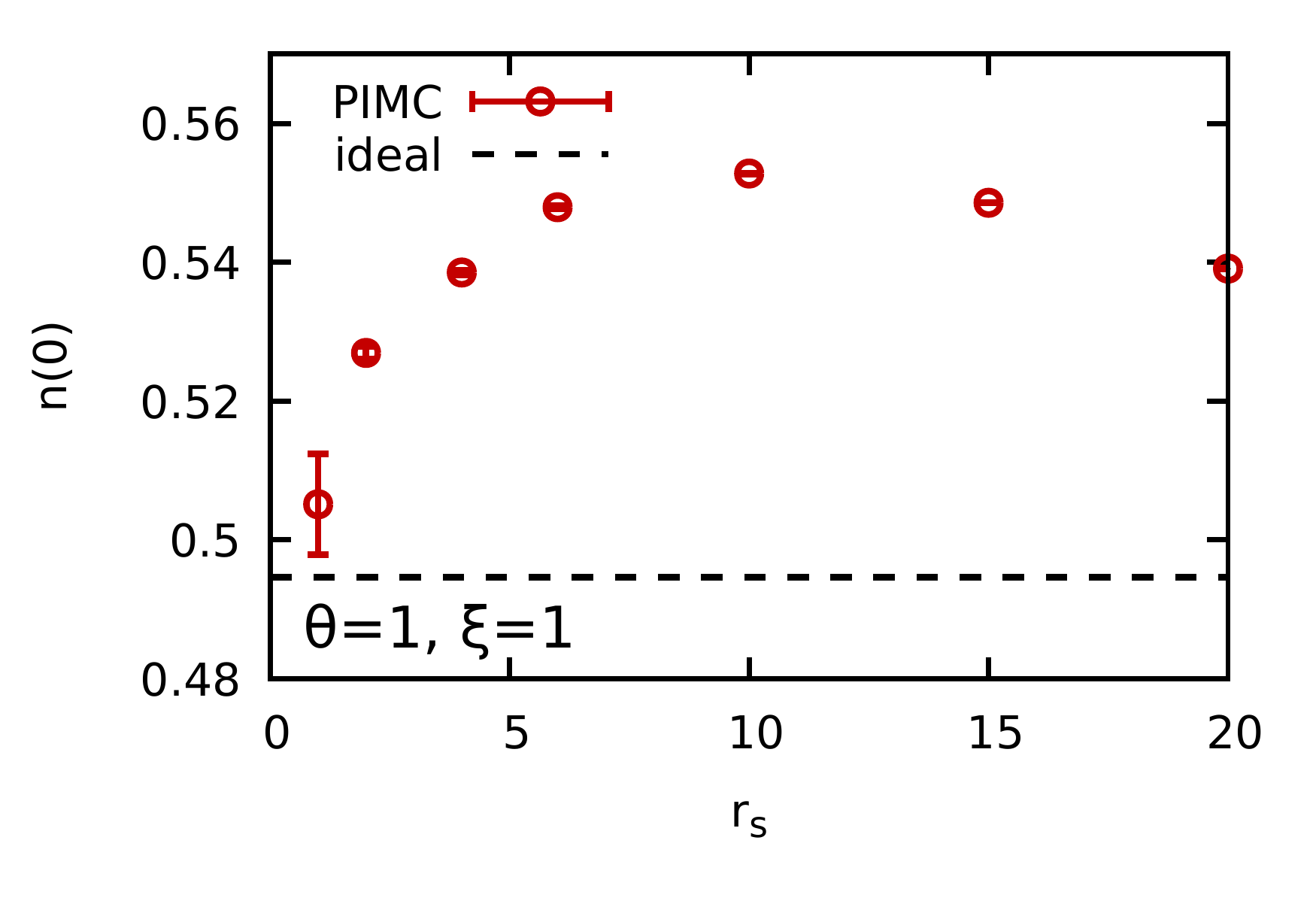}\\\vspace*{-1.01cm}\includegraphics[width=0.475\textwidth]{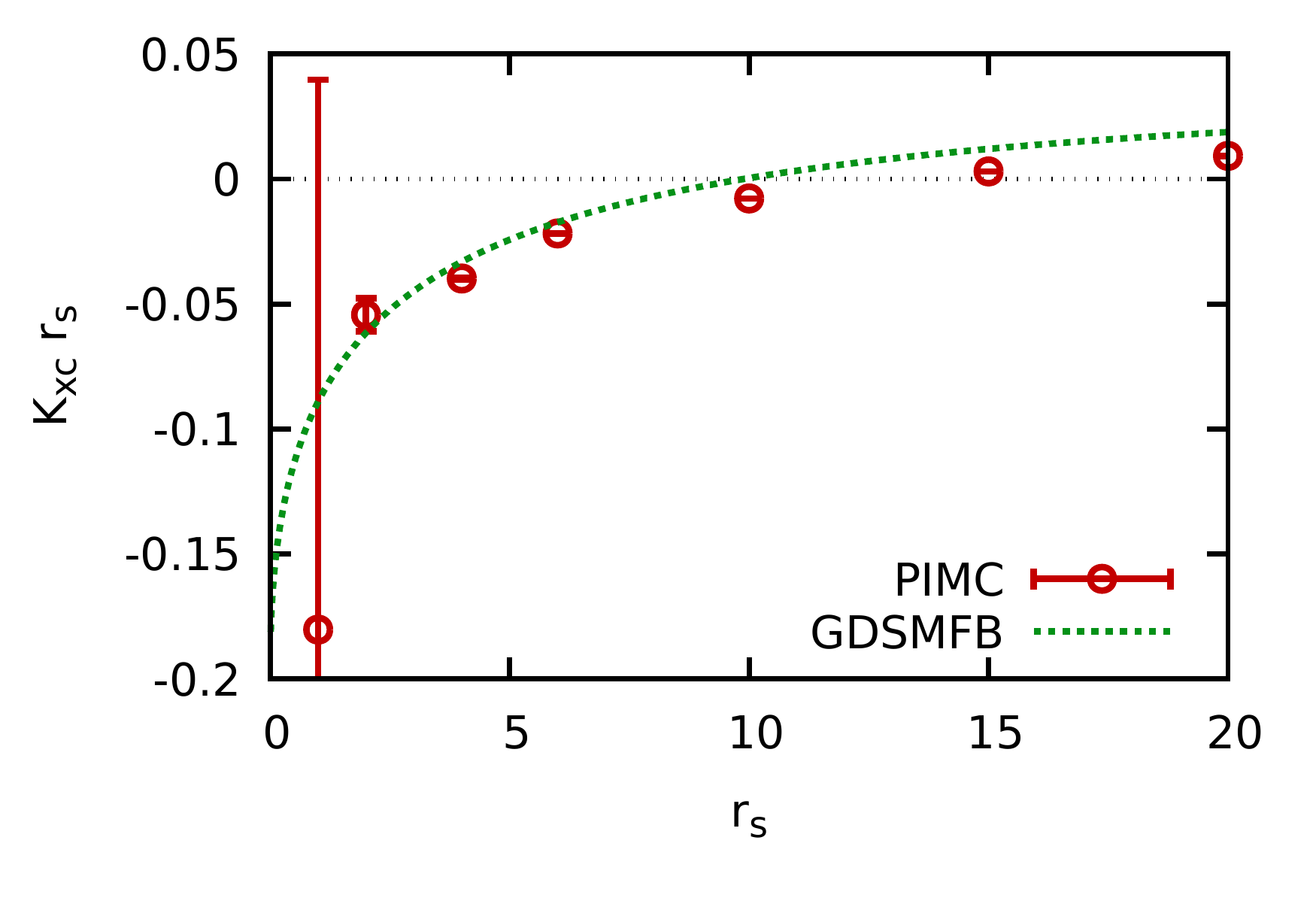}
\caption{\label{fig:polarized_rs_dependence}
Top: $r_s$ dependence of $n(0)$ for the spin-polarized UEG with $\theta^\uparrow=1$. Bottom: $r_s$ dependence of the exchange--correlation part to the kinetic energy (per particle) $K_\textnormal{xc}$ [Eq.~(\ref{eq:Kxc_definition})], with the red circles being our new PIMC results and the dotted green line having been computed from Eq.~(\ref{eq:Kxc}) using the parametrization by Groth \textit{et al.}~\cite{groth_prl}.
}
\end{figure}

Let us next get back to the topic of the counter-intuitive, interaction-induced increase of $n(0)$, which we analyze in detail in the top panel of Fig.~\ref{fig:polarized_rs_dependence}. More specifically, the horizontal dashed black line corresponds to the ideal Fermi gas, which does not depend on $r_s$. In addition, the red circles show our new PIMC data for different densities. Firstly, we note the increasing error bars towards small $r_s$, which are a direct consequence of the fermion sign problem. More specifically, a decrease in the coupling strength leads to an increase in the frequency of permutation cycles within the PIMC simulation, and cycles of adjacent length contribute with a different sign. The resulting cancellation of positive and negative terms then leads to a decreasing signal-to-noise ratio; see Refs.~\cite{Dornheim_JCP_permutation,dornheim_sign_problem} for more detailed information.

In the limit of $r_s\to0$, the UEG becomes ideal~\cite{review} and the PIMC data approach the horizontal line. With increasing coupling strength, the occupation at $k=0$ systematically increases, and attains a maximum at $r_s\approx10$ at these conditions. Increasing the density parameter even further leads to the opposite trend and, eventually, $n(0)$ will even become smaller than $n_0(0)$ as the electrons are pushed to larger $k$ by the strong repulsion. The particular comparison between $n(\mathbf{k})$ and $n_0(\mathbf{k})$ strongly depends on the value of $\xi$, which is explained in detail in the discussion of Fig.~\ref{fig:xi_n0_rs2} below.

Let us next consider the connection between the correlation-induced increase in the momentum distribution at zero momentum to the exchange--correlation part of the kinetic energy,
\begin{eqnarray}\label{eq:Kxc_definition}
K_\textnormal{xc} = K - E_0\ ,
\end{eqnarray}
where $K$ and $E_0$ are the total kinetic energy of the interacting and noninteracting system, respectively. For completeness, we mention that $K_\textnormal{xc}$ is directly related to the asymptotic behaviour of the electronic local field correction at large wave numbers~\cite{dornheim_ML,farid,holas_limit}. The $r_s$ dependence of Eq.~(\ref{eq:Kxc_definition}) is shown in the bottom panel of Fig.~\ref{fig:polarized_rs_dependence} for the same conditions as $n(0)$. More specifically, the red circles have been obtained by taking our direct PIMC results for $K$ for $N=33$ electrons and subsequently subtracting $E_0$ (taken from Ref.~\cite{Groth_PRB_2016}) for the same system size. The large error bars for small $r_s$ are again a direct consequence of the fermion sign problem, which is exacerbated by the definition of $K_\textnormal{xc}$ as the difference between two quantities that are more or less comparable in magnitude. Furthermore, the dotted green curve has been computed from the parametrization of the exchange--correlation free energy $f_\textnormal{xc}$ by Groth \textit{et al.}~\cite{groth_prl} by evaluating the relation~\cite{review}
 \begin{eqnarray}\label{eq:Kxc}
 K_\textnormal{xc} 
 &=& - f_\textnormal{xc}(r_s,\theta)
 -\theta \frac{\partial f_\textnormal{xc}(r_s,\theta)}{\partial\theta}\Bigg|_{r_s}\\ \nonumber
& & - r_s \frac{\partial f_\textnormal{xc}(r_s,\theta)}{\partial r_s}\Bigg|_\theta \quad .
 \end{eqnarray}
First and foremost, we note the good qualitative agreement between Eq.~(\ref{eq:Kxc}) and the PIMC data, although there appear small but significant differences towards large $r_s$. The most likely explanation for the latter are finite-size effects in the red circles in either (or both) $K$ or $E_0$, whereas the green curve has been constructed in the thermodynamic limit, i.e., in the limit of an infinite number of particles taken at a constant number density. A second, somewhat less likely explanation is the finite accuracy of the utilized parametrization of $f_\textnormal{xc}$, which might be exacerbated by the evaluation of the derivatives in Eq.~(\ref{eq:Kxc}).

From a physical perspective, we do indeed find a lowering in the kinetic energy due to electronic exchange--correlation effects for $r_s\lesssim10$, whereas the total kinetic energy is increased for stronger coupling. Therefore, the negative values of $K_\textnormal{xc}$ are certainly related, but not equal to the increase in $n(0)$ discussed above, which is consistent to recent findings for the unpolarized electron gas~\cite{dornheim2021ab}.

\subsubsection{Comparison to restricted PIMC\label{sec:RPIMC}}

\begin{figure*}\centering
\includegraphics[width=0.475\textwidth]{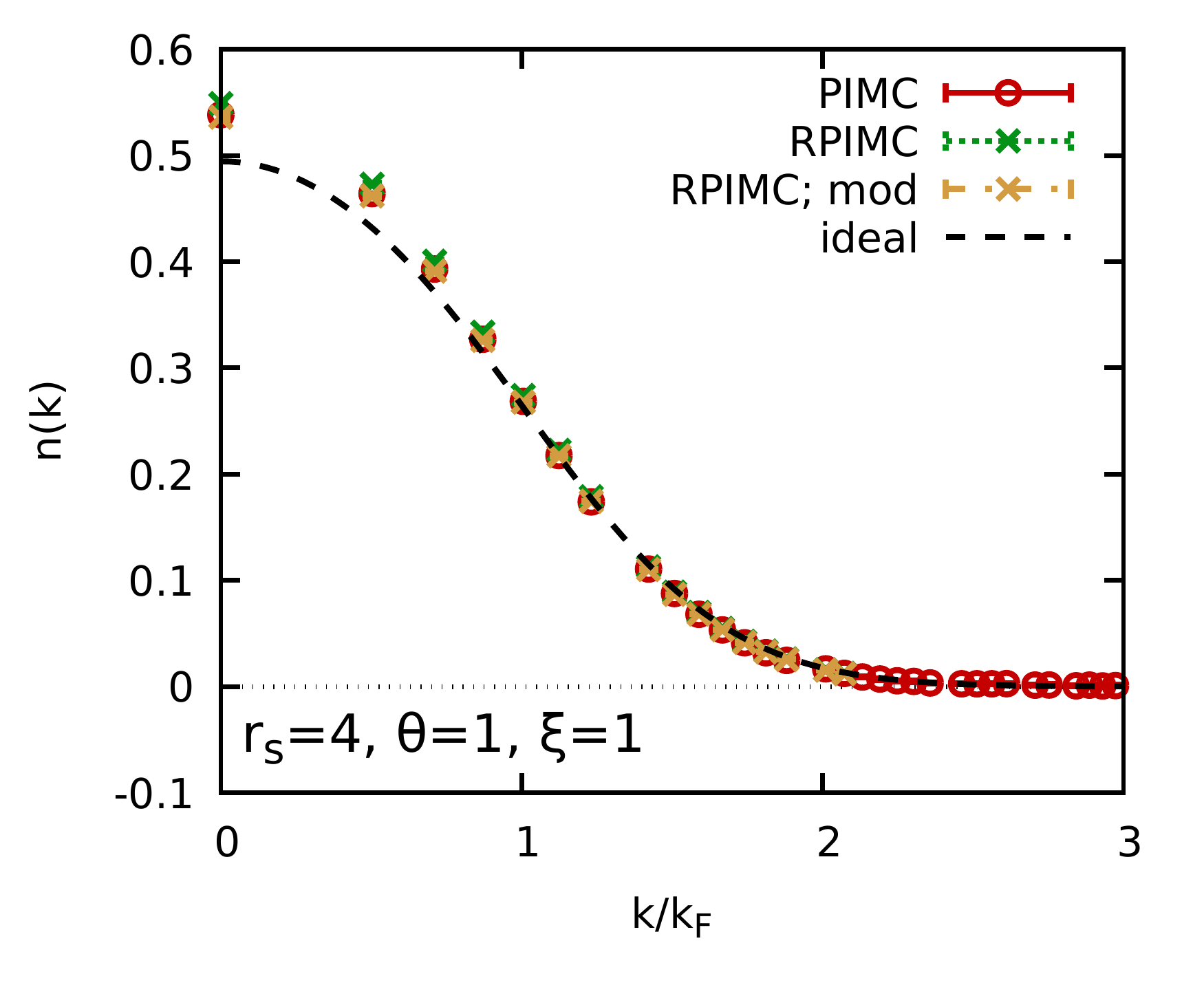}\includegraphics[width=0.475\textwidth]{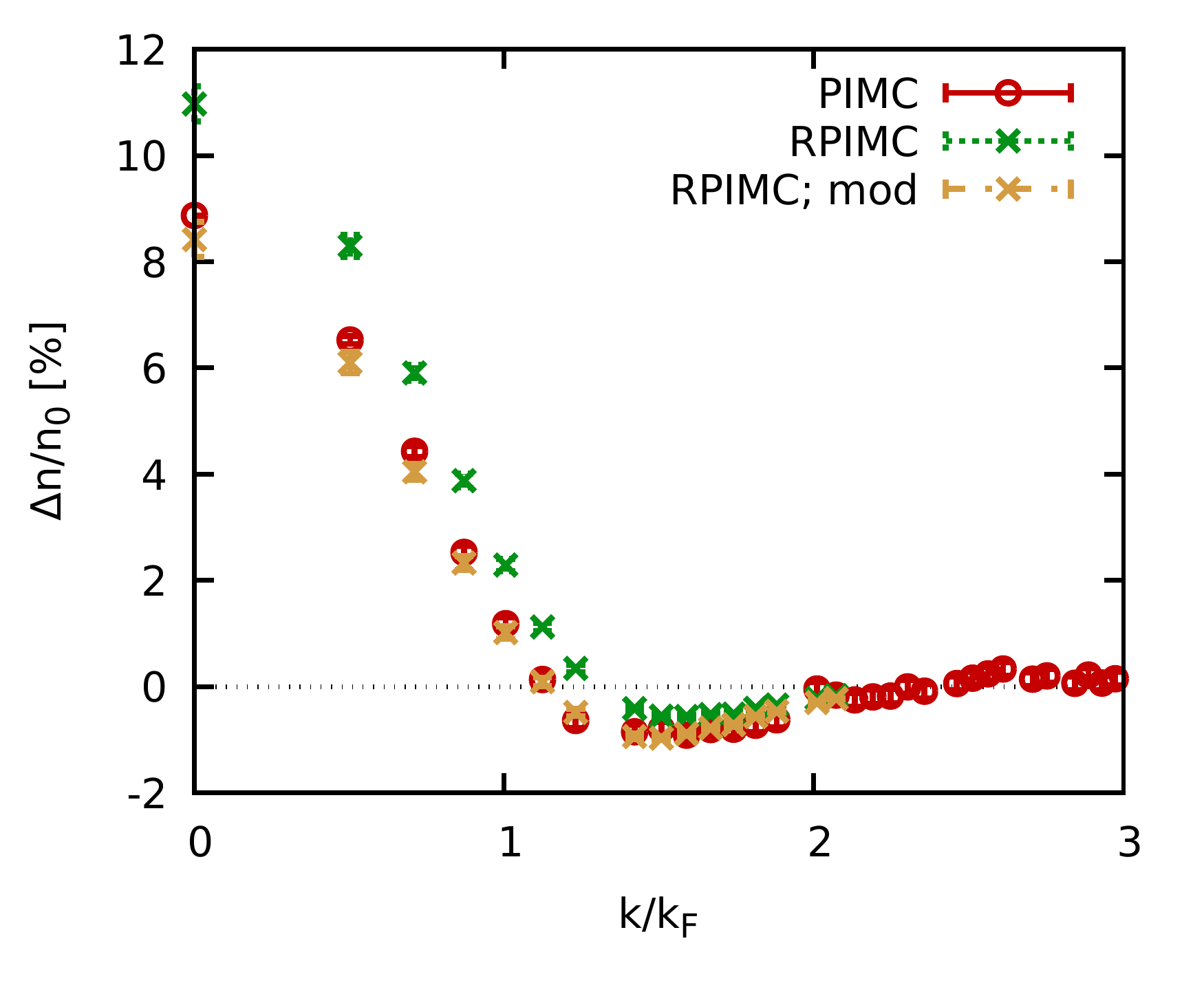}\\\vspace*{-1.01cm}\includegraphics[width=0.475\textwidth]{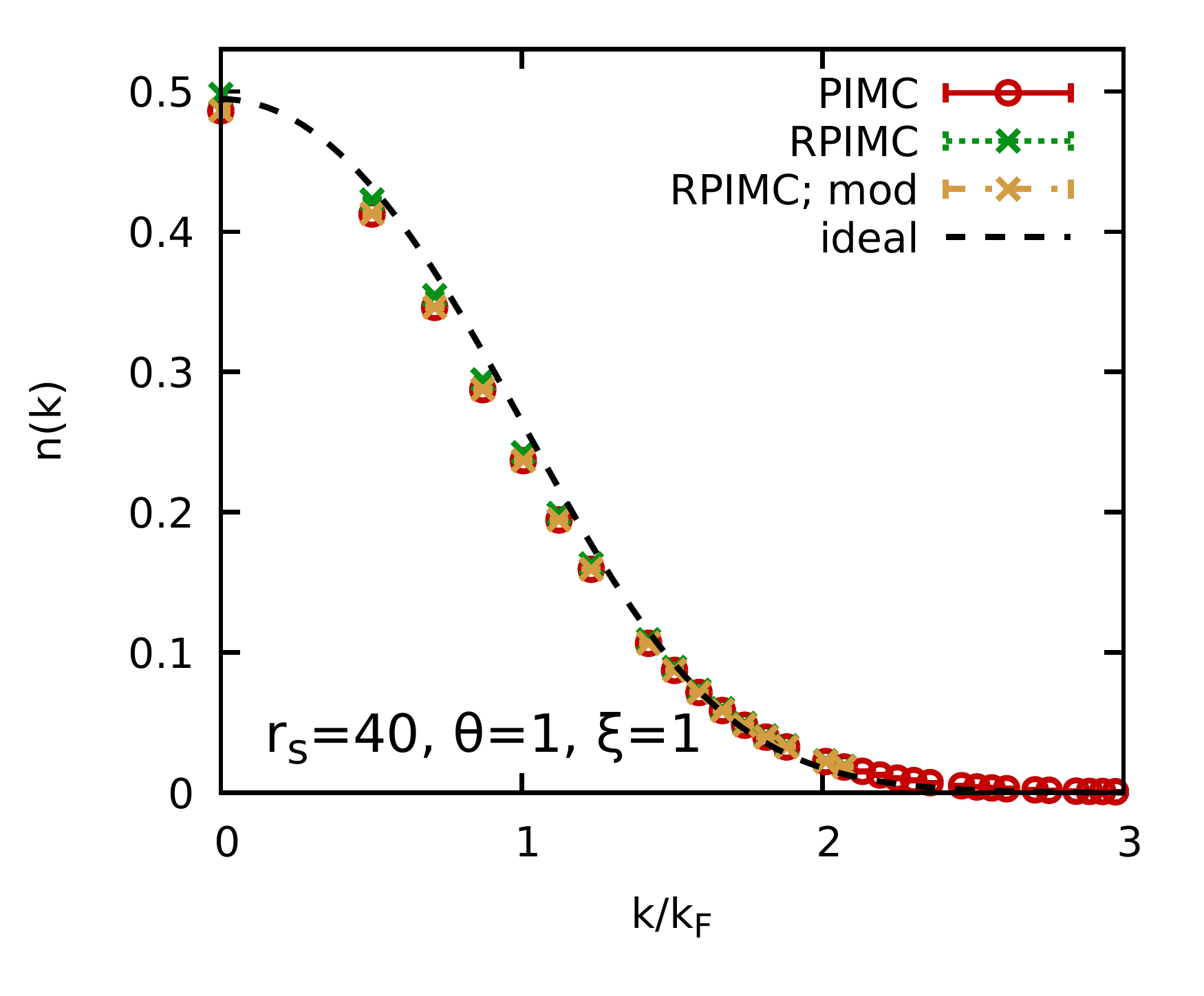}\includegraphics[width=0.475\textwidth]{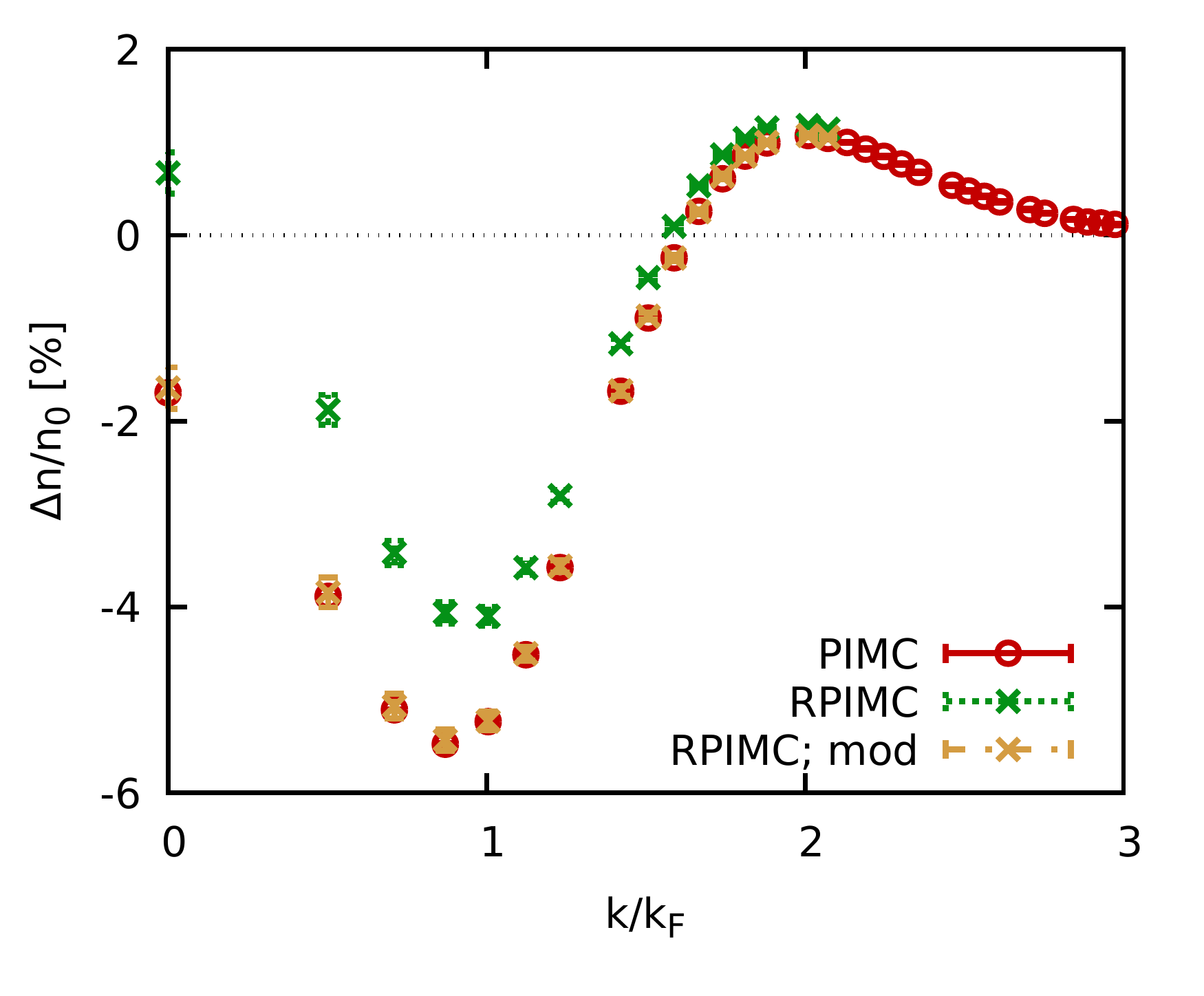}\\\vspace*{-1.01cm}\includegraphics[width=0.475\textwidth]{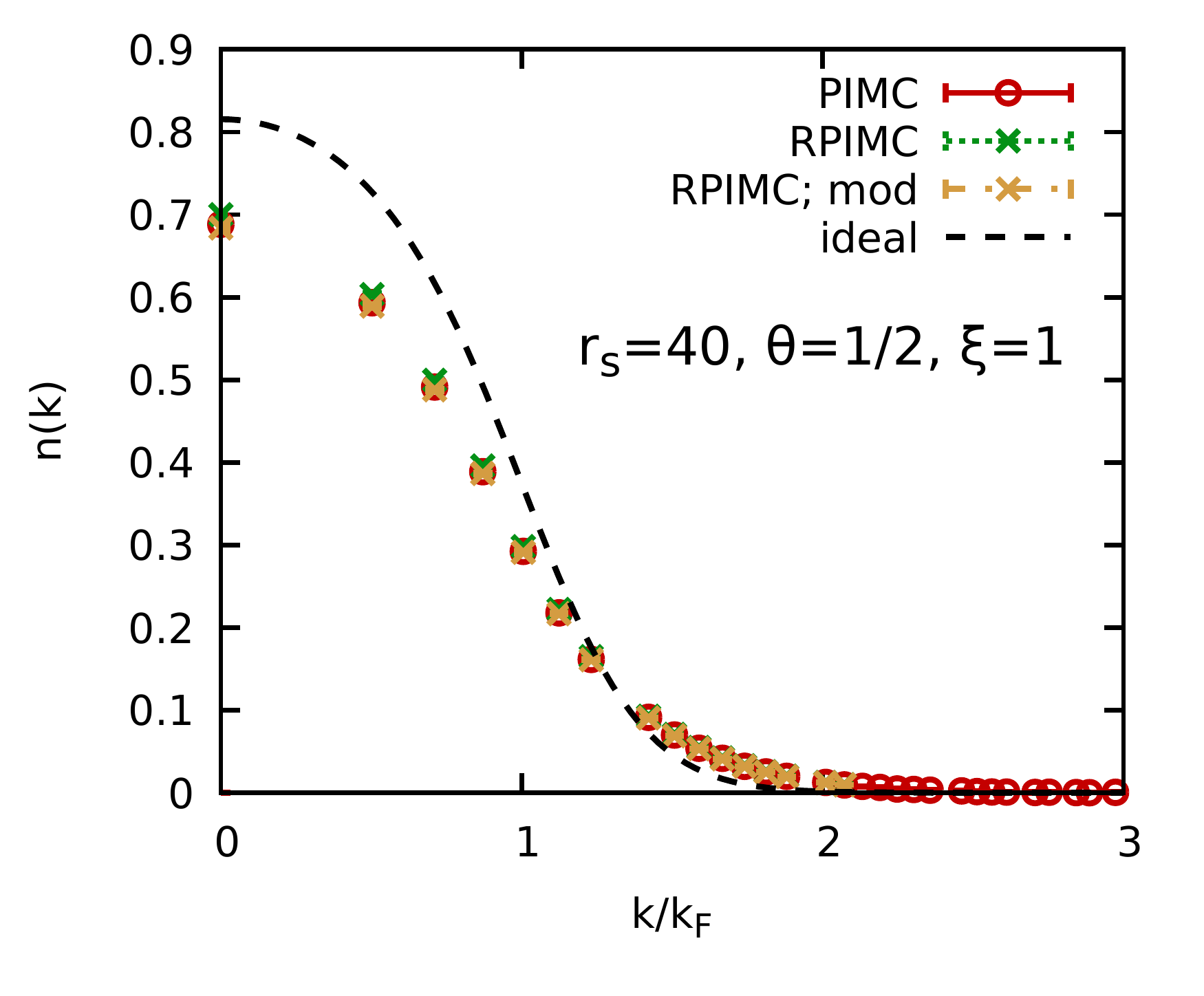}\includegraphics[width=0.475\textwidth]{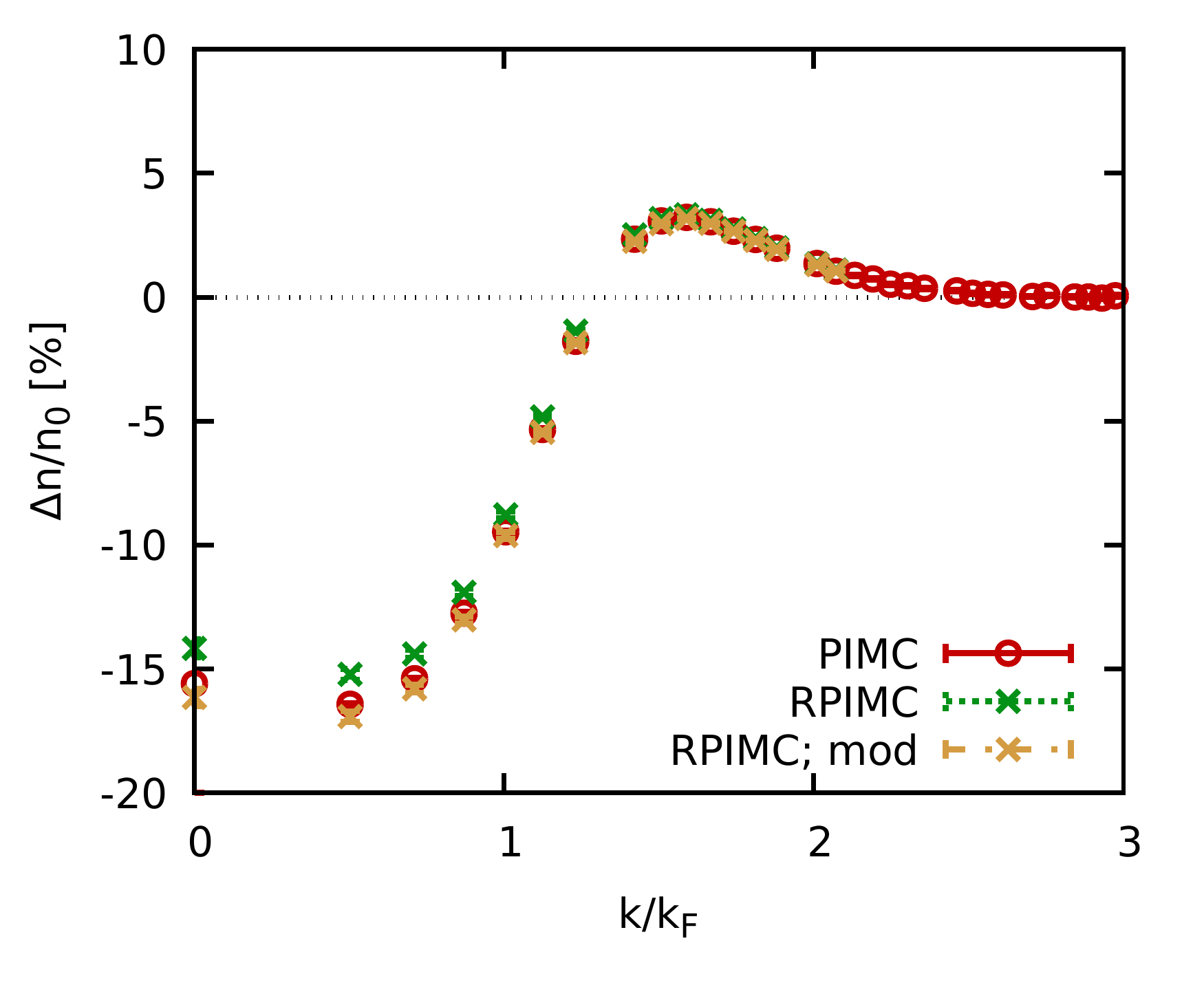}
\caption{\label{fig:polarized_rs4_theta1}
Momentum distribution function of the spin-polarized ($\xi=1$) UEG for $N=33$ electrons. Top row:
 $r_s=4$ and $\theta^{\xi=1}=1$;
center row:
 $r_s=40$ and $\theta^{\xi=1}=1$;
 bottom row:
 $r_s=40$ and $\theta^{\xi=1}=0.5$. Left column: $n(\mathbf{k})$, with the red circles, green crosses, and yellow crosses corresponding to our new direct PIMC results, restricted PIMC results from Ref.~\cite{Militzer_momentum_HEDP_2019}, and modified RPIMC results (see text), respectively. Right column: Relative difference between $n(\mathbf{k})$ and $n_0(\mathbf{k})$ in per cent, cf.~Eq.~(\ref{eq:percent}).
}
\end{figure*}

Let us conclude this study of the purely spin-polarized UEG with a comparison between our new direct PIMC results for $n(\mathbf{k})$ and the restricted PIMC data by Militzer \textit{et al.}~\cite{Militzer_momentum_HEDP_2019}. This is shown in Fig.~\ref{fig:polarized_rs4_theta1} where the top row corresponds to $r_s=4$ and $\theta^\uparrow=1$, i.e., a metallic density that is of relevance to contemporary WDM research~\cite{new_POP}. The left panel shows results for $n(\mathbf{k})$ itself, and the ideal Fermi function (dashed black line) has been included as a reference. Further, the red circles and green crosses depict the PIMC and RPIMC data, respectively, which are in qualitative though not quantitative agreement. More specifically, the RPIMC data are systematically higher than the PIMC data, which is consistent to the recent findings by Dornheim \textit{et al.}~\cite{dornheim2021ab} for the unpolarized UEG. In the latter study, this discrepancy was explained by the normalization of $n(\mathbf{k})$, which is computed exactly within our extended ensemble formalism, but has to be inferred from the off-diagonal density matrix $n(\mathbf{r},\mathbf{r}')$ in the case of RPIMC. Therefore, multiplying the RPIMC data by a constant factor of $Q=0.977$ led to perfect agreement between direct PIMC and restricted PIMC for $\xi=0$.

Following the same procedure (with the same factor) in the present case leads to the yellow crosses, which, indeed, are in excellent agreement to the PIMC data over the entire range of wave numbers. This can be seen particularly well in the right panel, where we show the relative deviation between the momentum distribution of the UEG and the ideal Fermi gas (in per cent) with respect to $n_0(0)$,
\begin{eqnarray}\label{eq:percent}
\frac{\Delta n}{n(0)} [\%] = \frac{n(\mathbf{k})-n_0(\mathbf{k})}{n_0(\mathbf{0})}\times 100\ .
\end{eqnarray}
In addition, this depiction also allows one to gain a more vivid insight into the behaviour of $K_\textnormal{xc}$
shown in Fig.~\ref{fig:polarized_rs_dependence} above: for small momenta ($k\lesssim k_\textnormal{F}$), the momentum distribution function of the interacting electron gas plainly exceed the ideal Fermi distribution $n_0(k)$, whereas the relative occupation is decreased in the range of $k_\textnormal{f} \lesssim k \lesssim 2k_\textnormal{F}$. Since the total kinetic energy is simply given by the integral
\begin{eqnarray}
K = \frac{1}{2} \int \textnormal{d}\mathbf{k}\ n(\mathbf{k})\ \mathbf{k}^2\ ,
\end{eqnarray}
the observed deviation profile directly indicates the relation $K[n(\mathbf{k})]<K[n_0(\mathbf{k})]$ at these conditions, thus resulting in the negative values of $K_\textnormal{xc}$ shown above.

We next consider the central row of Fig.~\ref{fig:polarized_rs4_theta1}, where we show the same analysis for $r_s=40$ and $\theta^\uparrow=1$. Physically, these conditions are located within the strongly coupled electron liquid regime~\cite{dornheim_dynamic,dornheim_electron_liquid}, where the strong Coulomb repulsion between the electrons predominates over quantum degeneracy effects such as Pauli blocking or quantum diffraction. In particular, this regime is expected to give rise to a \emph{negative dispersion relation}~\cite{dornheim_dynamic,dynamic_folgepaper,Hamann_CPP_2020,Hamann_PRB_2020}, which is of a potentially incipient excitonic nature~\cite{Takada_PRL_2002,Takada_PRB_2016}. At these conditions, the momentum distribution function of the interacting UEG compares markedly different to the ideal Fermi function, and the direct PIMC method predicts a depletion in the occupation at zero momentum, which holds for $k\lesssim1.5 k_\textnormal{F}$. While the raw RPIMC data (green crosses) actually predict an increase in $n(0)$ compared to the ideal system, this effect is most likely spurious. Specifically, multiplying the RPIMC data by the same factor $Q$ as in the previous case leads to the yellow crosses, which, again, results in a perfect agreement to the direct PIMC data.

Let us conclude this comparison between the direct and restricted PIMC methods by investigating a lower temperature, $\theta^\uparrow=0.5$, shown in the bottom row of Fig.~\ref{fig:polarized_rs4_theta1}. For completeness, we mention that such low values of the reduced temperature cannot be accessed by the direct PIMC method at metallic densities due to the aforementioned fermion sign problem. At these conditions, the occupation of momentum states at small $k$ is substantially depleted compared to $n_0(k)$ as the electrons are pushed out to large momenta, resulting in a positive value of $K_\textnormal{xc}$. The comparison between direct and restricted PIMC reveals the same issue with the normalization as in the two previous cases, and the thus modified yellow crosses agree with the red circles over the entire depicted $k$-range.

In summary, the only systematic error that we can find in the RPIMC data both at $r_s=4$ and $r_s=40$ are due to the normalization, and not a consequence of the fixed-node approximation that has been used to deal with the fermion sign problem. This is certainly encouraging, as the extended ensemble approach introduced in Ref.~\cite{dornheim2021ab} can easily be incorporated into RPIMC as well, which would completely overcome this problem.

\subsection{Intermediate polarizations and spin dependence\label{sec:intermediate}}

In the following section, we explicitly go beyond the purely ferromagnetic case to more closely isolate the effect of the spin-polarization itself.

\subsubsection{Analysis of finite-size effects\label{sec:FSC}}

\begin{figure*}\centering
\includegraphics[width=0.475\textwidth]{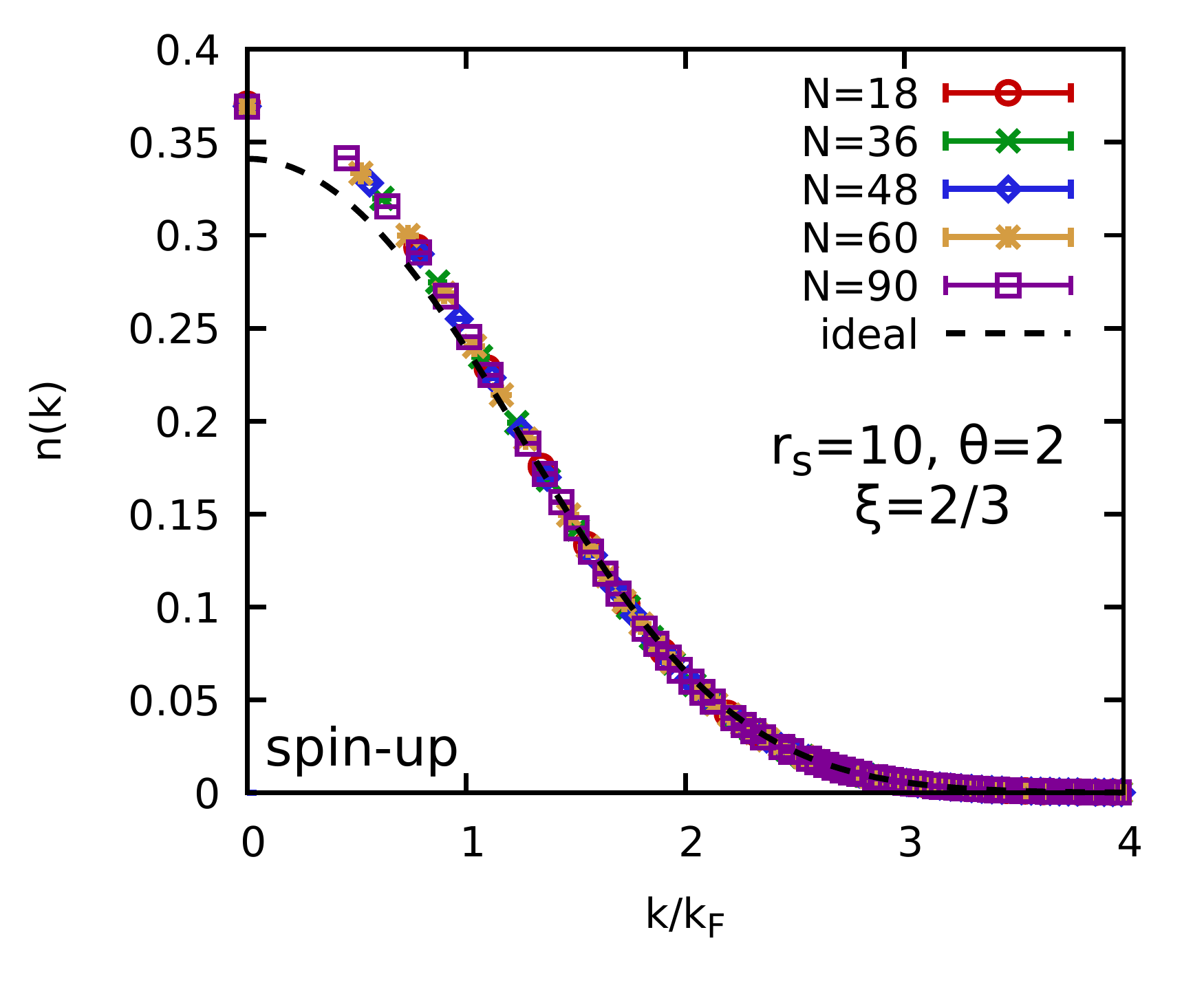}\includegraphics[width=0.475\textwidth]{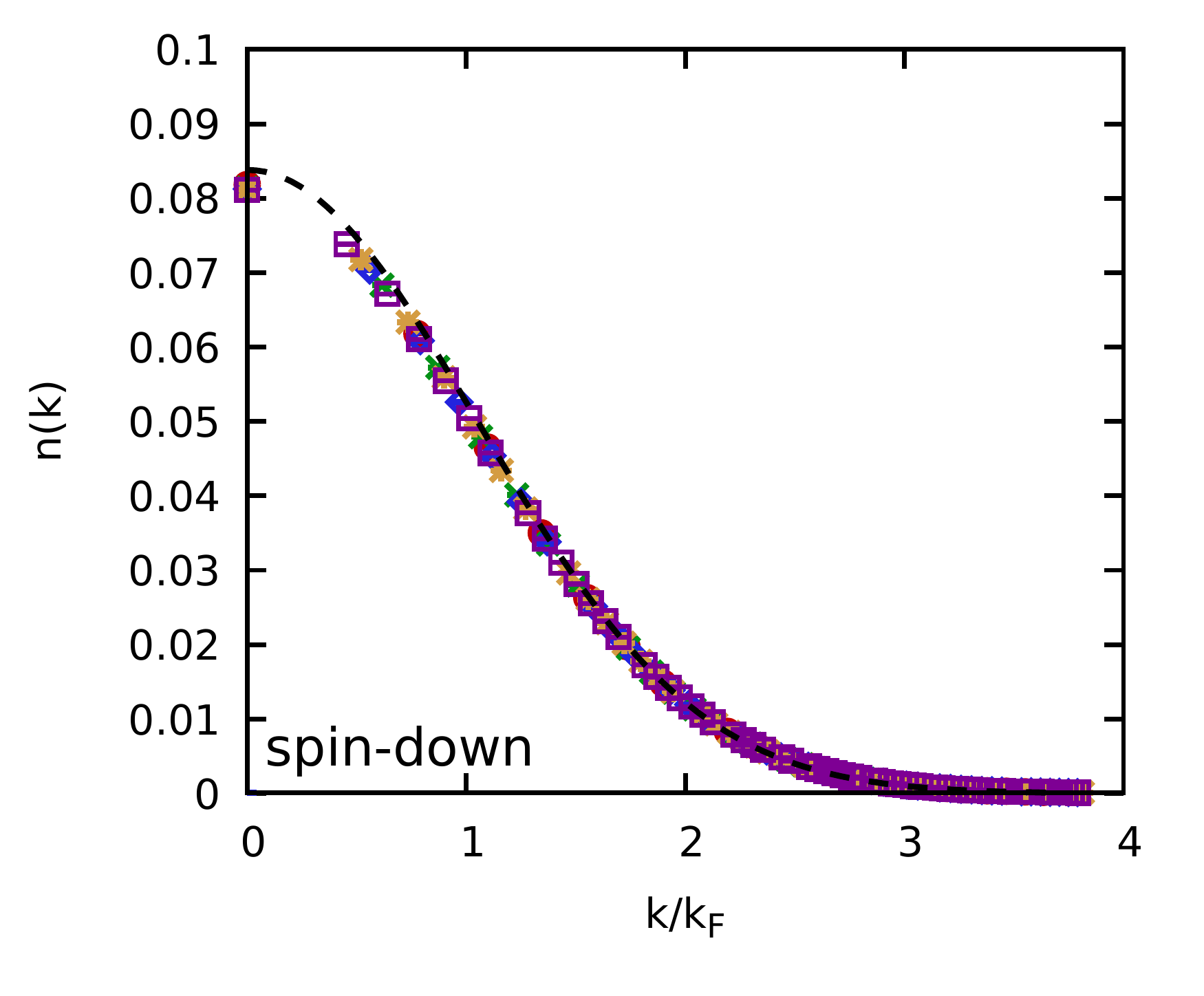}\\\vspace*{-1.02cm}
\hspace*{0.025\textwidth}\includegraphics[width=0.456\textwidth]{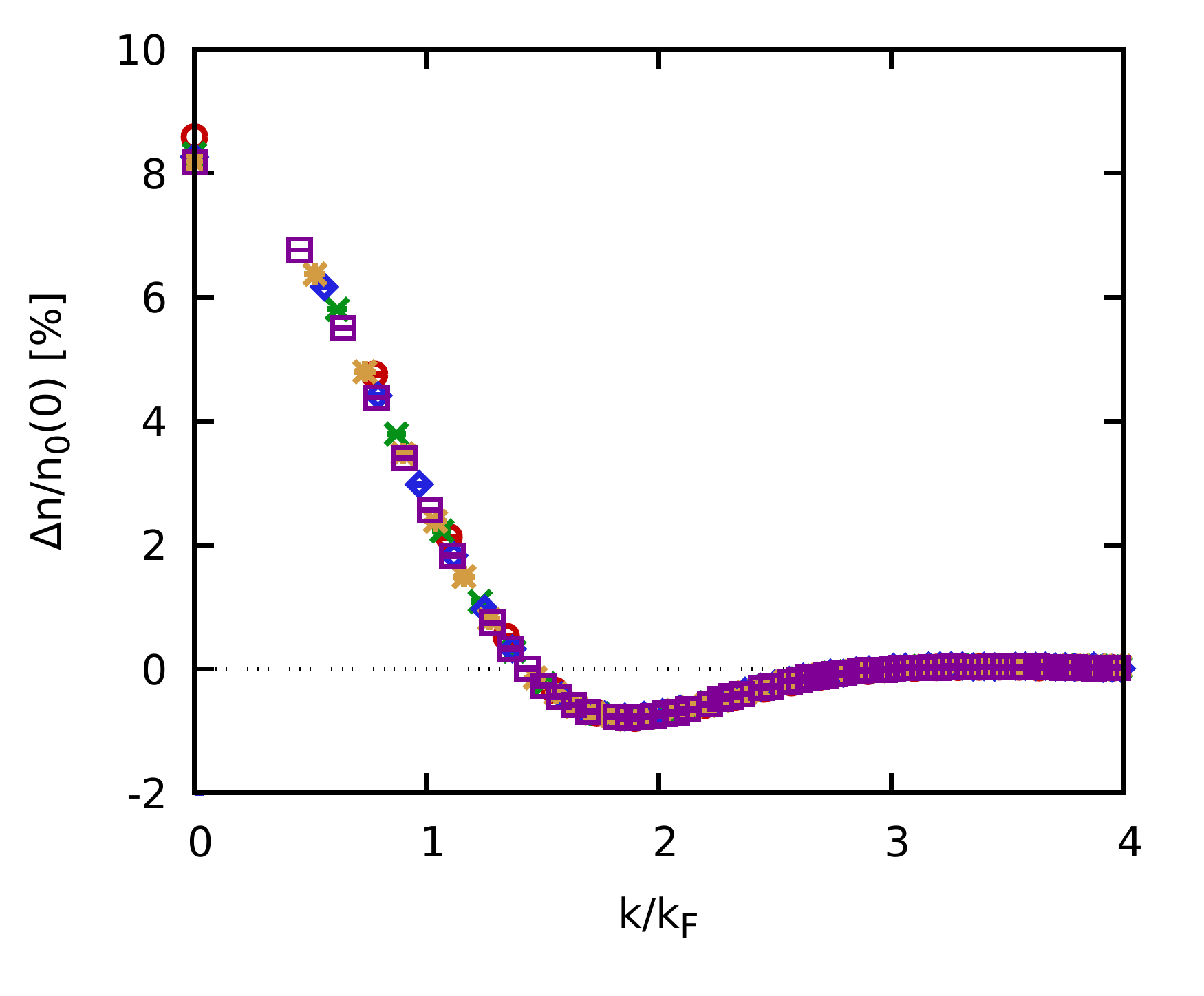}\hspace*{0.01\textwidth}\includegraphics[width=0.462\textwidth]{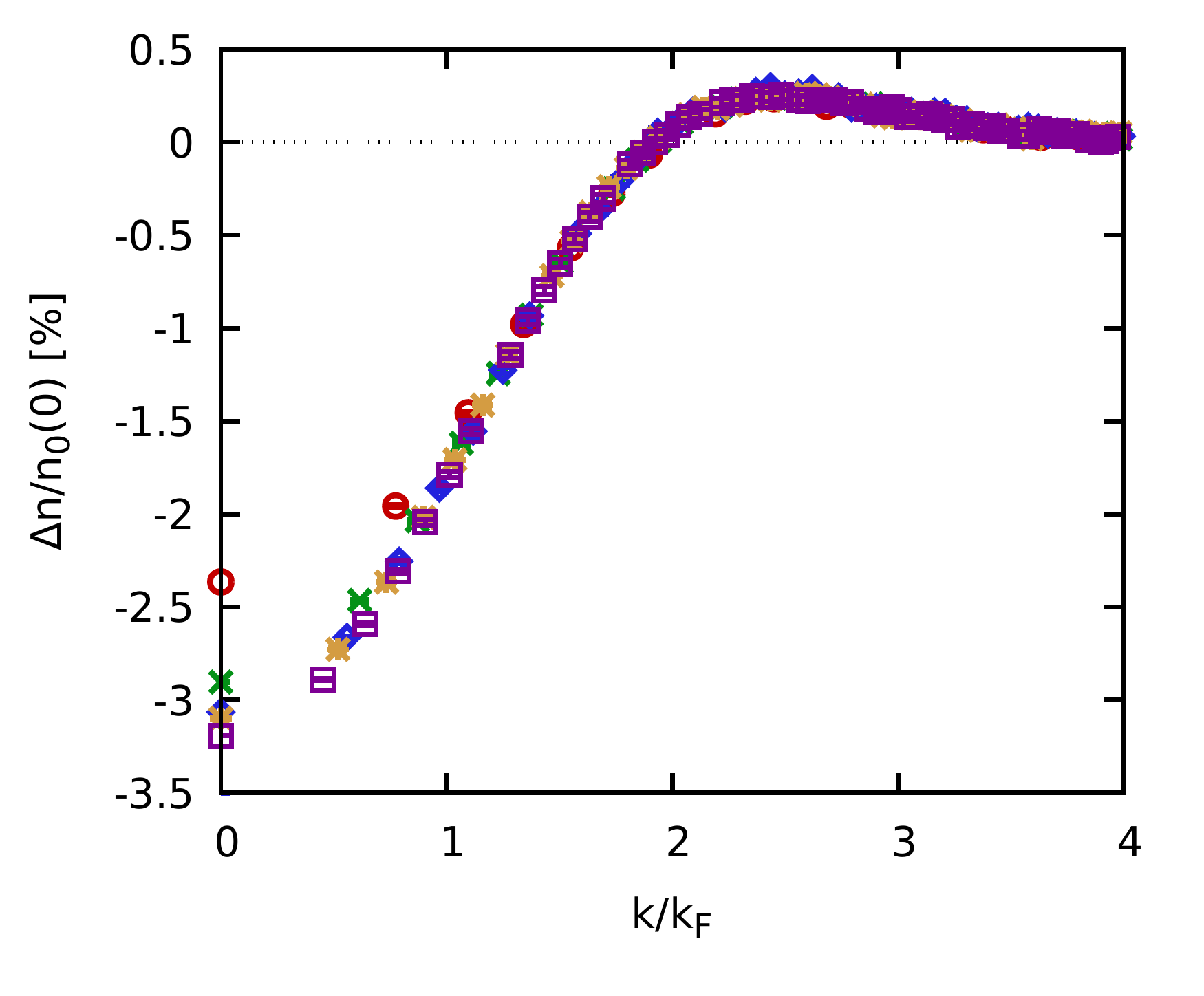}
\caption{\label{fig:FSC_rs10_theta2}
Top: System-size dependence of the momentum distribution of the UEG at $r_s=10$, $\theta=2$, and $\xi=2/3$. The left and right columns correspond to the spin-up and -down electrons, respectively. Bottom: Relative deviation between PIMC data for $n^\sigma(\mathbf{k})$ and the ideal Fermi distribution $n^\sigma_0(\mathbf{k})$ in per cent, cf.~Eq.~(\ref{eq:percent}).
Note that we use the Fermi wave number of the unpolarized system $k_\textnormal{F}^{\xi=0}$ as a reference.
}
\end{figure*}

Being restricted to the description of a finite number of particles, PIMC results are in general afflicted with so-called finite-size effects~\cite{dornheim_prl,Chiesa_PRL_2006,Dornheim_JCP_2021,dornheim_POP,Holzmann_PRB_FSC_2016}. Therefore, a careful analysis of the dependence of the results on the system size is usually indispensable. In the present work, this is shown in Fig.~\ref{fig:FSC_rs10_theta2} for $r_s=10$ and $\theta_{\xi=0}=2$ for $\xi=2/3$. Note that we always use the reduced temperature $\theta_{\xi=0}$ of the unpolarized system as a reference throughout the remainder of this work. We further mention that  the selected polarization $\xi=2/3$ constitutes a particularly relevant choice for the study of finite-size effects, as there are, by definition, always five times the number of spin-up compared to spin-down electrons. For example, we have $N^\uparrow=50$ but only $N^\downarrow=10$ for a total system size of $N=60$ in this case. Since only electron of the same spin-orientation exchange with each other, it is therefore reasonable to expect a different manifestation of finite-size effects in $n^\uparrow(\mathbf{k})$ and $n^\downarrow(\mathbf{k})$, which needs to be checked.

Let us start our investigation by considering results for $n^\uparrow(\mathbf{k})$ shown in the left column of Fig.~\ref{fig:FSC_rs10_theta2}, where the top and bottom panels show results for the momentum distribution itself and for the relative deviation to the ideal Fermi function [cf.~Eq.~(\ref{eq:percent})], respectively. More specifically, the different data points show our new direct PIMC data that have been obtained for different values of $N$. Remarkably, we find hardly any dependence of $n^\uparrow(\mathbf{k})$ on the system size even for as few as $N=18$ electrons. This can be seen particularly well in the bottom panel, where small deviations between the different data sets are noticeable only for small momenta. Still, even here these differences are clearly below $1\%$.
From a physical perspective, we find a pronounced interaction-induced increase in $n^\uparrow(\mathbf{k})$ compared to $n_0^\uparrow(\mathbf{k})$, with a maximum of $\sim8\%$ at zero momentum.

Let us next consider the top right panel of Fig.~\ref{fig:FSC_rs10_theta2}, where we show our direct PIMC results for $n^\downarrow(\mathbf{k})$. Again, hardly any dependence on $N$ can be resolved on this scale even for $N=18$, where there are only $N^\downarrow=3$ spin-down electrons within the simulation. The bottom panel of the same figure does reveal some systematic deviations for $N=18$ and $N=36$, but even the maximum finite-size effect is below $1\%$. In addition, we find that the occupation at small momenta is actually depleted compared to $n^\downarrow_0(\mathbf{k})$, which is in stark contrast to the behaviour of the spin-up electrons observed in the left column of the same figure. A more detailed investigation of this effect is presented in Fig.~\ref{fig:xi_n0_rs2} below. 

\begin{figure*}\centering
\includegraphics[width=0.475\textwidth]{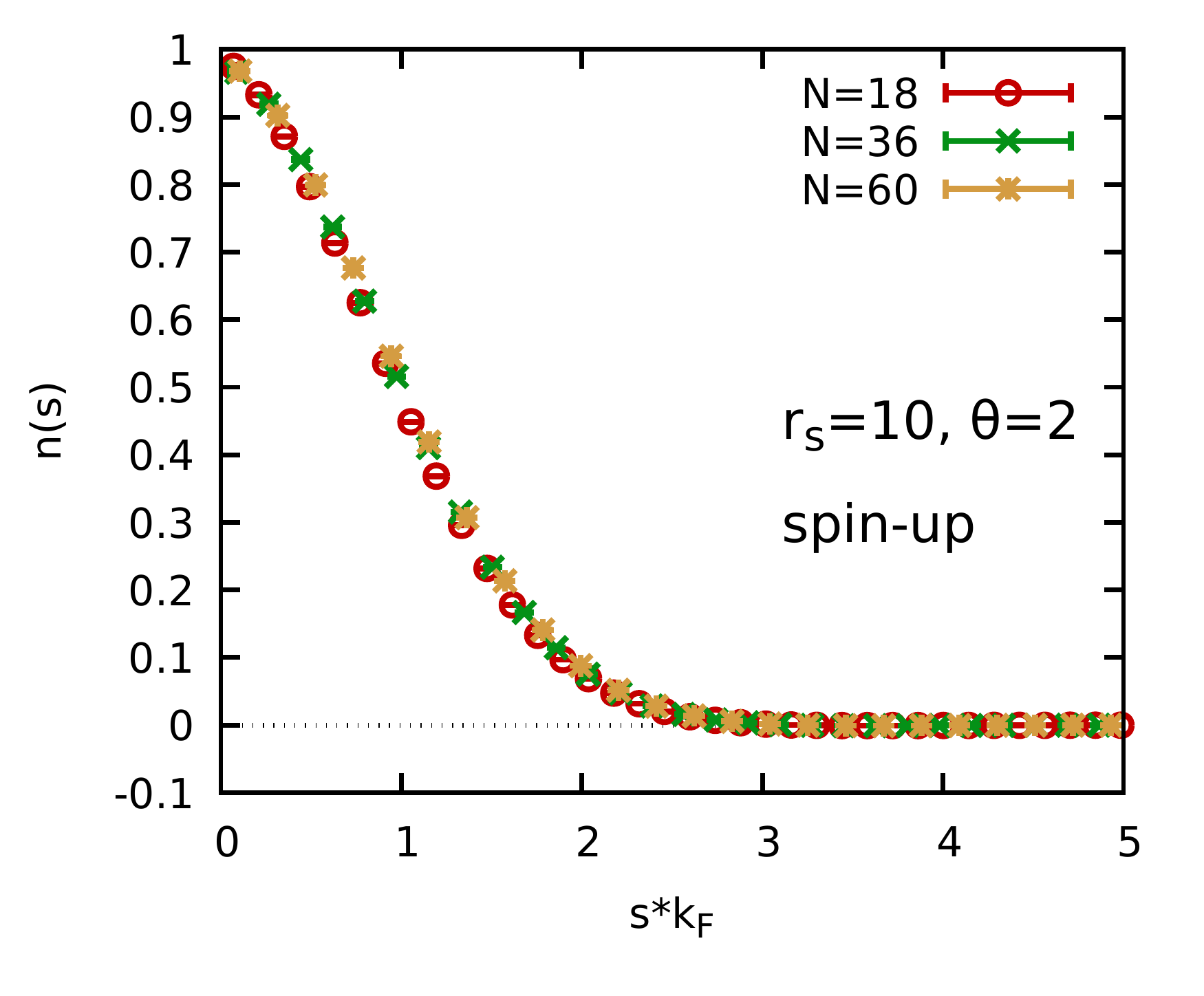}\includegraphics[width=0.475\textwidth]{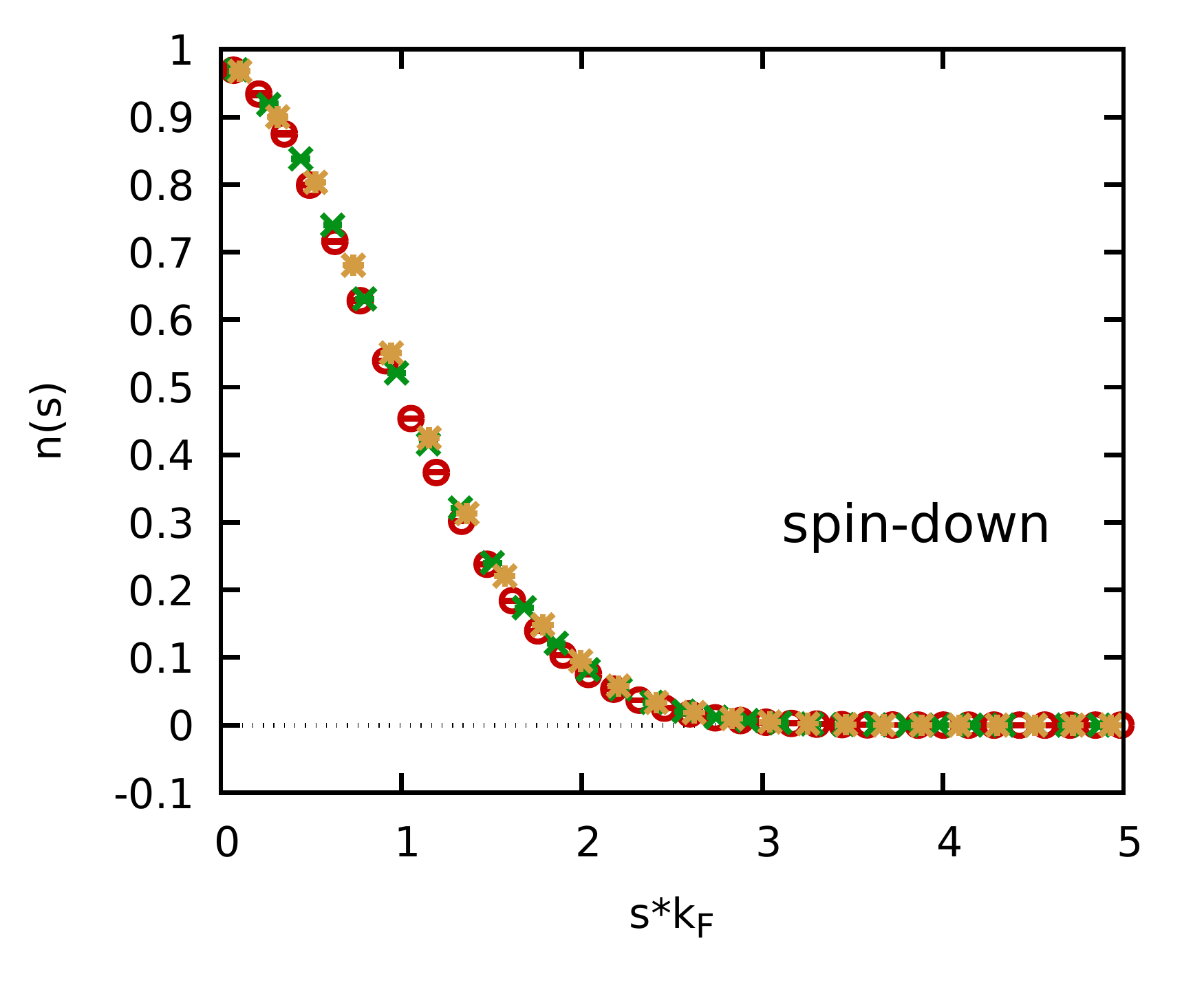}
\caption{\label{fig:FSC_Offdiagonal}
Offdiagonal density matrix $n(s)$ [cf.~Eq.~(\ref{eq:offdiagonal})] of the UEG at $r_s=10$, $\theta=2$, and $\xi=2/3$. The left and right panels show results for spin-up and spin-down electrons, and the red circles, green crosses, and yellow stars have been obtained for $N=18$, $N=36$, and $N=60$, respectively.
}
\end{figure*}

Lastly, we show PIMC results for the offdiagonal density matrix in coordinate space $n(s)$ [cf.~Eq.~(\ref{eq:offdiagonal}) above] in Fig.~\ref{fig:FSC_Offdiagonal}. Specifically, the left and right panels show results for spin-up and spin-down electrons, and the red circles, green crosses, and yellow starts correspond to $N=18$, $N=36$, and $N=60$, respectively. Evidently, both $n^\uparrow(s)$ and $n^\downarrow(s)$ converge towards unity in the limit of $s\to0$, as it is expected~\cite{cep,Militzer_momentum_HEDP_2019,dornheim2021ab}. Furthermore, the offdiagonal density matrix is of an approximately Gaussian shape, and no oscillations can be found at large separations $s$. This negligible impact of permutation cycles involving a large number of particles further helps to explain the small manifestation of finite-size effects at these conditions.

\begin{figure*}\centering
\includegraphics[width=0.475\textwidth]{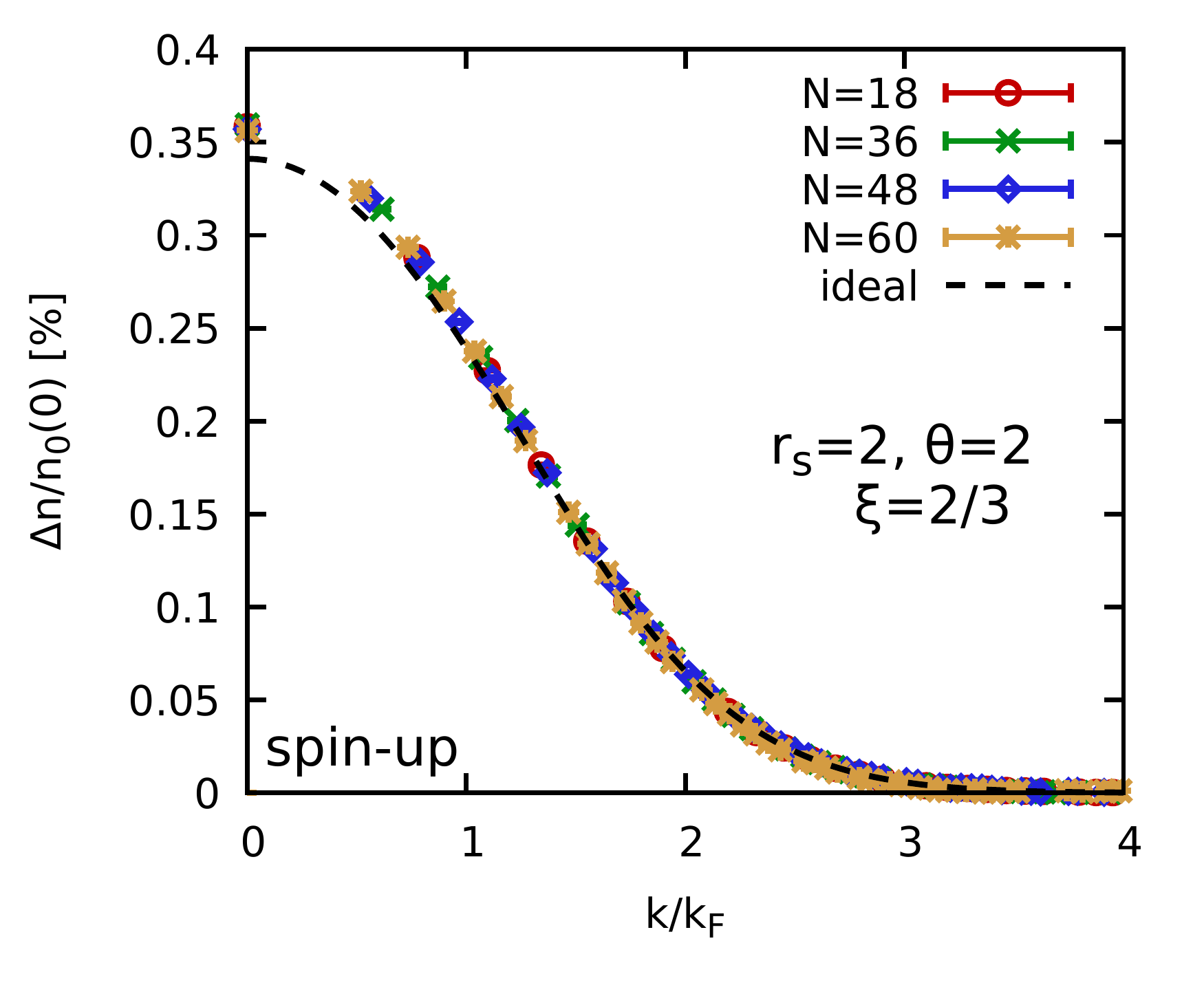}\includegraphics[width=0.475\textwidth]{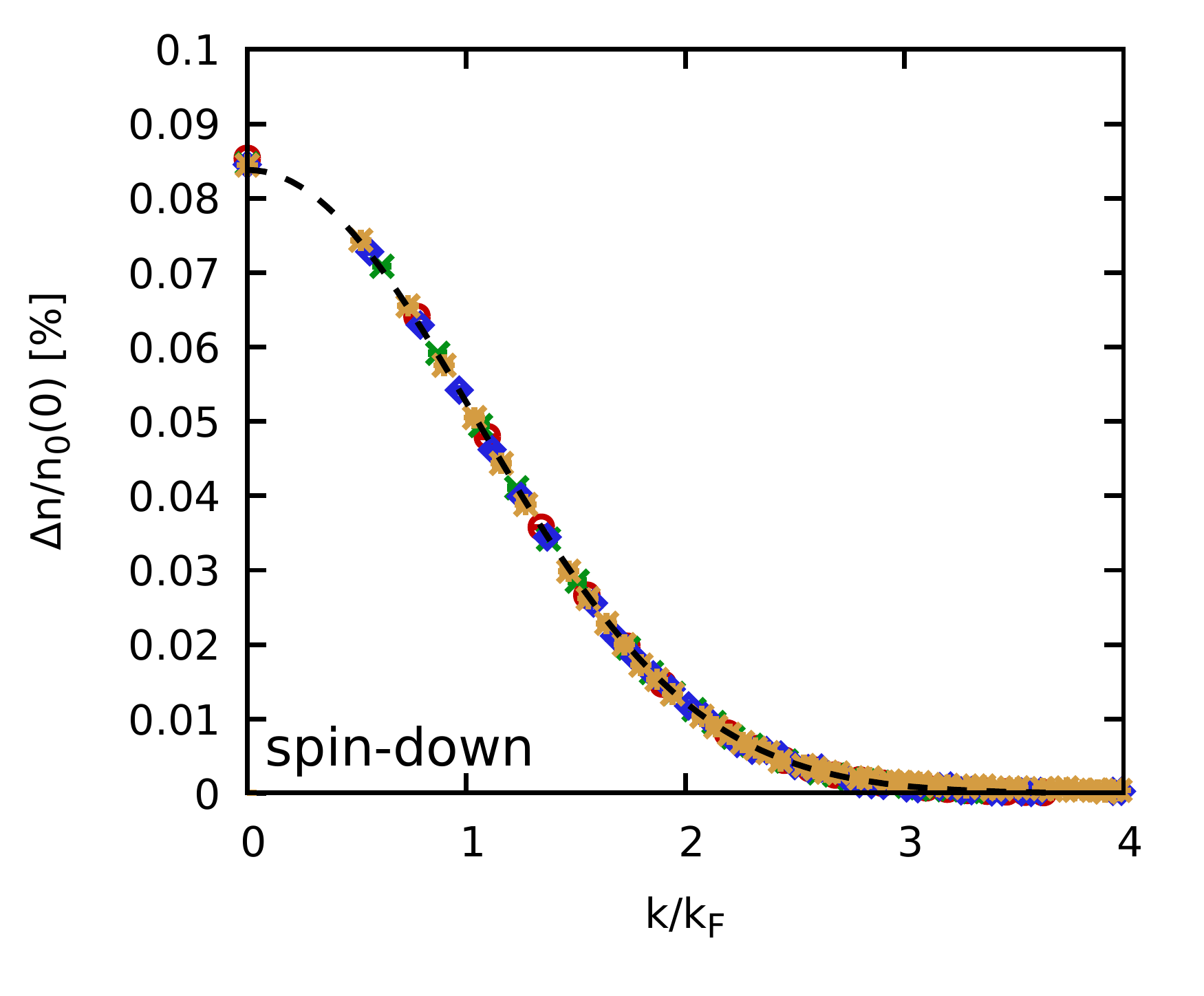}\\\vspace*{-1.03cm}
\hspace*{0.035\textwidth}\includegraphics[width=0.44\textwidth]{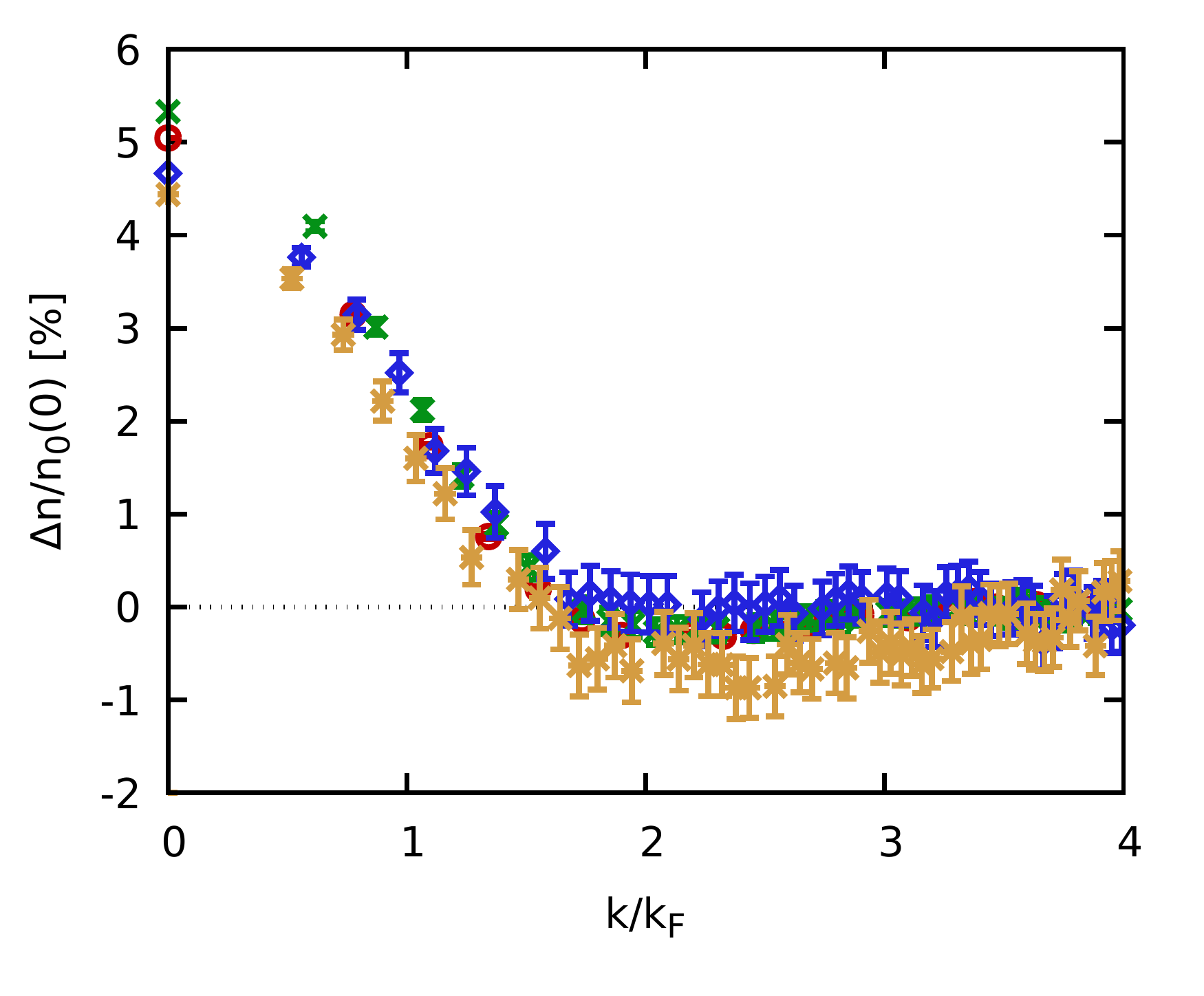}\hspace*{0.01\textwidth}\includegraphics[width=0.462\textwidth]{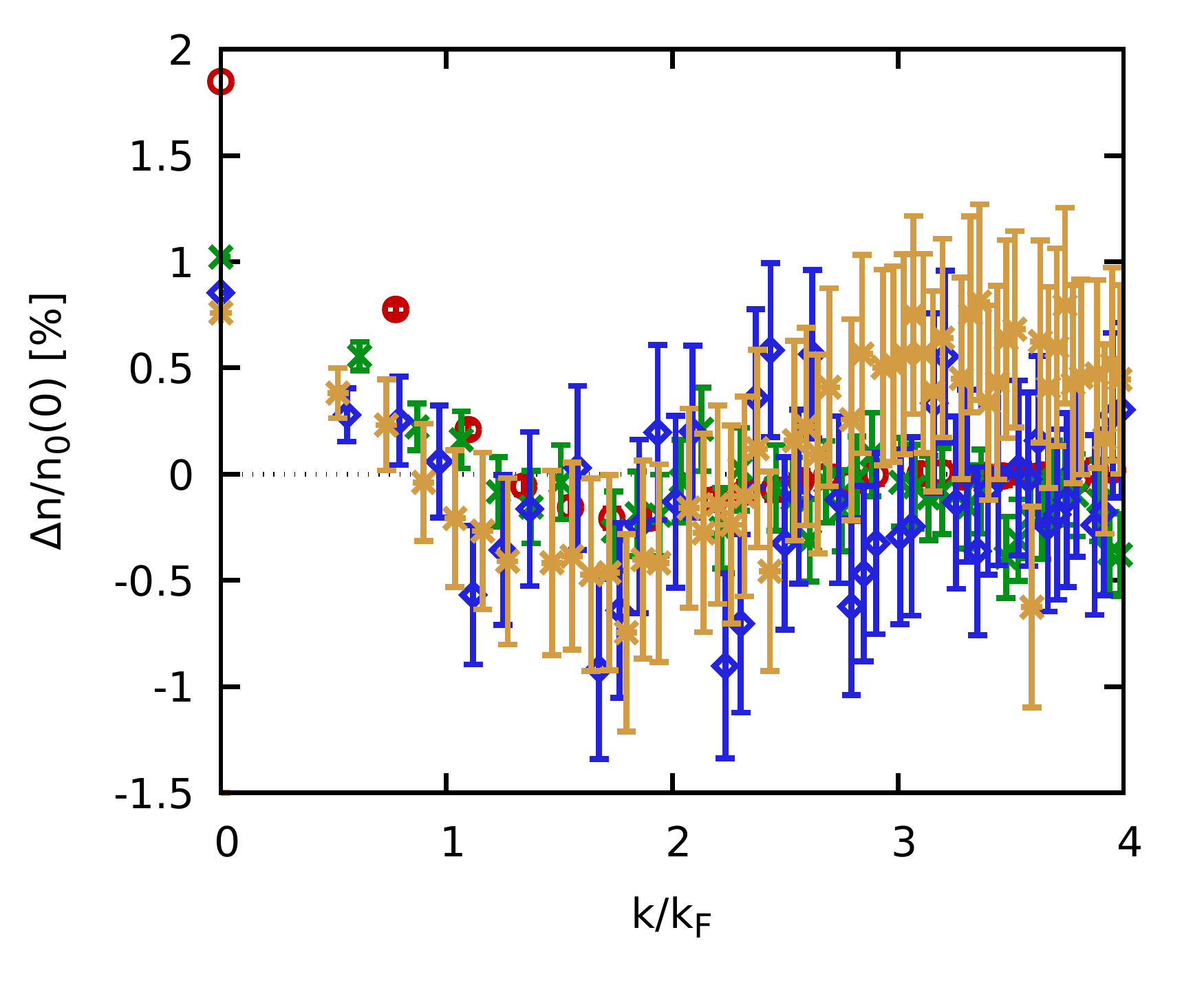}
\caption{\label{fig:FSC_rs2_theta2}
Top: System-size dependence of the momentum distribution of the UEG at $r_s=2$, $\theta=2$, and $\xi=2/3$. The left and right columns correspond to the spin-up and -down electrons, respectively. Bottom: Relative deviation between PIMC data for $n^\sigma(\mathbf{k})$ and the ideal Fermi distribution $n^\sigma_0(\mathbf{k})$ in per cent, cf.~Eq.~(\ref{eq:percent}).
Note that we use the Fermi wave number of the unpolarized system $k_\textnormal{F}^{\xi=0}$ as a reference.
}
\end{figure*}

While the small magnitude of the finite-size effects observed in Fig.~\ref{fig:FSC_rs10_theta2} are certainly encouraging, it is expected from previous studies~\cite{dornheim_prl,Dornheim_JCP_2021,review} that they might increase for higher densities. For this reason, we present a similar analysis for the same conditions at $r_s=2$ in Fig.~\ref{fig:FSC_rs2_theta2}. Physically, this corresponds to a metallic density that is highly relevant to contemporary WDM research, and can be realized experimentally for example with aluminum~\cite{Sperling_PRL_2015,Ramakrishna_PRB_2021,Dornheim_PRL_2020_ESA}. In addition, we note that the smaller value of the coupling parameter $r_s$ leads to a more pronounced impact of quantum degeneracy effects, which, in turn, results in a more severe fermion sign problem. This is conveniently characterized by the so-called \emph{average sign} $S$ (see, e.g., Ref.~\cite{dornheim_sign_problem} for details), which constitutes a straightforward measure for the amount of cancellations of positive and negative terms within the simulation. In particular, the required computation time scales as $1/S^2$, such that a value of $S=0.1$ would increase the CPU time by a factor of $100$ compared to a PIMC simulation without a sign problem.
For the system at hand, we find an average sign of $S\approx0.01$ in the extended ensemble for $N=60$ and $r_s=2$, whereas it is $S\approx0.4$ for $r_s=10$. Therefore, PIMC simulations of $N=90$ electrons are at present not computationally feasible at the higher density.

Returning to the topic of finite-size effects, again no deviations between the PIMC data sets for different $N$ can be seen with the naked eye in the top row of Fig.~\ref{fig:FSC_rs2_theta2} for either $n^\uparrow(\mathbf{k})$ or $n^\downarrow(\mathbf{k})$. The relative deviation to the ideal Fermi function shown in the bottom row allows for a more detailed perspective, where small differences between $N=18$ and $N=60$ of approximately $1\%$ can be resolved for small momenta. In any case, it is safe to conclude that direct PIMC simulations with $N\sim60$ electrons allow to reliably estimate the main physical features of the momentum distribution function as finite-size effects are small at the conditions that are considered in this work. For completeness, we mention that this changes at very low temperatures, where both the application of twisted boundary conditions~\cite{Lin_PRE_2001,Spink_PRB_2013} and an additional finite-size correction are required~\cite{Holzmann_PRL_2011}.

\subsubsection{Interplay of spin-polarization with density and temperature}

\begin{figure*}\centering
\includegraphics[width=0.495\textwidth]{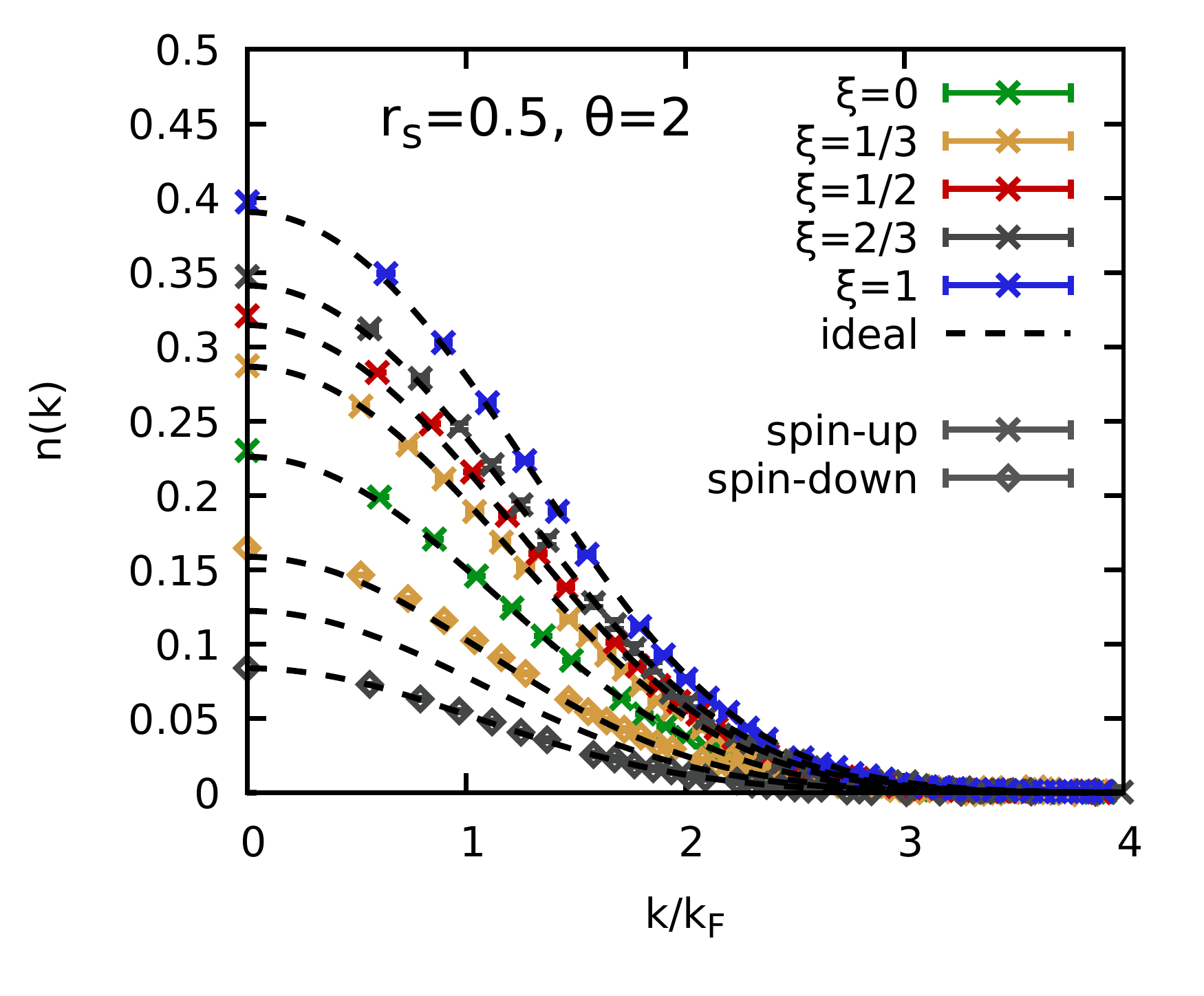}\includegraphics[width=0.495\textwidth]{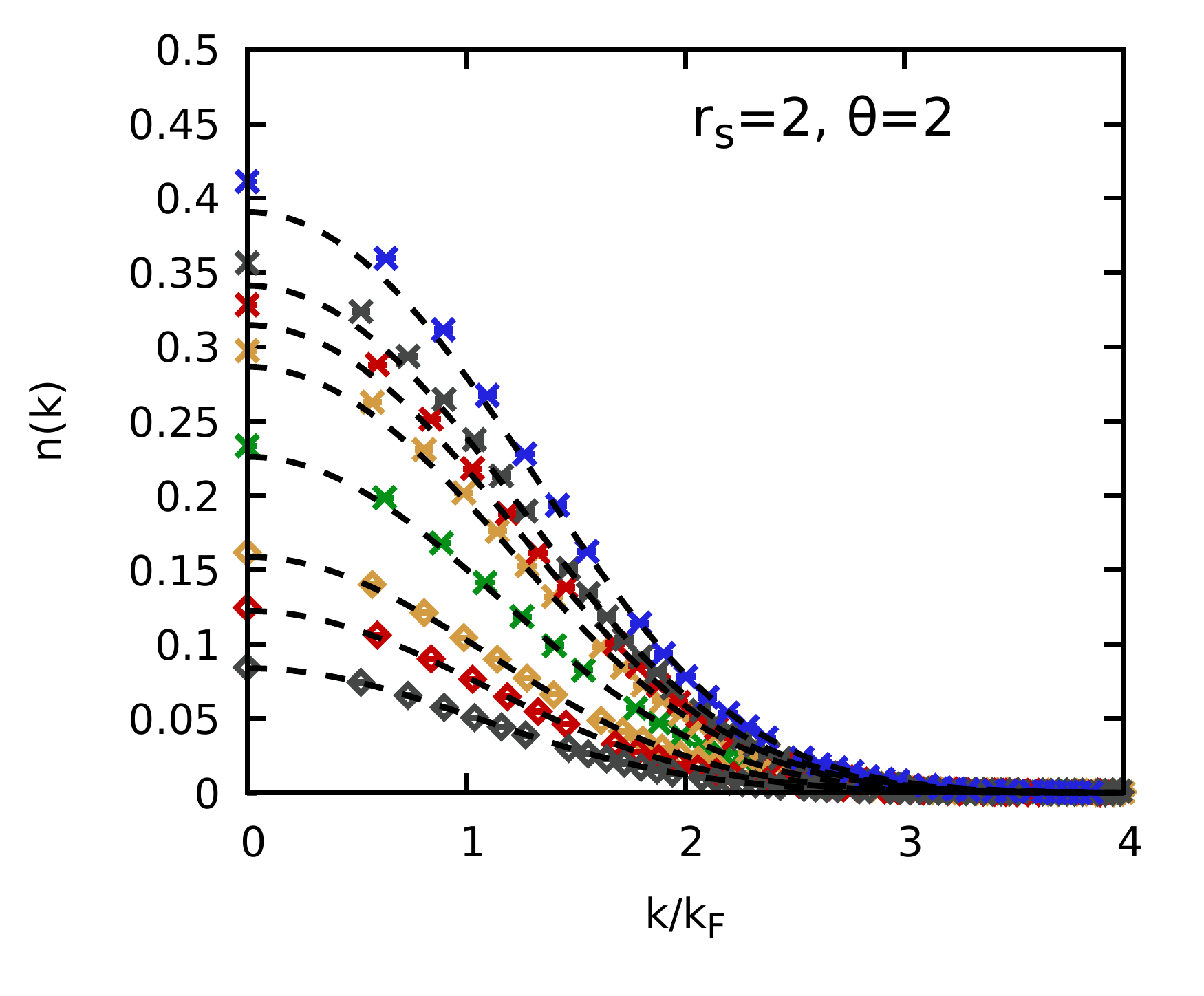}\\\vspace*{-1.02cm}\includegraphics[width=0.495\textwidth]{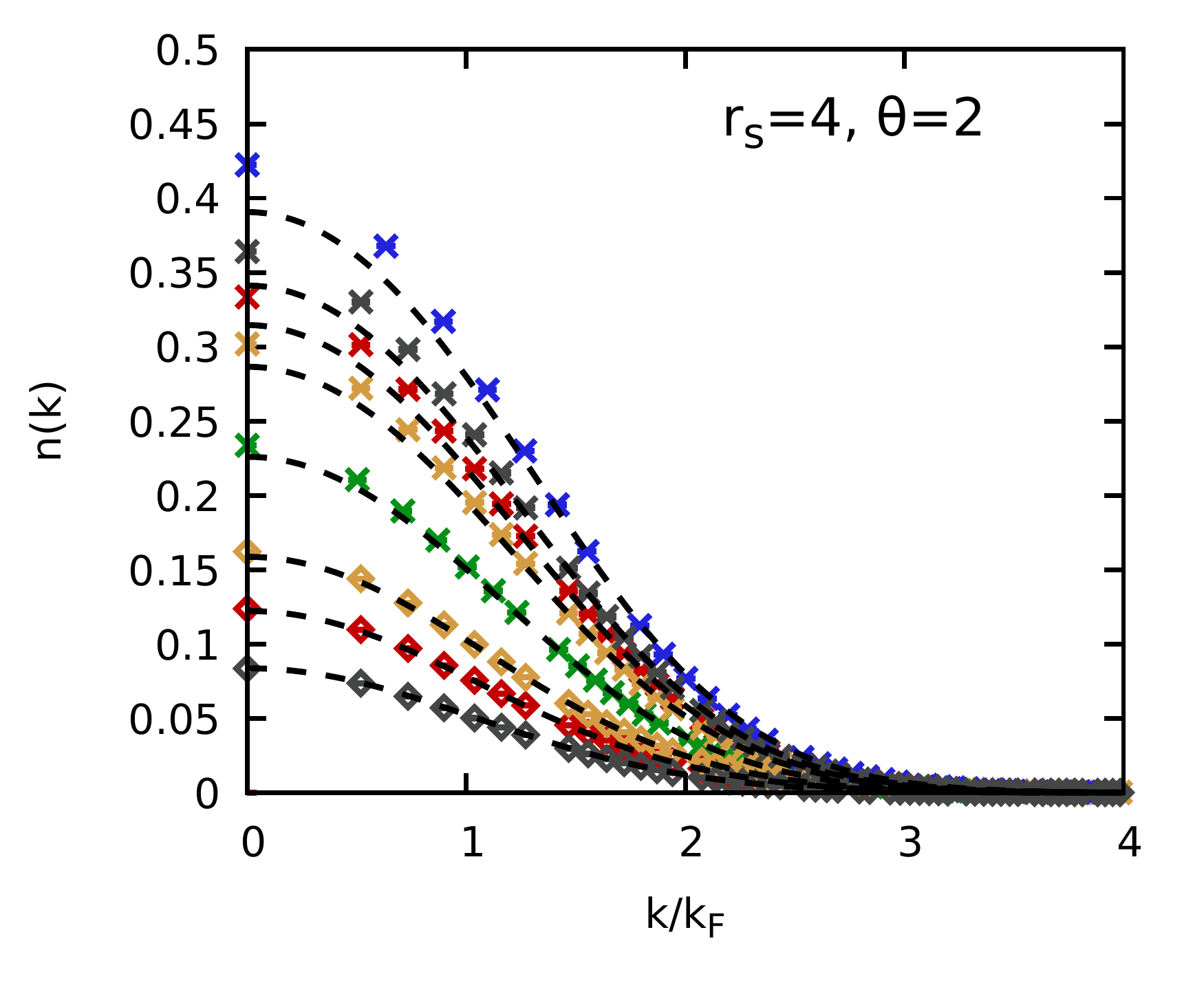}\includegraphics[width=0.495\textwidth]{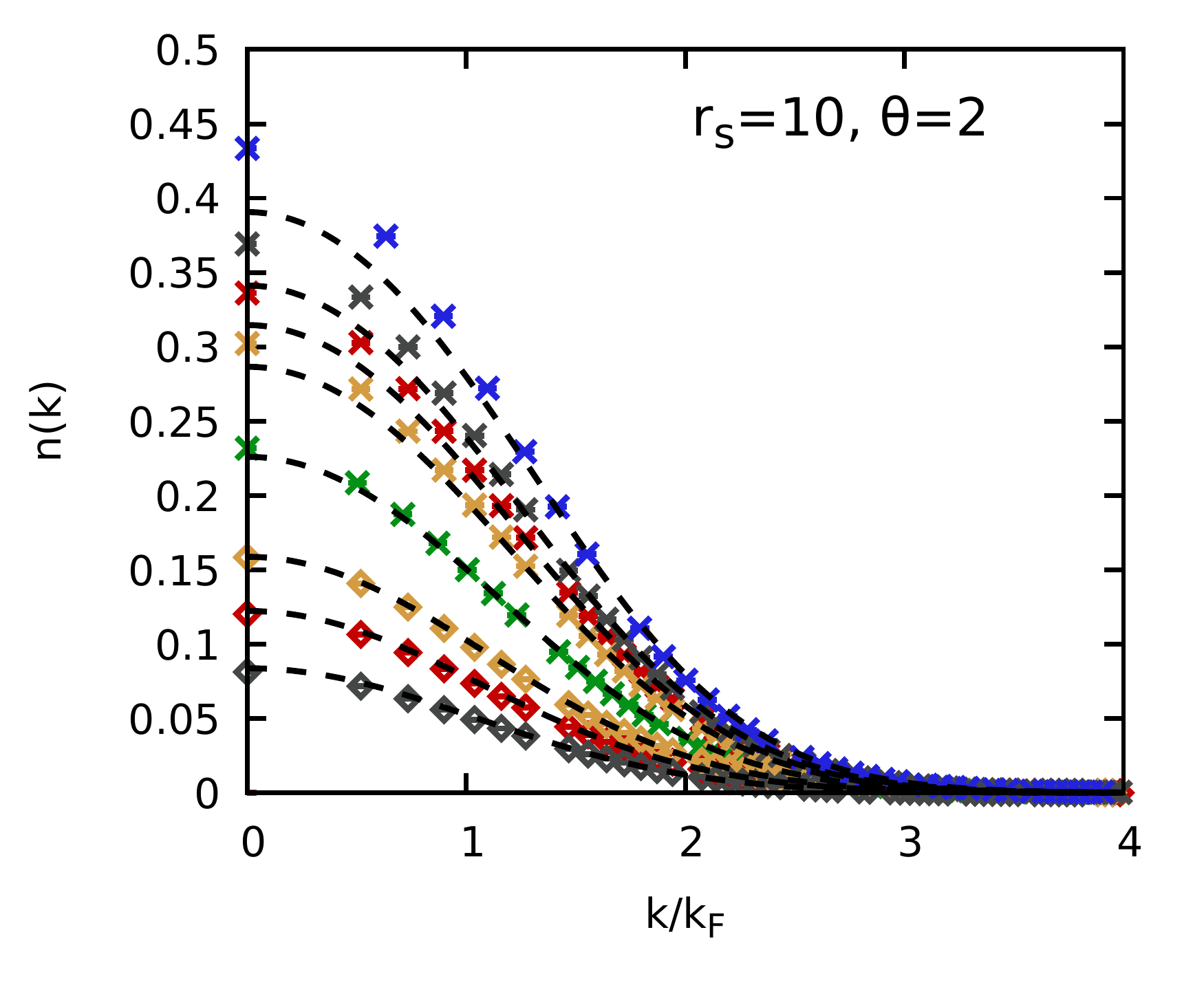}
\caption{\label{fig:rs2_theta2_xi_dependence}
Momentum distribution function $n^\sigma(\mathbf{k})$ for $\theta_{\xi=0}=2$. Top left: $r_s=0.5$; top right: $r_s=2$; bottom left: $r_s=4$; bottom right: $r_s=10$. The colours distinguish different values of the spin-polarization $\xi$, and the dashed black curves show the corresponding results for the ideal Fermi gas $n_0^\sigma(\mathbf{k})$.
Finally, the crosses and diamonds correspond to spin-up and spin-down electrons, respectively. Note that we use the Fermi wave number of the unpolarized system $k_\textnormal{F}^{\xi=0}$ as a reference. 
}
\end{figure*}

In the following, we will explicitly investigate the impact of the spin-polarization on the momentum distribution $n(\mathbf{k})$ and its spin-resolved components $n^\uparrow(\mathbf{k})$ and $n^\downarrow(\mathbf{k})$. To ensure a better comparability, we will always compare results for the same temperature $T$ for all $\xi$, thus resulting in different values of $\theta^\sigma$ [cf.~Eq.~(\ref{eq:theta_up_down})]. As a reference, we always give both $\theta$ and $k_\textnormal{F}$ for the case of a fully unpolarized system.

In Fig.~\ref{fig:rs2_theta2_xi_dependence}, we show the momentum distribution function of the UEG at $\theta=2$ for different values of $\xi$
for $r_s=0.5$ (top left), $r_s=2$ (top right), $r_s=4$ (bottom left), and $r_s=10$ (bottom right). More specifically, the crosses and diamonds show our new direct PIMC results for $n^\uparrow(\mathbf{k})$ and $n^\downarrow(\mathbf{k})$, and the dashed black lines depict the corresponding data for the ideal Fermi gas at the same conditions. For $r_s=0.5$, the PIMC data closely follow the ideal curves for all $\xi$ as electronic correlation effects are comparably small. In addition, we observe the following monotonous ordering of $n^\sigma(\mathbf{k})$: starting at $n^\uparrow(\mathbf{k})=n^\downarrow(\mathbf{k})$ for the unpolarized case (green crosses), both the ideal curves and the PIMC data for the spin-up electrons monotonically increase with $\xi$. This is mainly a consequence of the increase in the number density $n^\uparrow$ (or, equivalently, the decrease in the density parameter $r_s^\uparrow$, cf.~Fig.~\ref{fig:xi_scale}). Conversely, we find the opposite behaviour for the spin-down electrons, with $n^\downarrow(\mathbf{k})=0$ in the limit of $\xi=1$.

\begin{figure*}\centering
\includegraphics[width=0.475\textwidth]{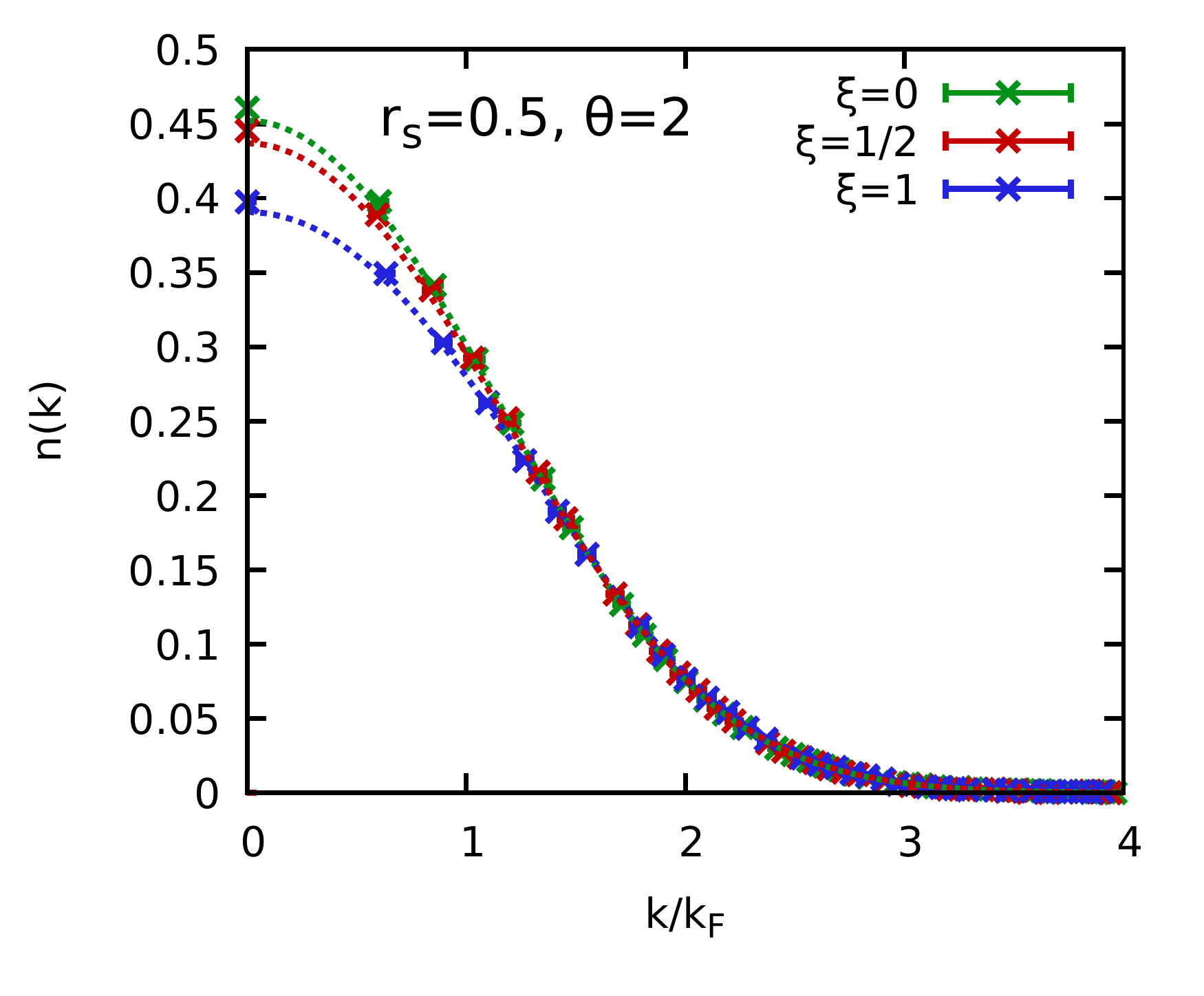}\includegraphics[width=0.475\textwidth]{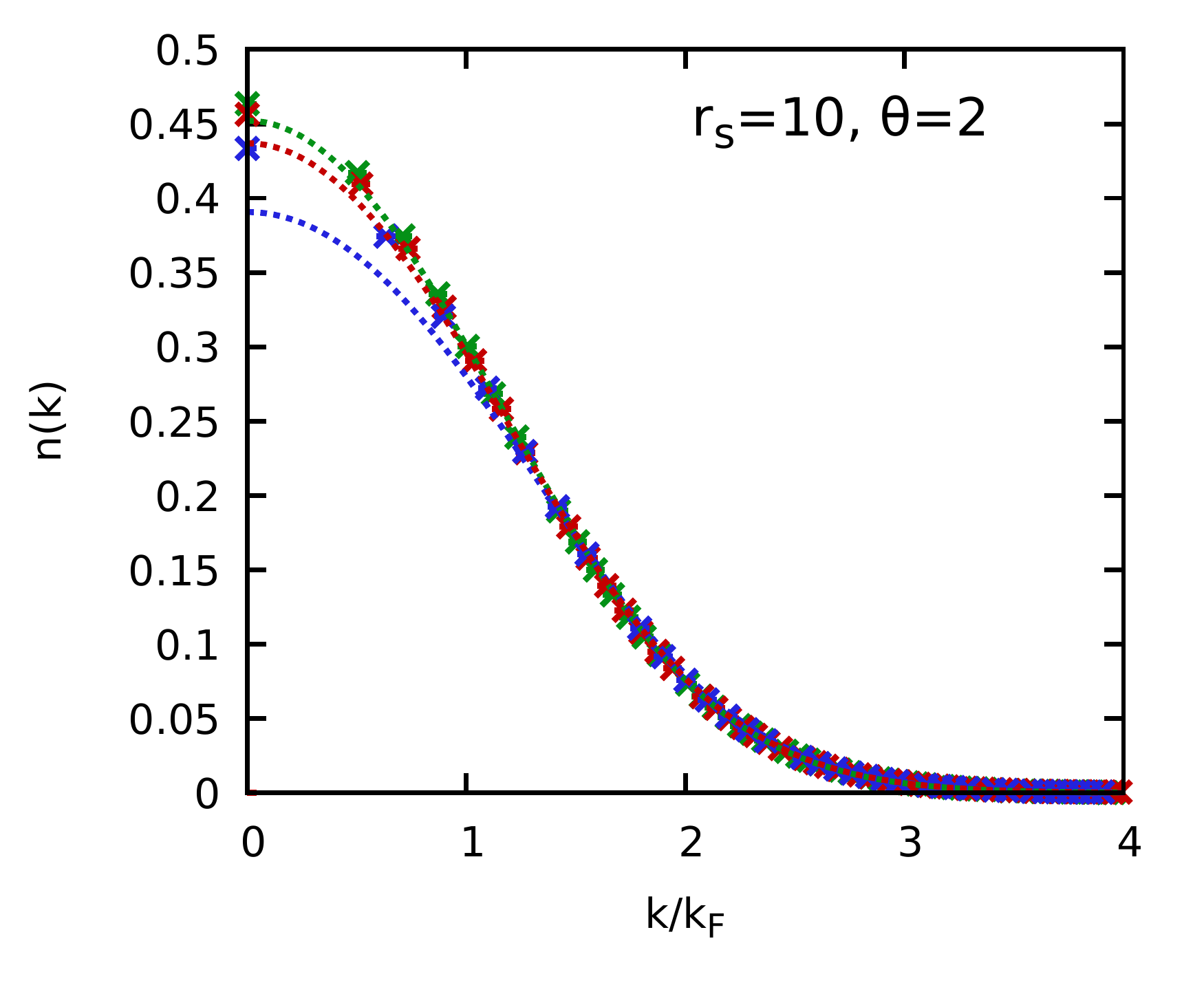}
\caption{\label{fig:xi_total}
Total momentum distribution function $n(\mathbf{k})$ at $\theta_{\xi=0}=2$. Left: $r_s=0.5$; right: $r_s=10$. The colours distinguish different values of the spin-polarization $\xi$, and the corresponding dotted curves show results for the ideal Fermi gas $n_0^\sigma(\mathbf{k})$. Note that we use the Fermi wave number of the unpolarized system $k_\textnormal{F}^{\xi=0}$ as a reference.
}
\end{figure*} 

A less obvious question is the behaviour of the total momentum distribution function $n(\mathbf{k})=n^\uparrow(\mathbf{k})+n^\downarrow(\mathbf{k})$, which we show in the left panel of Fig.~\ref{fig:xi_total} for these conditions. We note that the curves and data points for $\xi=1/3$ and $\xi=2/3$ have been omitted for better visibility. Interestingly, we observe the opposite ordering compared to $n^\uparrow(\mathbf{k})$ shown above, i.e., the largest value around zero momentum occurs for $\xi=0$, both in the PIMC data and the ideal Fermi distribution function. This is a direct consequence of the fermionic anti-symmetry under particle exchange and the resulting Pauli blocking, and can be understood as follows: In the limit of $\xi=1$ all electrons in the system are mutually affected by their common fermionic nature, which effectively pushes them towards larger momenta. In the opposite limit of $\xi=0$, only half of the electrons mutually affect each other, and, consequently, the fermionic push towards larger momenta is weaker. Naturally, $\xi=1/2$ is located between these two extremes, and, thus, located somewhere in the middle.

Let us next consider the impact of an increasing density parameter $r_s$ on the spin-resolved components $n^\uparrow(\mathbf{k})$ and $n^\downarrow(\mathbf{k})$. This is shown in the top right panel of Fig.~\ref{fig:rs2_theta2_xi_dependence} for the case of $r_s=2$. Evidently, the effect of the coupling strength is most pronounced for the spin-up electrons, where in particular the occupation of the zero-momentum state is substantially increased compared to the ideal Fermi distribution. Further, we observe that this effect increases with $\xi$. The spin-down electrons, on the other hand, can hardly be distinguished from $n_0^\downarrow(\mathbf{k})$ with the naked eye on the depicted scale.

The further increase of the coupling strength to $r_s=4$ (bottom left) leads to a further increase in the PIMC data for $n^\uparrow(0)$ compared to $n_0^\uparrow(0)$, whereas the spin-down electrons, again, remain hardly affected by correlations. Finally, the bottom right panel shows results for $r_s=10$, which is located at the margins of the strongly coupled electron liquid regime~\cite{dornheim_electron_liquid,dornheim_dynamic}. First and foremost, we note that the correlation-induced increase in $n^\uparrow(0)$ compared to $n_0^\uparrow(0)$ strongly depends on $\xi$ in this case, and is most pronounced at $\xi=1$. In addition, we observe a correlation-induced decrease for the spin-down electrons, although it is substantially smaller. 

In order to get a more complete picture of the physics at play, we analyse the total momentum distribution $n(\mathbf{k})$ for $r_s=10$ in the right panel of Fig.~\ref{fig:xi_total}. As a direct consequence of the increased coupling strength, the electrons are more strongly spatially separated, and quantum degeneracy effects are less pronounced. Therefore, spin-effects play a substantially smaller role in the direct PIMC data for the UEG than for the ideal (dotted) curves. The large correlation-induced increase in $n^\uparrow(0)$ for large $\xi$ is thus caused by the substantial spin dependence of $n_0(0)$, which is masked for the UEG by the Coulomb repulsion at low densities. Furthermore, the data for the UEG are closer to the unpolarized ideal curve than its polarized analogue, as spin effects are less important for the ideal Fermi gas at $\xi=0$ as well.

\begin{figure*}\centering
\includegraphics[width=0.475\textwidth]{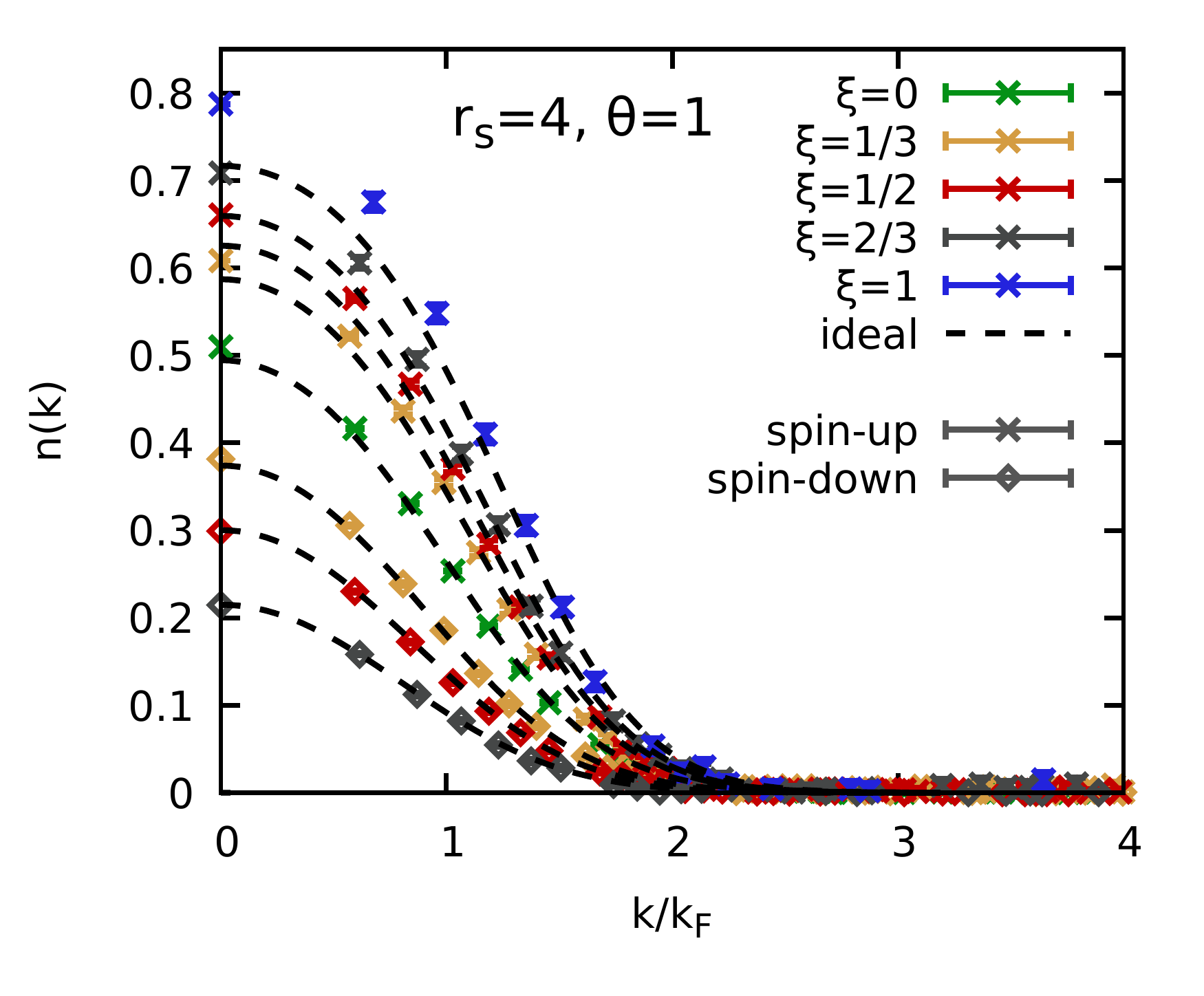}\includegraphics[width=0.475\textwidth]{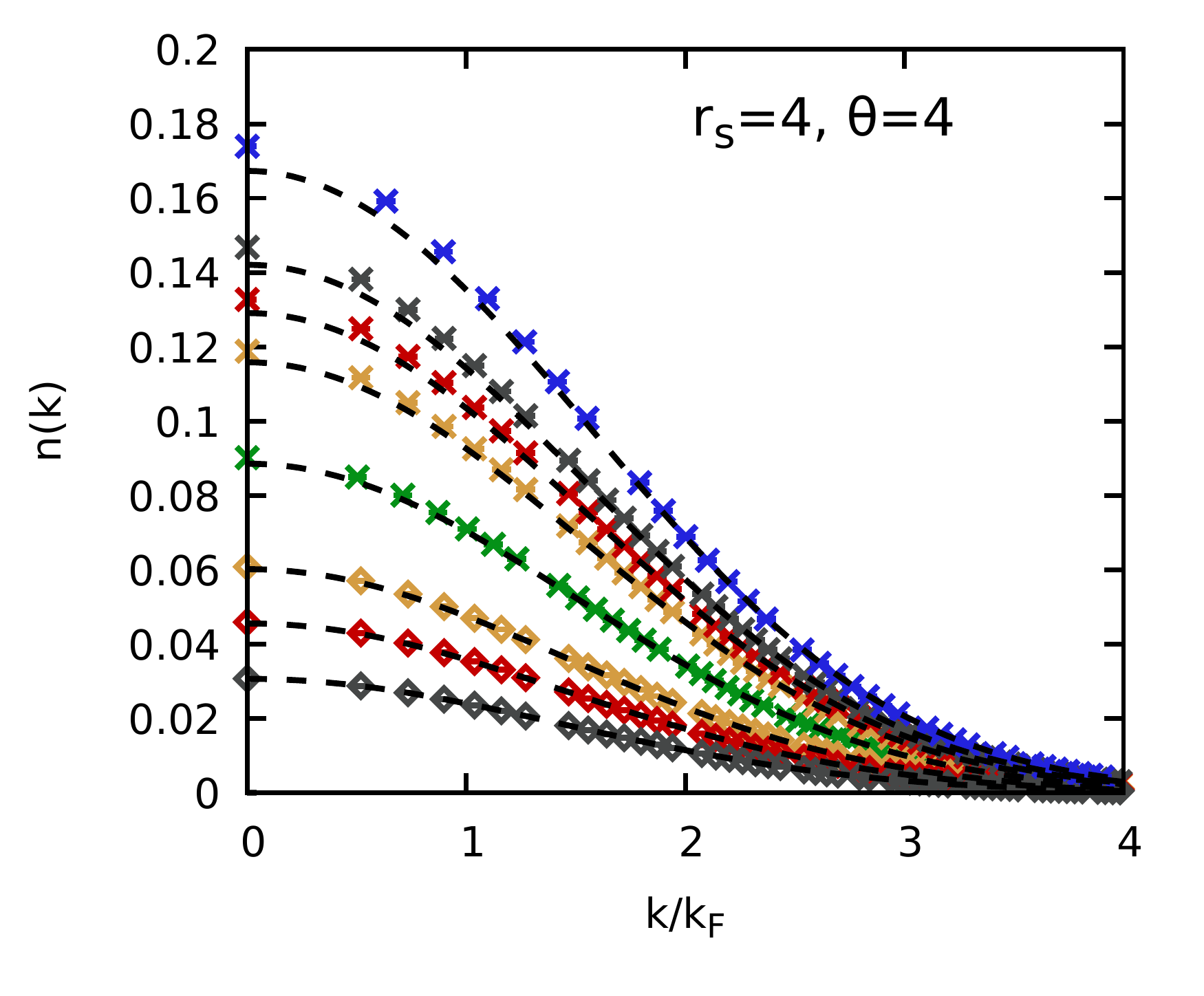}\\\vspace*{-1.01cm}\includegraphics[width=0.475\textwidth]{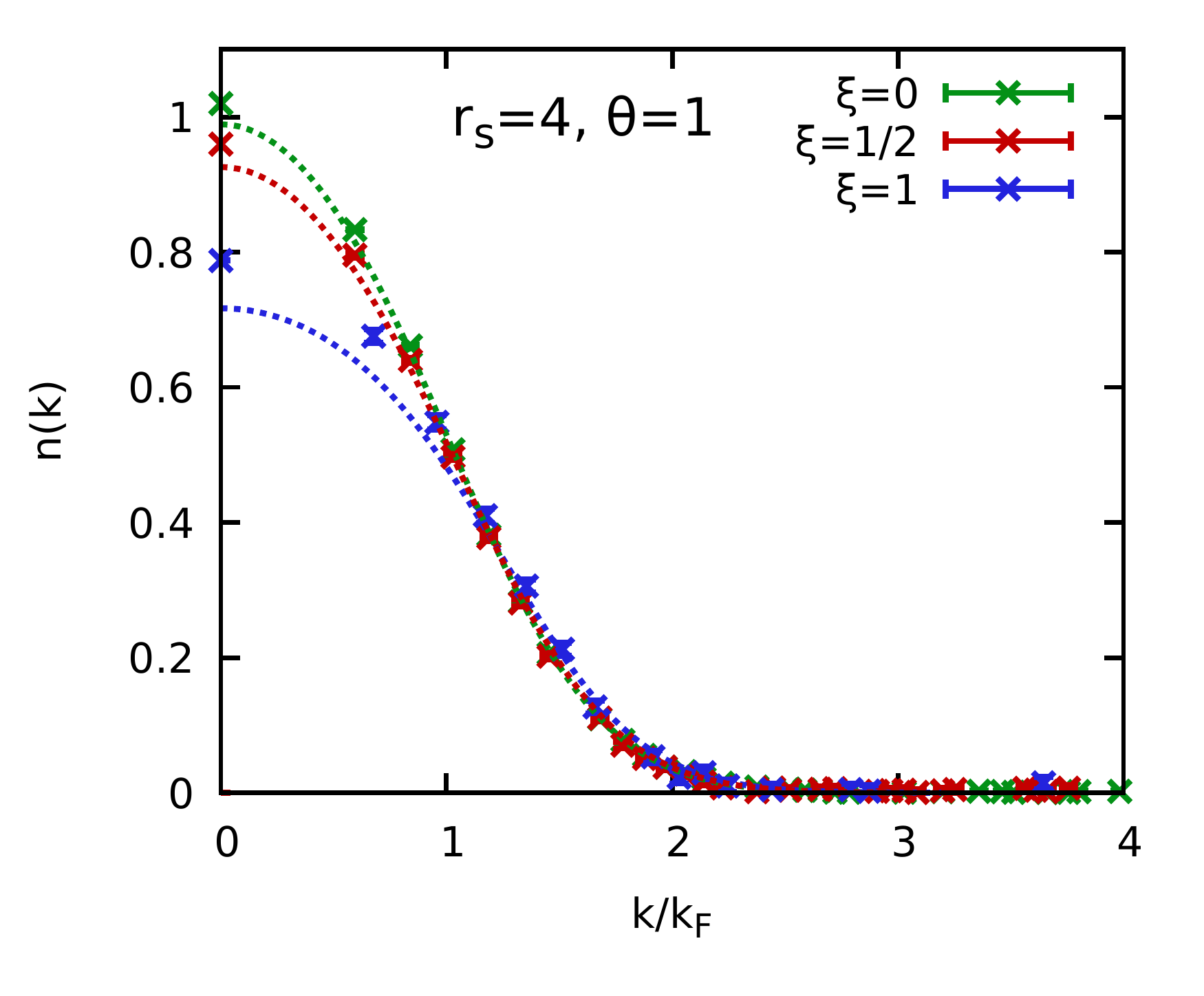}\includegraphics[width=0.475\textwidth]{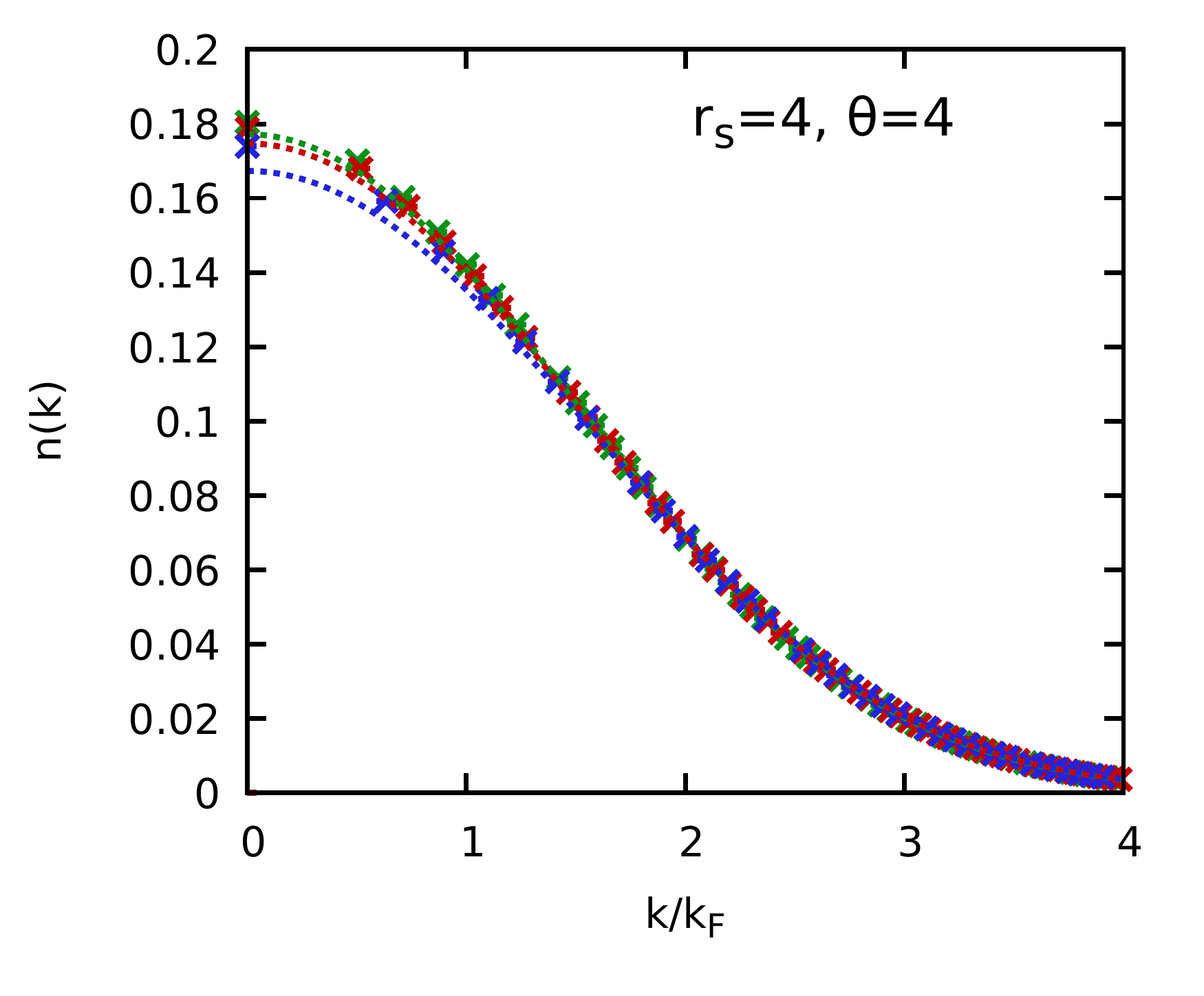}
\caption{\label{fig:xi_T}
Top: Spin-resolved momentum distribution function $n^\sigma(\mathbf{k})$ for $r_s=2$. Left: $\theta_{\xi=0}=1$; right: $\theta_{\xi=0}=4$. The colours distinguish different values of the spin-polarization $\xi$, and the dashed black curves show the corresponding results for the ideal Fermi gas $n_0^\sigma(\mathbf{k})$.
Finally, the crosses and diamonds correspond to spin-up and spin-down electrons, respectively.
Bottom: Corresponding total momentum distributions $n(\mathbf{k})=n^\uparrow(\mathbf{k})+n^\downarrow(\mathbf{k})$.
Note that we use the Fermi wave number of the unpolarized system $k_\textnormal{F}^{\xi=0}$ as a reference.
}
\end{figure*}

Let us next investigate the effect of the temperature on the spin dependence of the momentum distribution, which is analyzed in Fig.~\ref{fig:xi_T} for the metallic density of $r_s=4$. The left column corresponds to $\theta_{\xi=0}=1$, and the top panel shows results for the spin-resolved components $n^\uparrow(\mathbf{k})$ (crosses) and $n^\downarrow(\mathbf{k})$ (diamonds). Evidently, the reduction of the temperature by a factor of one half compared to Fig.~\ref{fig:rs2_theta2_xi_dependence} leads to a more pronounced difference in $n^\uparrow(0)$, in particular for the fully ferromagnetic case. The spin-down electrons, on the other hand, are hardly affected by the Coulomb repulsion for $\xi>0$. The bottom panel shows the corresponding results for the total momentum distribution $n(\mathbf{k})$, where again the impact of the spin-effects is less pronounced for the UEG compared to the ideal Fermi gas.

The right column of Fig.~\ref{fig:xi_T} shows the same investigation for a higher temperature, $\theta_{\xi=0}=4$. Firstly, we note that all curves are substantially broadened by thermal excitations, as it is expected. Furthermore, the correlation-induced increase in the zero momentum state is less pronounced than at lower temperatures, and will eventually completely vanish in the limit of large $T$ when the system becomes increasingly ideal. Furthermore, $n^\downarrow(\mathbf{k})$ can hardly be distinguished from the corresponding ideal curves with the naked eye. Considering the total momentum distribution function $n(\mathbf{k})$ depicted in the bottom panel, we find that the deviations between the curves for the different values of $\xi$ are substantially smaller compared to the cases of $\theta_{\xi=0}=2$ and $\theta_{\xi=0}=1$ both in the PIMC data for the UEG, and in the ideal results. This is expected as spin-effects, too, will completely vanish in the limit of large temperatures, where the system becomes classical. In particular, $n(\mathbf{k})$ will converge towards the well-known Boltzmann distribution in this regime.

\subsubsection{Zero-momentum occupation and exchange--correlation kinetic energy\label{sec:n0_kxc}}

\begin{figure}\centering
\includegraphics[width=0.475\textwidth]{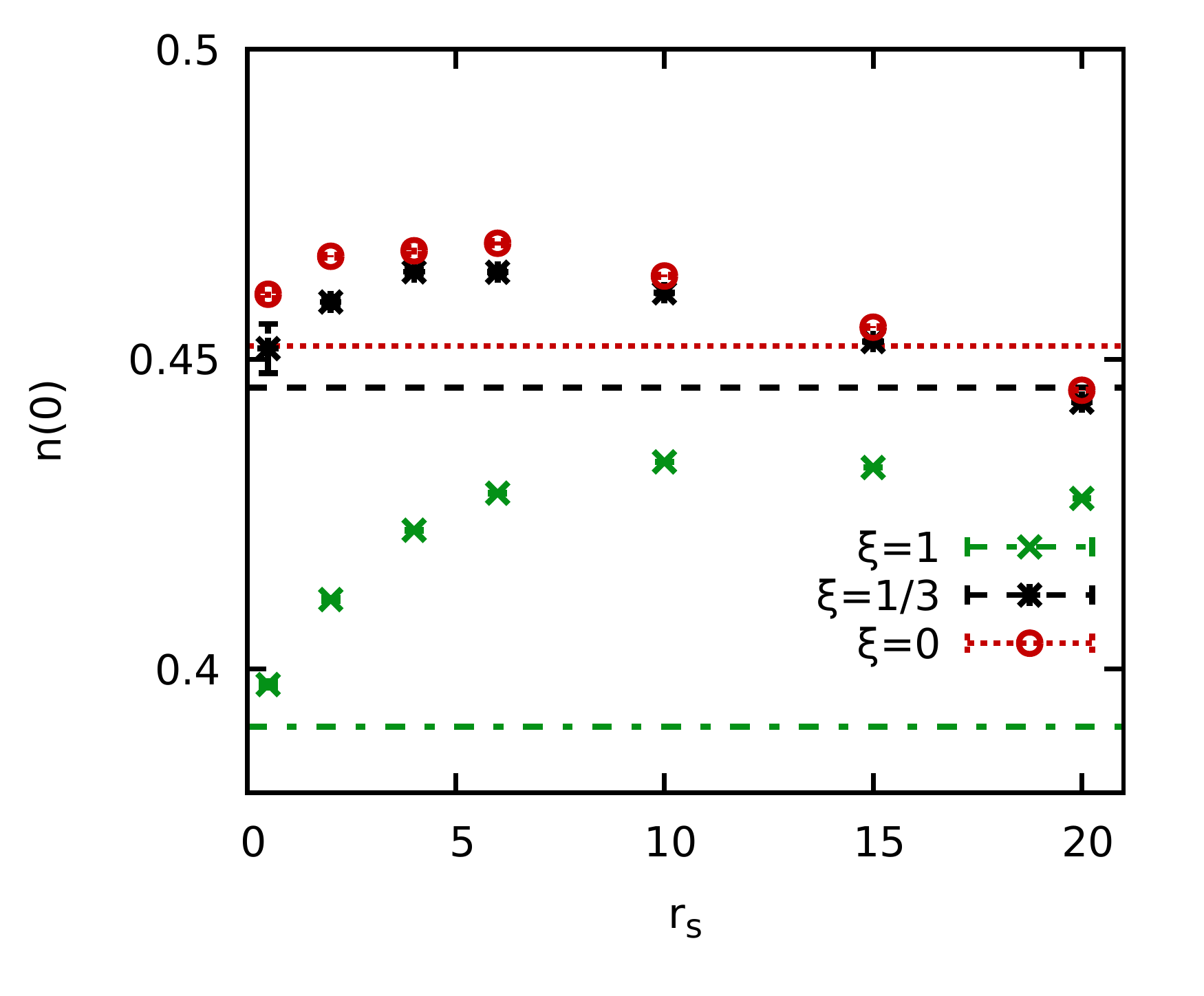}\\\vspace*{-1.01cm}
\includegraphics[width=0.475\textwidth]{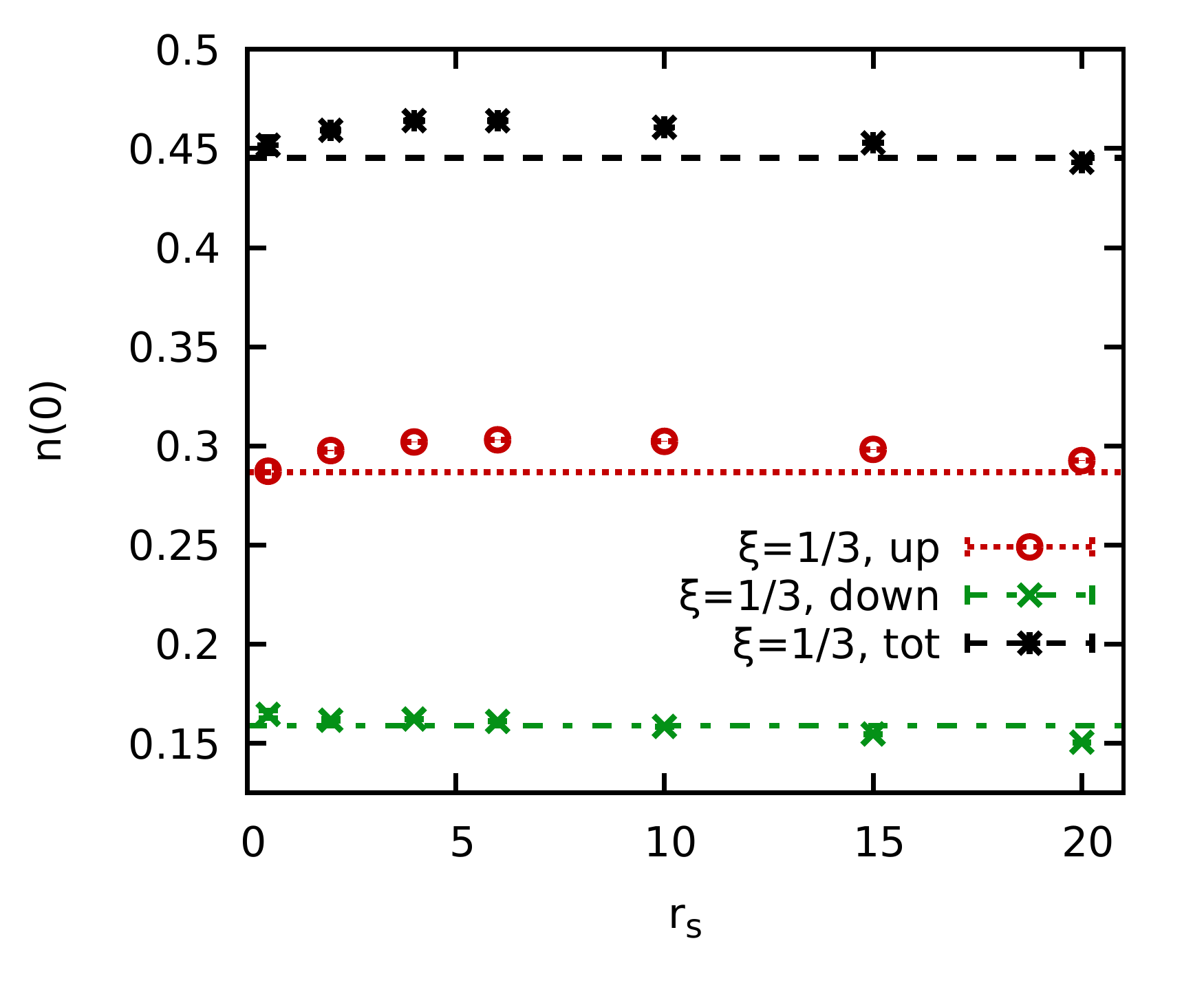}
\caption{\label{fig:xi_n0_theta2_rs}
Density dependence of the occupation at zero momentum $n(0)$ of the UEG at $\theta_{\xi=0}=2$. Top panel: direct PIMC data for the total momentum distribution $n(0)=n^\uparrow(0)+n^\downarrow(0)$ for $\xi=0$ (red circles), $\xi=1/3$ (black stars), and $\xi=1$ (green crosses). The horizontal lines depict the corresponding ideal values $n_0(0)$. Bottom panel: direct PIMC results for $\xi=1/3$, black stars: $n(\mathbf{k})$; red circles: $n^\uparrow(\mathbf{k})$; green crosses: $n^\downarrow(\mathbf{k})$.
}
\end{figure}

Let us conclude our investigation with a more detailed study of the interaction-induced change in the occupation at zero momentum and the related lowering of the kinetic energy. As a first step, we show the density dependence of the total momentum distribution $n(0)$ at $\theta_{\xi=0}$ for three different values of the spin-polarization $\xi$ in the top panel of Fig.~\ref{fig:xi_n0_theta2_rs}.
More specifically, the red circles, black stars, and green crosses show our new direct PIMC data for the UEG for $\xi=0$, $\xi=1/3$, and $\xi=1$, respectively, and the horizontal lines depict the corresponding ideal values $n_0(0)$ that do not depend on the density. 
For small $r_s$, all three data sets exhibit a qualitatively similar behaviour and monotonically increase starting from the ideal value at $r_s=0$. In addition, the data points for $\xi=0$ and $\xi=1/3$ remain close to each other over the entire depicted $r_s$-range and almost agree with each other at $r_s=20$. This is, of course, expected, as spin-effects will eventually completely vanish at large $r_s$ due to the increased coupling strength~\cite{dornheim_electron_liquid}. Furthermore, the PIMC results for $n(0)$ at $r_s=20$ is below the ideal value for both of these spin-polarization, as the electrons are pushed towards larger momenta by the Coulomb coupling. 

In contrast, the green crosses exhibit a related, but clearly distinct progression. In particular, the interaction-induced increase in $n(0)$ is substantially larger compared to the other data sets and attains a maximum for $r_s\sim10$. This has already been explained above by the more pronounced spin dependence of $n_0(0)$ compared to $n(0)$ of the UEG, and is thus directly caused by the large gap between the respective Fermi functions for the different values of $\xi$. 

For completeness, we also show the $r_s$ dependence of the spin-resolved components of the momentum distribution function $n^\uparrow(0)$ and $n^\downarrow(0)$ at $\xi=1/3$ in the bottom panel of Fig.~\ref{fig:xi_n0_theta2_rs}. This plot further substantiates our previous findings that the correlation-induced increase in the occupation of the zero-momentum state is mostly caused by the spin-up electrons. In addition, we find that the PIMC data for $n^\downarrow(0)$ are actually smaller than $n^\downarrow_0(0)$ for $r_s\gtrsim10$, whereas the opposite still holds for $n^\uparrow(0)$. A possible explanation for this effect is given by the comparably increased spin-resolved density parameter $r_s^\downarrow$ shown in Fig.~\ref{fig:xi_scale} above.

\begin{figure}\centering
\includegraphics[width=0.475\textwidth]{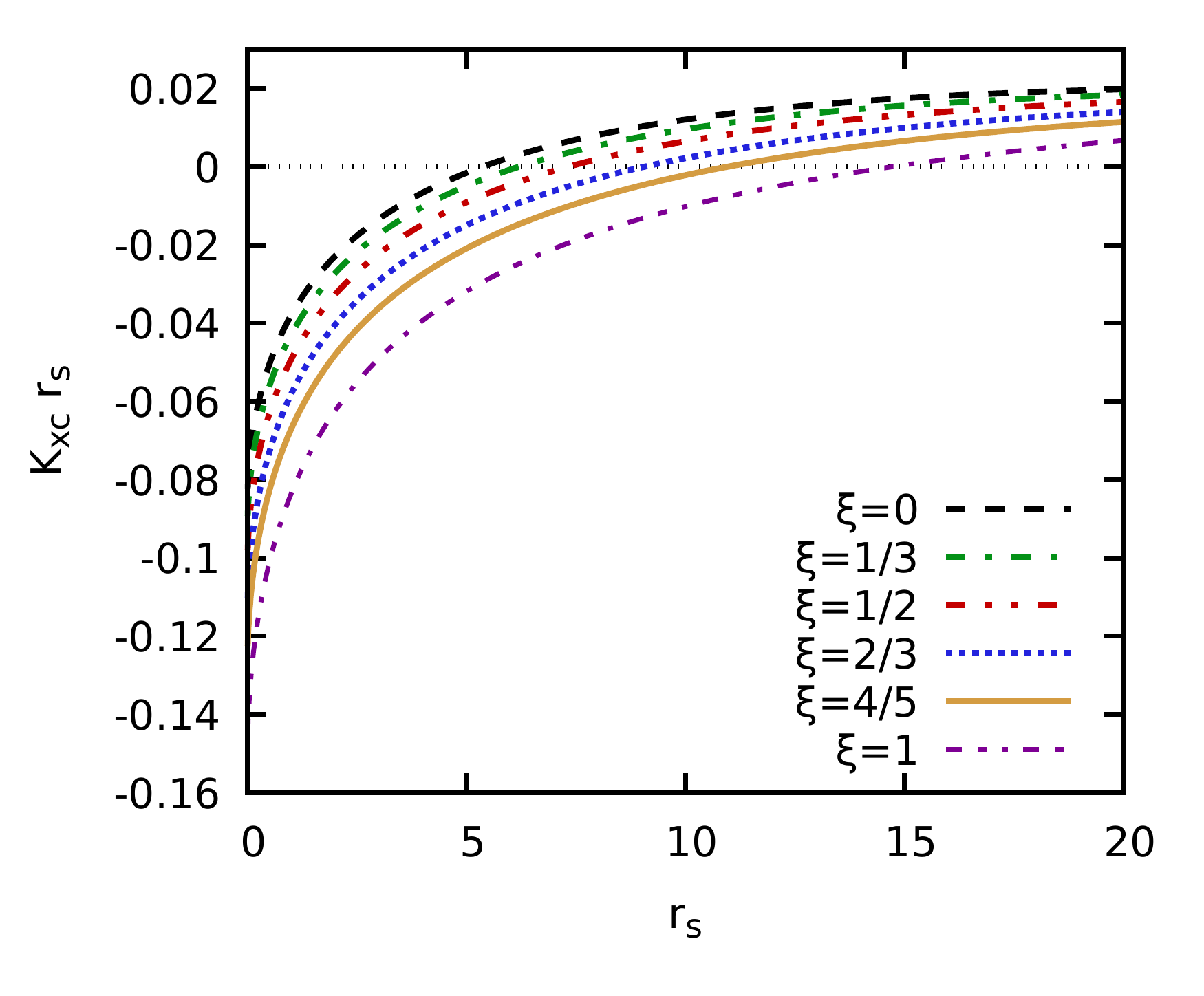}
\caption{\label{fig:Kxc_rs}
Density dependence of the exchange--correlation part of the kinetic energy $K_\textnormal{xc}$ at $\theta_{\xi=0}=2$ for different values of the spin-polarization $\xi$ evaluated from the parametrization of $f_\textnormal{xc}$ by Groth \textit{et al.}~\cite{groth_prl} via Eq.~(\ref{eq:Kxc}).
}
\end{figure}

Let us next consider Fig.~\ref{fig:Kxc_rs}, where we show the density dependence of the exchange--correlation part to the kinetic energy $K_\textnormal{xc}$ (obtained by evaluating Eq.~(\ref{eq:Kxc}) using as input the parametrization of $f_\textnormal{xc}$ by Groth \textit{et al.}~\cite{groth_prl}) for different values of the spin-polarization $\xi$. First and foremost, we note the similar progression of $K_\textnormal{xc}\cdot r_s$ for all $\xi$, which attain a finite negative value in the limit of $r_s\to0$, monotonously increase with $r_s$, and eventually become positive. In addition, we find a strict ordering of these curves with $\xi$, and the distancing between individual curves increases with the spin-polarization. This is again directly caused by the comparably larger spin dependence of the ideal energy $E_0$ for large $\xi$, whereas the actual kinetic energy $K$ of the UEG is less affected by the spin, and thus more closely resembles $E_0$ for the unpolarized case. 
From a physical perspective, we note that the behaviour observed in Fig.~\ref{fig:Kxc_rs} might indicate a substantial \emph{negative tail} at large momenta in the static local field correction of the ferromagnetic UEG at WDM conditions. This prediction can be verified by independent PIMC simulations of the spin-polarized UEG, which constitutes an interesting project for future research; see also the outlook in Sec.~\ref{sec:summary} for more details.

\begin{figure}\centering
\includegraphics[width=0.475\textwidth]{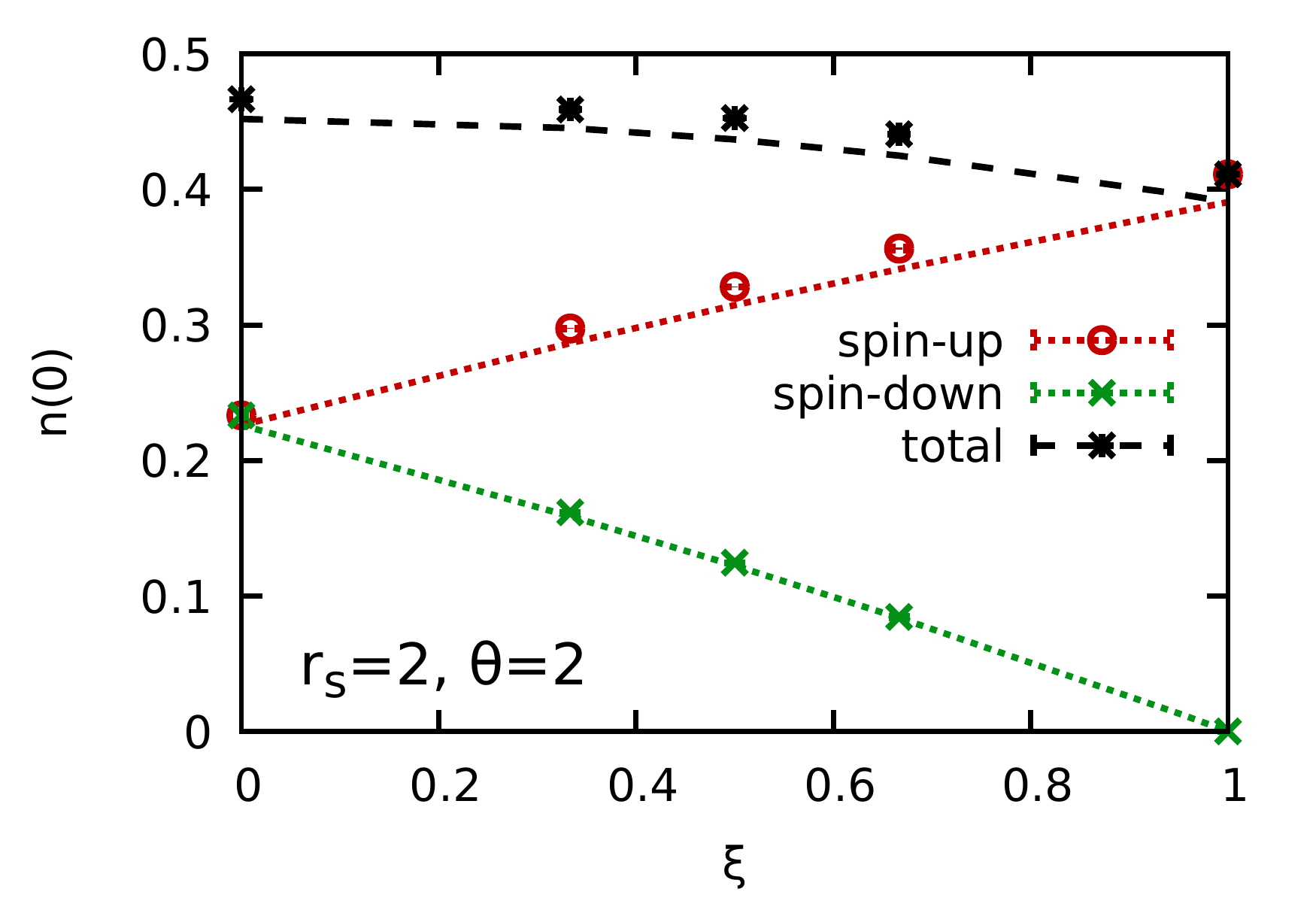}\\\vspace*{-1.01cm}\includegraphics[width=0.475\textwidth]{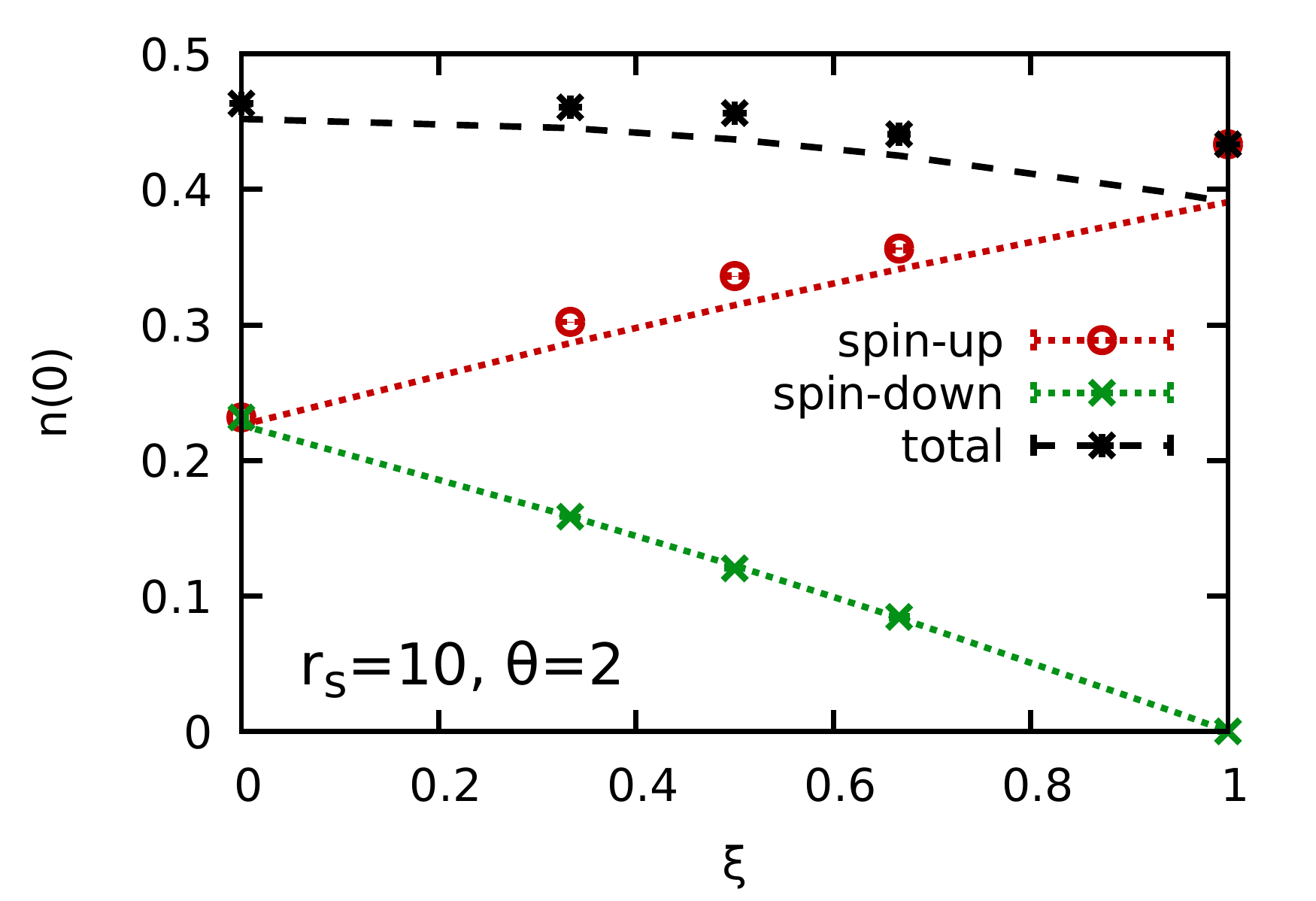}\\\vspace*{-1.01cm}\includegraphics[width=0.475\textwidth]{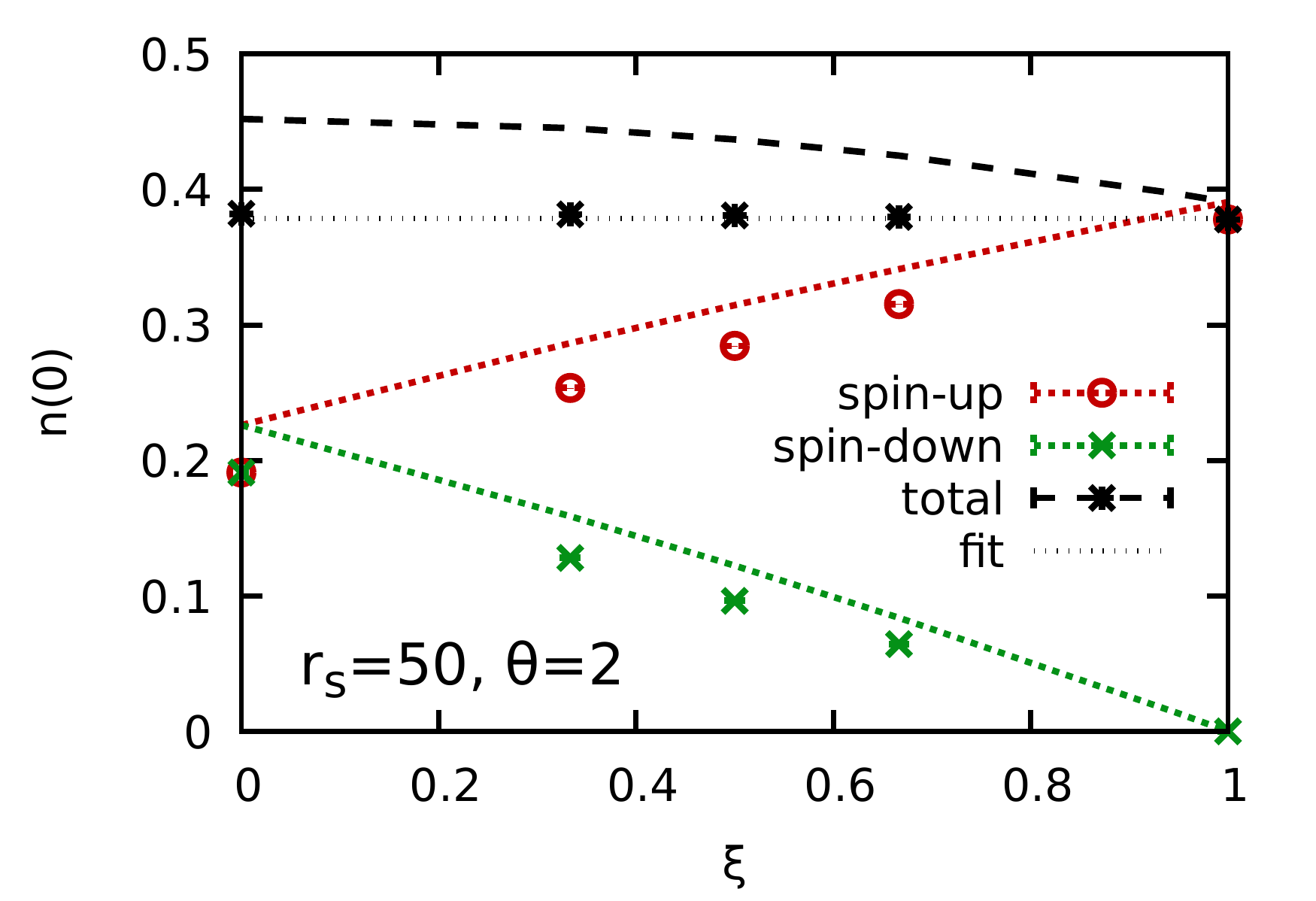}
\caption{\label{fig:xi_n0_rs2}
Polarization dependence of the occupation at zero momentum for $r_s=2$ (top), $r_s=10$ (center), and $r_s=50$ (bottom) at $\theta_{\xi=0}=2$. The red circles, green crosses, and black stars correspond to the spin-up component $n^\uparrow(0)$, spin-down component $n^\downarrow(0)$, and the total distribution function $n(0)$, respectively. The corresponding lines show the respective ideal Fermi distribution function. 
}
\end{figure}

The final investigation to be presented in this work is the dependence of the occupation at zero momentum on the spin-polarization with both the density and the temperature being kept constant. This is shown in Fig.~\ref{fig:xi_n0_rs2} for $\theta_{\xi=0}$ and three different values of the density parameter $r_s$. 
In particular, the top panel shows results for $r_s=2$ and the red circles, green stars, and black crosses correspond to the spin-up component $n^\uparrow(0)$, spin-down component $n^\downarrow(0)$, and the total momentum distribution $n(0)$. Most obviously, both the individual spin-up and spin-down components strongly depend on $\xi$, which is a direct consequence of the corresponding relative shift in the number densities $n^\uparrow$ and $n^\downarrow$. Furthermore, the total distribution $n(0)$ also noticeably depends on $\xi$ both for the ideal and the interacting case, and monotonically decreases with $\xi$. The physical origin of this effect is the increased impact of the Pauli blocking between electrons of the same species, which pushes the occupation towards large momenta.
Finally, we again find that the correlation-induced increase in $n(0)$ is nearly exclusively due to $n^\uparrow(0)$, as it is by now expected.

The center panel of Fig.~\ref{fig:xi_n0_rs2} shows the same information for a larger value of the coupling strength, $r_s=10$. Overall, the results are qualitatively quite similar to the $r_s=2$ case shown in the top panel, but the difference between $n(0)$ and $n_0(0)$ is significantly increased at $\xi=1$ as the spin-effects are effectively masked in the UEG by the Coulomb repulsion.

Lastly, the bottom panel shows results for strongly coupled electron liquid regime, $r_s=50$. In this case, the strong Coulomb coupling almost completely removes the dependence of the PIMC data for the total momentum distribution function on the spin-polarization, and the corresponding 
black stars are well reproduced by a constant fit (dotted grey line). In addition, both spin-components exhibit a decreased occupation at zero momentum compared to the ideal Fermi function over the entire $\xi$-range.

\begin{figure}\centering
\includegraphics[width=0.475\textwidth]{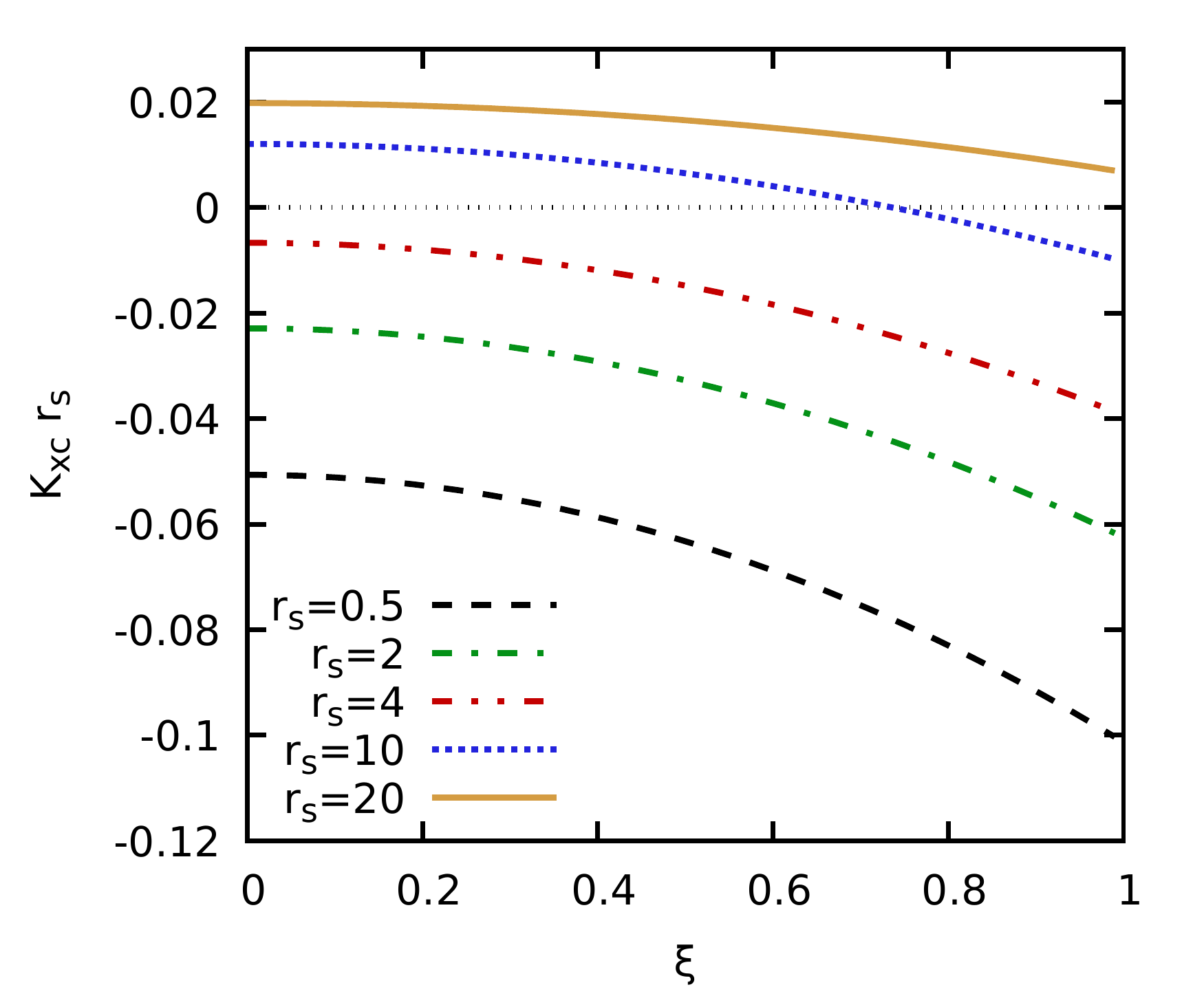}
\caption{\label{fig:xi_Kxc}
Polarization dependence of the exchange--correlation contribution to the kinetic energy $K_\textnormal{xc}$ at $\theta_{\xi=0}=2$ evaluated from the parametrization of $f_\textnormal{xc}$ by Groth \textit{et al.}~\cite{groth_prl} via Eq.~(\ref{eq:Kxc}). The different lines correspond to various values of the density parameter $r_s$.
}
\end{figure}

Let us conclude our investigation with an analysis of the $\xi$ dependence of the exchange--correlation kinetic energy $K_\textnormal{xc}$. To this end, we again compute $K_\textnormal{xc}$ via Eq.~(\ref{eq:Kxc}) from the accurate parametrization by Groth \emph{et al.}~\cite{groth_prl}, and the results are shown in Fig.~\ref{fig:xi_Kxc}. Overall, the relative spin dependence is substantial for all depicted values of the density parameter $r_s$ and is of the order of $100\%$. Even at $r_s=20$, where most physical observables of the UEG like the total energy $K$ only weakly depend on $\xi$, the strong $\xi$ dependence of $E_0$ is directly reflected in Eq.~(\ref{eq:Kxc_definition}).
In the WDM regime ($r_s=0.5,2,4$), we find substantially larger values for $K_\textnormal{xc}$ for the ferromagnetic case, which might be directly reflected in important material properties like the static local field correction~\cite{holas_limit,dynamic_folgepaper,dornheim_ML}. In particular, a pronounced \emph{negative tail} of the local field correction at intermediate to large wave numbers could have a noticeable impact on the spectral properties~\cite{dornheim_dynamic,Hamann_PRB_2020,Hamann_CPP_2020,dynamic_folgepaper,Dornheim_PRE_2020} of the system, with potentially important implications for X-ray Thomson scattering applications~\cite{siegfried_review}.

\section{Summary and Discussion\label{sec:summary}}

In summary, we have presented an extensive study of the impact of spin-effects on the momentum distribution function $n(\mathbf{k})$ and related properties. This has been achieved on the basis of extensive new direct PIMC simulations for different densities $r_s$, temperatures $\theta$, and degrees of the spin-polarization $\xi$. 

As a first step, we have considered the ferromagnetic UEG ($\xi=1$), and have found a substantially larger correlation-induced increase in the occupation of the zero-momentum state $n(\mathbf{0})$ compared to the paramagnetic case studied previously~\cite{dornheim2021ab}. At the Fermi temperature, this effect is present over the entire range of metallic densities and persists well into the electron-liquid regime, $r_s=20$. The physical origin of this effect can be understood as follows: in the ideal Fermi gas, spin-effects are more pronounced for large values of $\xi$, as they only manifest between electrons of the same species. For the UEG, on the other hand, spin-effects are increasingly suppressed by the Coulomb repulsion. Therefore, the UEG is more similar to an unpolarized ideal Fermi gas, which, in turn, leads to an increase in exchange--correlation properties such as $n(\mathbf{0})-n_0(\mathbf{0})$ with increasing $\xi$.

This first part of our investigation was concluded by a comparison of our new direct PIMC data to previous results using the restricted PIMC method by Militzer \emph{et al.}~\cite{Militzer_momentum_HEDP_2019}. In particular, we have found that the RPIMC data are afflicted with the same systematic error in the normalization that has been reported in our previous study of the momentum distribution of the unpolarized UEG~\cite{dornheim2021ab}. It is important to note that this factor is not due to the application of the fixed-node approximation, which was employed to mitigate the fermion sign problem, but is a direct consequence of the indirect determination of the normalization of $n(\mathbf{k})$ from the off-diagonal density matrix $n(\mathbf{r},\mathbf{r}')$. It is, thus, not an inherent feature of the RPIMC method and can be avoided by adopting the extended ensemble approach introduced in Ref.~\cite{dornheim2021ab}.

The second part of our work was devoted to the explicit analysis of spin-effects. To this end, we have performed direct PIMC simulations of the UEG for intermediate values of the spin-polarization $\xi$, beginning with an investigation of finite-size effects for two representative values of the density parameter, $r_s=10$ and $r_s=2$. More specifically, we have intentionally chosen the intermediate polarization $\xi=2/3$ (with $N^\uparrow=5N^\downarrow$), as this might result in different degrees of severity of the $N$ dependence within the two species of electrons. Remarkably, we have found that finite-size effects only manifest at small $k$ and even here hardly exceed $1\%$ for both $n^\uparrow(\mathbf{k})$ and $n^\downarrow(\mathbf{k})$, and for both considered $r_s$-values for as few as $N=18$ (i.e., $N^\uparrow=15$ and $N^\downarrow=3$) electrons.

Subsequently, we have presented new results for the spin-resolved momentum distribution $n^\sigma(\mathbf{k})$ for different values of $\xi$. First and foremost, we have found that the interaction-induced increase in $n(\mathbf{0})$ is mainly facilitated by the spin-up electrons (majority electrons). This is a direct consequence of the spin-resolved reduced parameters $r_s^\downarrow$ and $\theta^\downarrow$, which indicate a larger density parameter and higher reduced temperature for the spin-down electrons. This, in turn, leads to less pronounced spin-effects in $n_0^\downarrow(\mathbf{0})$, resulting in a smaller difference to $n^\downarrow(\mathbf{0})$ of the interacting UEG.
Extending this analysis to different absolute values of the temperature has given the expected trends, i.e., a stronger impact of spin-effects for lower temperatures, and the opposite trend in the high-temperature regime.

Finally, we have found that both the shift in the occupation of the zero-momentum state and the exchange--correlation part of the kinetic energy strongly depend on the spin-polarization even in the limit of the strongly correlated electron liquid ($r_s=50$). In fact, both the momentum distribution $n(\mathbf{k})$ and the full kinetic energy $K$ of the UEG are completely independent of $\xi$ in this regime due to the strong Coulomb repulsion. The large $\xi$ dependence of the corresponding exchange--correlation properties is thus directly caused by the large magnitude of spin-effects in the ideal Fermi gas at the same conditions.

Let us conclude this discussion by outlining a few directions for future investigations. Firstly, we mention that our extensive set of new direct PIMC data are freely available online~\cite{repo} and can be used as an accurate benchmark for the development of new methods and approximations, or as input for parametrizations. Furthermore, we re-iterate the high importance of the momentum distribution function of electrons for the description of transport properties of WDM in an external magnetic field~\cite{Haensel}. With respect to the UEG itself, the numerical investigation of the large-momentum tail of $n(\mathbf{k})$ [cf.~Eq.~(\ref{eq:g0_k10})] in the fully polarized case will further advance or current understanding of this fundamental model system, but remains out of reach for PIMC methods operating in coordinate space. In contrast, Hunger \emph{et al.}~\cite{Hunger_PRE_2021} have recently demonstrated that the configuration PIMC method (which directly operates in $\mathbf{k}$-space) is capable to resolve the required 8-10 orders of magnitude in $n(\mathbf{k})$, and its application to the spin-polarized case, thus, constitutes an enticing possibility. 
Finally, the substantially negative values of $K_\textnormal{xc}$ for large $\xi$ that have been reported in this work potentially indicate a pronounced negative tail for large wave numbers in the static local field correction of the ferromagnetic UEG. Yet, previous PIMC studies of such linear-response properties have been limited to $\xi=0$, and an extension of these efforts to other values of $\xi$ is highly desirable.

$ $

\section*{Acknowledgments}

This work was partly funded by the Center of Advanced Systems Understanding (CASUS) which is financed by Germany's Federal Ministry of Education and Research (BMBF) and by the Saxon Ministry for Science, Culture and Tourism (SMWK) with tax funds on the basis of the budget approved by the Saxon State Parliament.
The PIMC calculations were carried out at the Norddeutscher Verbund f\"ur Hoch- und H\"ochstleistungsrechnen (HLRN) under grant shp00026, and on a Bull Cluster at the Center for Information Services and High Performace Computing (ZIH) at Technische Universit\"at Dresden.

\bibliography{bibliography}

\begin{thebibliography}{116}
\bibitem [{rep()}]{repo}%
  \BibitemOpen
  \href@noop {} {}\bibinfo {note} {A link to a repository containing all PIMC
  raw data will be made available upon publication.}\BibitemShut {Stop}%
\end{thebibliography}
\end{document}